# Graft Copolymers of Maleic Anhydride and Its Isostructural Analogues: High Performance Engineering Materials

Zakir M. O. Rzayev

*Institute of Science & Engineering, Division of Nanotechnology & Nanomedicine,*
*Hacettepe University, Beytepe 06800 Ankara, Turkey*
*E-mail: zmo@hacettepe.edu.tr*

**Abstract** – This review summarizes the main advances published over the last 15 years outlining the different methods of grafting, including reactive extruder systems, surface modification, grafting and graft copolymerization of synthetic and natural polymers with maleic anhydride and its isostructural analogues such as maleimides and maleates, and anhydrides, esters and imides of citraconic and itaconic acids, derivatives of fumaric acid, etc. Special attention is spared to the grafting of conventional and non-conventional synthetic and natural polymers, including biodegradable polymers, mechanism of grafting and graft copolymerization, *in situ* grafting reactions in melt by reactive extrusion systems, in solutions and solid state (photo- and plasma-induced graftings), and H-bonding effect in the reactive blend processing. The structural phenomena, unique properties and application areas of these copolymers and their various modifications and composites as high performance engineering materials have been also described.

**Contents**

## 1. Introduction

Maleic anhydride (MA) and its isostructural analogues (maleic, fumaric, citraconic and itaconic acids and their amide, imide, ester and nitril derivatives) as polyfunctional monomers are being widely used in the synthesis of reactive macromolecules with linear, hyperbranched and self-assembled structures to prepare high performance engineering, bioengineering and nano engineering materials. Functional copolymers of these monomers with donor-acceptor type organic and organometallic (Si, Sn, Ge, B etc.), cycloolefin and heterocyclic (O, N, S and etc.) comonomers are synthesized by the radical copolymerization [1-26], complex-radical alternating copolymerization [27-40], terpolymerization [41-67], cyclocopolymerization [68-79], photopolymerization [80-90], interlamellar co-(ter)polymerization [91-97] and controlled/living radical copolymerization such as nitroxy-mediated [98-103], ATRP [104-111] and RAFT [112-119] methods. These monomers are also succesively utilized for the graft modification of various thermoplastic polymers (polyolefins, polystyrene, polyamides, etc.), biodegradable polymers, polysaccharides, natural and synthetic rubber, biopolymers, etc. Introduction of MA on the non-polar backbone of polyolefins and rubbers has overcome the disadvantage of low surface energy of these polymers, improving their surface hydrophilicity for the benefit of printing and coating applications, and adhesion with polar polymers (polyamides), metal, and glass fibers [1,2].

In the last decade, grafting of MA onto various thermoplastic polymers (predominantly polyolefins) and preparation of high performance engineering materials and nanocomposites by using reactive extruder systems and *in situ* compatibilization of polymer blends have been significantly developed, some results of which are employed in commercial applications. Fenouillot et al. [120] described the fundamental aspect of the reactive processing of thermoplastic polymers, including polymer graft grafting and/or functionalization using MA and vinylsilanes, bulk polymerization of urethane, lactams, acrylate and ε-caprolactone and new copolymer synthesis. This review also covers the state of the art in domains of rheology (specifically modelling of rheo-kinetics), diffusion and mixing viscous reactive media.

Among the chemical modification methods used in reactive extrusion system, free radical grafting of reactive polyfunctional monomers involving reaction of a polymer with monomers (grafting reaction) or a mixture of monomers (graft copolymerization) are probably most important. One of the most common monomers in the polymer modification is MA and its isostructural analogues such as *N*-substituted maleimides, fumaric, citraconic and itaconic acids and their esters, amides, imides, and anhydrides of these dicarboxylic acids. Chemical structure of these monomers, which can be used in grafting and graft copolymerization reactions with synthetic and natural polymers, is represented in Figure 1.

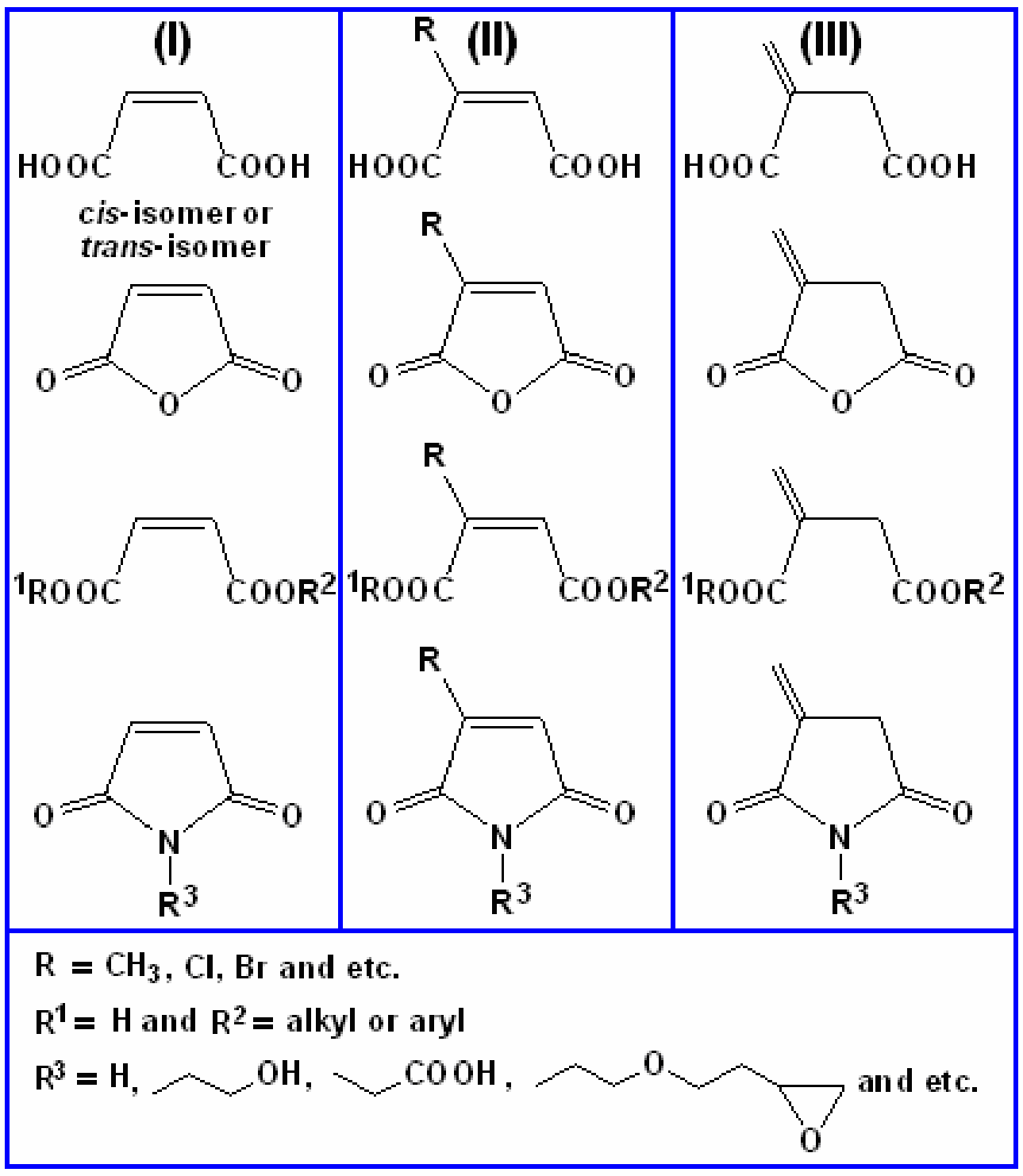

**Figure 1.** Chemical structures of functional monomers as effective grafting agents for the modification of synthetic and natural polymers.

Modification of conventional polymers by grafting and graft-(co)polymerization techniques has received much academic and practical interest. This method allows to imparting a variety of functional groups to a polymer. Bhattacharya and Misra [121] have documented graft copolymerization reactions initiated by chemical treatment, photo-irradiation, high-energy radiation technique, etc. as a versatile means to modify polymers. According to authors, there are several means to modify polymer properties such as blending, grafting, and curing. Among these methods of modification of polymers, grafting and graft copolymerization are one of the most promising methods.

In this review, different methods of grafting, including reactive extrusion systems, grafting and graft copolymerization of synthetic and natural polymers with MA and its isostructural analogues are described. Special attention is spared to the grafting of conventional and non-conventional synthetic and natural polymers, including biodegradable polymers, mechanism of grafting and graft copolymerization, *in situ* grafting reactions, and usage of MA alternating, random and graft copolymers as compatibilizers in the reactive polymer blends, various composites, as well as structure, unique properties and application areas of these copolymers as high performance engineering materials.

## 2. Grafting of Polyolefins

History of graft modification of polyolefin, especially polypropylene (PP) with MA dates back to the 1969s [122] when a method was developed for reacting MA on a particular isotactic PP below its melting point. Since then studies have included the maleation of both isotactic and atactic PP under a variety of conditions. This reaction has been achieved in melt processes where the molten polymer is mixed with MA and with a peroxide initiator, either in an extruder or in an internal mixer at an elevated temperature. Alternatively, for the best understanding of mechanism grafting process, solution conditions have been used, where the polymer is dissolved in a suitable solvent at an appropriate temperature and MA is added with an initiator. Finally, the anhydride functionalization of PP has been achieved by a solid phase grafting process. Since the 1960s, dibutyl maleate and acrylic acid have been grafted onto polyolefins in screw extruders [123]. Ide and Hasegawa [124] have reported grafting of MA onto isotactic PP in the molten state using benzoyl peroxide as an initiator and a Brabender plastograph. This graft copolymer was used in blends of polyamide-6 and PP as a reactive compatibilizer [125]. Grafting of MA onto low density polyethylene (PE) backbone in the presence of dicumyl peroxide in the melt in a bath reactor have been studied by Swiger and Mango [126]. They also prepared a reactive blend of MA grafted PE with polyamide 66 in a twin-screw extruder. Cha and White [127] have reported MA modification of polyolefins in an internal mixer (batch reactor) and a twin-extruder. They have presented a detailed kinetic model for grafting processes in a reactor and twin-screw extruder systems.

For the evaluation of grafting mechanism, functionalization of PP with MA have been carried out both in solution [125,128-132] and in the molten state [128,129,133-137] using various extruder systems. Taking into the consideration low costs and operating facility, such grafting reactions were preferably carried out in the melt via reactive processing.

In fact, MA and its isostructural analogoues, such as fumaric, citraconic anhydrides, are strong hydrophilic monomers. If unsaturated dicarboxylic acids and their anhydrides are grafted onto polymers they will carry a denser distribution of carbonyl or free carboxylic groups. These reactive groups can also serve as sites for further macromolecular reactions of copolymers and grafted polymers, especially for compatibilization of immicsible polymers and preparation of various reactive blends with higher engineering performance and controlled morphology and mechanical properties.

## 2.1. Grafting in solution

The grafting of PP with MA in xylene solution, using benzoyl peroxide as the initiator, has been reported by Ide et al. [124]. Little evidence of degradation of polymer product was found. The effect of solvent type and amount, catalyst type and amount, and the effect of initiator concentration on the MA grafting of PP were studied by Rengarajan et al. [138] It was shown that all of these factors had a significant effect on the grafting degree of the PP. Borsig and Hrckova [139] later compared the level of functionalization of isotactic PP using both solid phase and solution method. They found no significant effect on the differences for the grafting efficiency between the two methods. They also studied the reaction in solution but with atactic PP. The focus was on the influence of the separate components of the reaction system on the degradation of the PP. It was found that the grafting reaction of MA onto PP was accompanied by reactions, leading to degradation or reactions leading to an increase in molecular mass.

MA grafting of isotactic PP in a solution process and evaluation of the effects of monomer and initiator concentration, reaction time, and temperature on percentage grafting were investigated by Sathe et al. [140]. Grafting of MA onto thermoplastic elastomer, such as styrene–(ethylene-butylene)–styrene (SEBS) triblock copolymer, was carried out in xylene solution in the presence of dicumyl peroxide as an initiator. Authors identified the reaction products using liquid chromatography (LC), IR and $^{13}$C NMR. Side products from the graft reaction were analyzed by the LC. They found that xylene affected the graft reaction through its active methyl groups. Reaction mechanisms also studied by performing free radical kinetics analysis. According to the authors, a proper choice of the solvent might favor graft efficiency better. Sipos et al. [141] investigated the kinetics of grafting of MA to various hydrocarbon substrates (eicosane, squalene and PE) in the pure hydrocarbons and in 1,2-dichlorobenzene solution using 2,5-dimethyl-2,5-di(*t*-butylperoxy)-3-hexyne as an initiator (half life of about 1 h at a typical reaction temperature of 150$^{o}$C). The obtained results authors interpreted in terms of a chain mechanism, including a slow propagation step in which a succinic anhydride radical abstracts a hydrogen atom from the same or different chains. In this work, The same general mechanism was proposed for grafting of MA onto PE and the hydrocarbons in 1,2-dichlorobenzene solution.

Marconi et al. [142] esterificated (grafted) a commercial poly(ethylene-*co*-vinyl alcohol) with given monomer unit composition (40:60) with maleic, succinic and glutaric anhydrides in anhydrous DMF solution at 55 $^{o}$C under nitrogen flow for 48 h. Grafting degree was determined by titration of the side-chain acidic groups (–OOC-CH=CH-COOH, –OOC-$CH_2CH_2$-COOH and –OOC-$CH_2CH_2CH_2$-COOH), $^{1}$H NMR spectroscopy and elemental analysis. Structure of prepared terpolymers was confirmed by FTIR spectroscopy. The influence of these different types of carboxy derivatives on the biological activity of the polymer was also evaluated. According to the authors, introduction of ionic groups into the polymer backbone, which are able to increase its hydrophilicity, as well as the interfacial mobility of polymer surface, thus promote a favorable and selective adsorption of the plasma proteins. It was shown that these type of polymers exert an anti-adhesive action towards the blood platelets due to electrostatic repulsion between the polymer acidic groups and the negatively charged external platelet membrane [143].

Functionalization of isotactic PP with dimethyl itaconate (DMI) as functional polar monomer using 2,5-dimethyl-2,5-bis(*tert*-butylperoxy)hexane as a radical initiator was carried out in both boiling xylene and decalin as solvent media [144]. The effect of DMI and the initiator concentration on the extent of grafting was studied by varying reaction time and temperature. It was found that temperature affects the content of DMI grafted onto PP, which is slightly higher for reactions carried out in xylene than in decalin. The results also show that the amount of DMI incorporated is proportional to the initial DMI and initiator concentrations used in the grafting reaction up to certain concentrations, and thereafter a decrease in the content of grafting (in wt.-%) was found. The maximum value of grafting obtained was 0.7 %. The melt flow index (MFI) values increase with increasing initial amount of initiator used in the grafting reaction. The degradation of the PP chain is higher when xylene is used as solvent. MFI values of 20-100 were found for modified PP compared with 11.4 found for the unmodified polymer [145,146].

In the past years, as one of the most effective methods, solvothermal method has been widely used to prepare many kinds of new materials [147], including grafted thermoplastic elastomers [148,149]. In this process, the solvents are sealed in vessel (bomb, autoclave, etc.) and can be brought to temperatures well above their boiling points by the increase of aurtogenous pressures resulting from heating. The solvothermal method was successfully used by Qi et al. [149] for the preparation of MA grafted poly(acrylonitrile-*co*-butadiene-*co*-styrene)s, poly(ABS-*g*-MA)s. Authors also studied the effects of reaction time, temperature, MA, initiator and terpolymer concentrations, comonomer (styrene) and different solvents on grafting degree. The grafting reactions were performed in a sealed vessel with 1,2-dichloroethane as solvent and benzoyl peroxide as initiator at 120$^o$C under nitrogen atmosphere. By IR (presence of 1780 cm$^{-1}$ C=O stretching) and $^1$H NMR (new peak at 3.736 ppm for the anhydride protons) spectroscopy, it was confirmed that MA is successefully grafted onto ABS backbone. Authors demonstrated that to prepare the graft copolymers through solvothermal method grafting reaction can be carried out both in good solvent (1,2-dichloroethane) and poor (acetone or ethanol) solvent, which is much different from traditional solution grafting method, and a high grafting degree (~ 4.2 wt.%) can be obtained in 1,2-dichloroethane as a good solvent.

Some studies on enhancing grafting efficiency have involved the use of mixed monomer systems; in particular, the synergistic effects of comonomers may lead to more efficient grafting processes. This strategy involves choosing a monomer combination in which the comonomer is effective in trapping the radical formed on the ABS backbone and the resultant growing radicals are highly reactive toward the desired sites [150]. As a comonomer, styrene effectively improved the grafting reaction of MA onto ABS terpolymer [149]. The effect of binary system of MA and styrene with various compositions on the grafting onto ABS was studied in different solvents. It was found that the grafting degree increased with the increasing of styrene content, and reached a maximum (7.9 wt.%) at a monomer feed molar ratio of 0.42 in 1,2-dichloroethane. Then, it decreased with further increasing of styrene content. The maximum values of grafting degree were 2.1 and 1.0 wt.% in acetone and ethanol, respectively. According to the authors, observed effect showed that the elevated temperatures and aurtogenous pressures resulting from heating in sealed autoclave are favourable to increase solubility and reactivity of polymers, so the grafting reaction proceeding via graft copolymerization reaction still occured when poor solvents were used. This result showed that a synergistic effect, i.e., an activation of MA through acceptor-donor interaction (complexation) with styrene (Rzayev's comment), on the grafting process of styrene and MA binary mixture onto ABS occured, and the styrene content in the binary monomer mixtures had a considerable influence on the grafting degree in the grafting reaction [149].

In fact, anhydrides of unsaturated dicarboxylic acids such as maleic, citraconic and itaconic acids, are strong hydrophilic monomers. These reactive groups can serve as sites for further functionalization of grafted polymers. The synthesis of the graft copolymers of PP (powder and granular) with acrylic acid (AA), maleic and citraconic (CA) anhydrides, and poly[PP-g-(MA-*alt*-AA)] using grafting in solution and reactive extrusion techniques and their main characteristics, as well as results of the structure–composition–property relationship have been reported by Rzayev et al. [151-156]. Functionalization of isotactic polypropylene (*i*-PP) with CA and MA was carried out in 1,2,4-trichlorobenzene (TCB) solution with dicumyl peroxide (DCP) as an initiator at 160$^o$C under nitrogen atmosphere. Chemical and physical structures and thermal behavior of the synthesized graft copolymers with different anhydride units were determined by volumetric titration (acid number), FTIR and $^1$H-NMR spectroscopy, X-ray powder diffraction (XRD), DSC and TGA thermal analyses. It was shown that the crystallinity and thermal behavior of grafted *i*-PP's depend on anhydride unit concentration in grafted *i*-PP; grafting reaction proceeds selectively which is not accompanied by oligomerization of CA and degradation of the main chain as in known maleic anhydride/PP system. This fact was explained by inhibition effect of α-methyl group in CA grafted unit onto the chain by β-scission reactions and no homopolymerization of CA in the chosen grafting conditions. The effect of concentration on the anhydride content and grafting efficiency was also investigated. In this study the grafted monomer content varies in the range of 0.01-0.56 mol %. As a general behavior, the grafted monomer content first increases with the comonomer content in the feed, reaches a maximum value and then decreases. The maximum grafting is achieved at 7.5 wt. % for CA monomer content in the feed. It was shown that the glass-transition temperature ($T_g$) and crystalline properties of grafted *i*-PP were different from non-grafted *i*-PP. In addition, the melting point ($T_m$) of the functionalized *i*-PP's changed little. The structure, macrotacticity, crystallinity, crystallization and thermal behavior of synthesized *i*-PP grafts depend on the monomer unit concentration in polymers. Grafting reaction with CA proceeds more selectively than those for MA. That can be explained by relatively low electron-acceptor properties of citraconic double bond due to a $CH_3$ group in the α-position, as well as by steric effect of this group [156].

As a general behavior, the anhydride content initially increases with the monomer content in the feed, reaches a maximum value and then decreases. For instance, for CA grafting, the maximum grafting is achieved at 7.5% (weight monomer/weight polymer) monomer content in the feed. When the CA content in the feed is low, there are enough initial radicals to combine with CA molecules and to initiate *i*-PP macroradicals. Therefore, the grafting degree would increase with increasing CA content in the feed. But with a further increase in the initial CA content, more and more radicals would be consumed in formation of radicals from the thermal decomposition (homolytic scission) of the initiator molecules, which can combine with CA monomers or *i*-PP backbone chains. The number of initial radicals to induce above considered reactions would then decrease consistently. Therefore, the grafting degree of *i*-PP would decrease and a maximum value for the grafting degree of *i*-PP appears. Thermal oxidation may also be a possible cause of degradation of the *i*-PP chains; however, in our case all the reactions were conducted under pure nitrogen atmosphere, therefore this side reaction could be neglected. Grafting reaction proceeds selectively which is not accompanied by oligomerization of CA and degradation of the main chain [156] as in known maleic anhydride/PP system [151,154]. CA homopolymerization is additionally tried to perform under the same experimental conditions, however CA graft homopolymerization does not take place.

Although reactivity of CA radicals is lower than MA, grafting reaction of CA with *i*-PP is more selective than in the case of MA grafting reactions. This inhibition effect of chain scission of the α-methyl groups may be due to the formation of quaternary carbon atom in CA grafted linkage as shown in Figure 2 [45,46].

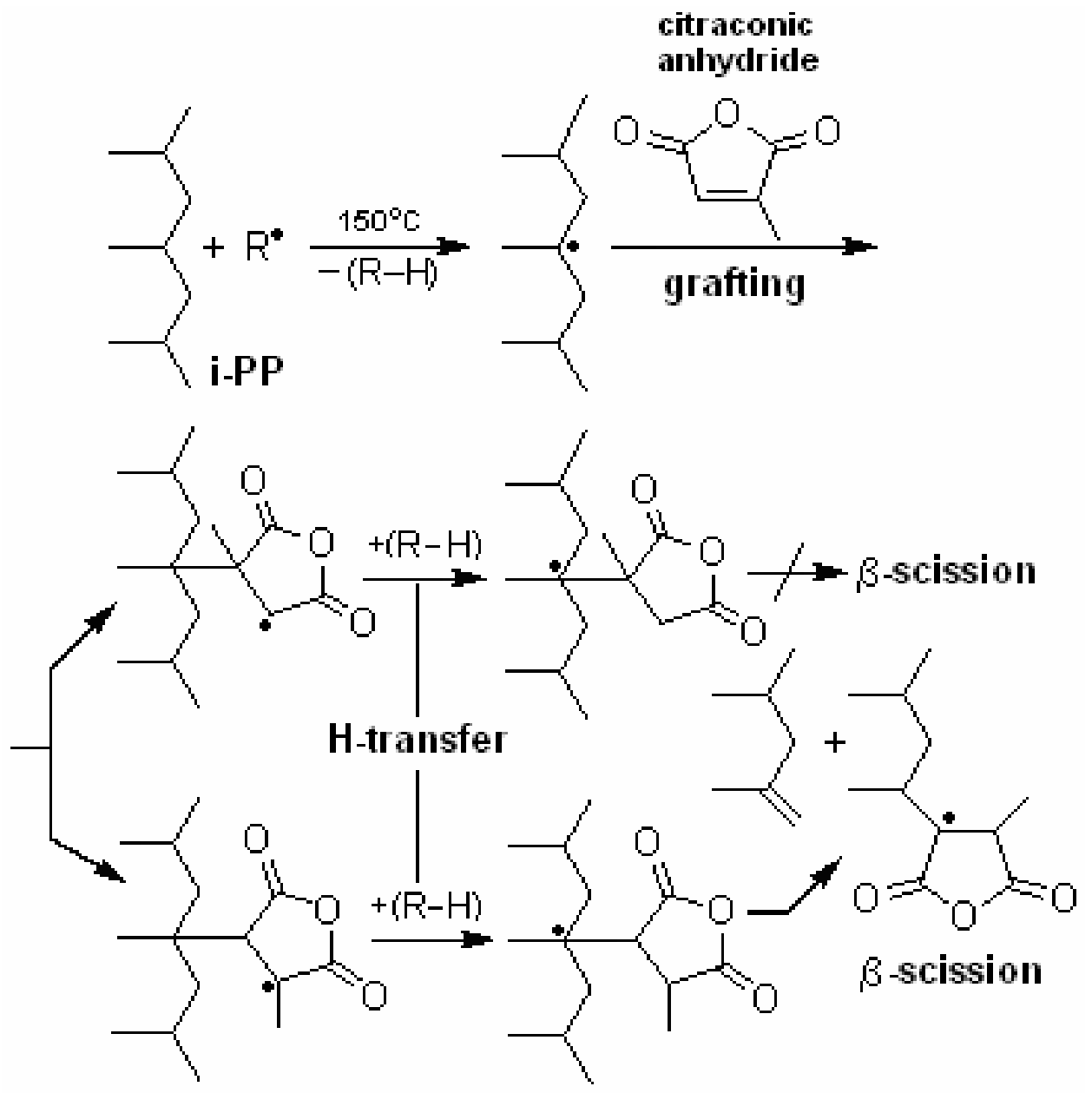

**Figure 2.** Proposed mechanism of grafting reactions of CA with i-PP through hemolytic chain degradation by β-scission.

Grafted anhydride linkages are characterized by FTIR and $^1$H NMR (400 MHz) spectroscopy [156]. The visible change of intensity of stretching and deformation bands for CH, $CH_2$ and $CH_3$ groups, which are dependent on the grafting degree, is also observed. The comparative analysis of spectra virgin *i*-PP and grafted polymers with different compositions indicated that the increasing in intensity of the characteristic bands at 2961, 2923, 2874 and 2840 cm$^{-1}$ (C–H stretching in CH, $CH_2$ and $CH_3$ groups), 1470 and 1380 ($CH_2$ and $CH_3$ deformations) and 1162, 998, 974, 898 and 808 cm$^{-1}$, which are associated with various vibrations of CH–$CH_3$ group conformation (998 cm$^{-1}$ helix band characterizes the fraction of isotactic spirals in *i*-PP structure and causes the interaction of $CH_3$ and $CH_2$ groups; the intensity of this band usually used as a criterion of regularity of *i*-PP macrospirals; 974 cm$^{-1}$ band assignment to isotactic form of macromolecules intensity of which visibly increases in increasing the content of grafted carboxylic fragments in PP.

**Table 1.** Tacticity, thermal ($T_g$ , $T_m$, $T_c$ and $T_d$) and crystallization ($T_c$) parameters of grafted *i*-PPs

| Grafted PPs | Grafted monomer unit (%) | Tacticity* (%) $A^{998}/A^{974}$ | DSC-TGA analysis (°C) $T_g$ | $T_m$ | $T_c$ | $T_d$ |
|---|---|---|---|---|---|---|
| PP-*g*-CA | 0.45 | 97.6 | 58.2 | 158 | 117 | 401 |
| PP-*g*-MA | 0.56 | 97.2 | 16.4 | 162 | 118 | 402 |
| PP | 0.00 | 95.0 | -28.3 | 166 | 115 | 379 |

*These values were determined by FTIR method.

The ratio of absorbance values for these two bands ($A^{998}/A^{974}$) are used for the determination of tacticity of the grafted *i*-PP's, the results of which are presented in Table 1. This is a well known method for the determination of tacticity of PP [127]. This observed effect also depends on the type of grafted monomer and grafting reaction. The properties of grafted *i*-PP in an engineering application are critically dependent on the extent of crystallinity and the nature of crystalline morphology of *i*-PP. The effects of the composition on the thermal behavior, crystallization parameters, and crystallinity of synthesized functional *i*-PPs were determined with DSC, differential thermo-gravimetry, and XRD analysis. The effect of variation of grafted monomer content on crystallization temperature ($T_c$) and melting temperature ($T_m$) is given in Tables 1 and 2.

DSC thermograms showed sharp exo- and endo-therms during cooling and heating, respectively. The grafted *i*-PP has a lower $T_m$ than the ungrafted PP. This may be due to the grafting, which destroys the ordered structure of *i*-PP crystals. The chain degradation of *i*-PP during grafting may also result in a reduction of $T_m$. The $\Delta H_m$ results show that the crystallinity of grafted PP is higher than that of virgin isotactic PP.

**Table 2.** Effects of monomer unit content on melting temperature and crystallinity ($\chi_c$) of the grafted *i*-PP's

| Monomer content in feed (wt %) | AN* | Grafted unit contents (mol %) | $T_m$ (°C) | $\chi_c$ (%) by DSC |
|---|---|---|---|---|
| *i*-PP-*g*-CA | | | | |
| 1 | 0.14 | 0.01 | 163.2 | 37.4 |
| 2.5 | 1.12 | 0.10 | 162.9 | 40.7 |
| 5 | 2.46 | 0.22 | 161.9 | 44.2 |
| 7.5 | 5.05 | 0.45 | 158.4 | 49.0 |
| 10 | 1.68 | 0.15 | 162.5 | 43.3 |
| *i*-PP-*g*-MA | | | | |
| 1 | 1.40 | 0.13 | 161.3 | 33.5 |
| 2.5 | 5.61 | 0.50 | 158.5 | 43.1 |
| 5 | 6.31 | 0.56 | 162.2 | 45.0 |
| 7.5 | 4.21 | 0.38 | 163.3 | 42.7 |
| 10 | 2.10 | 0.19 | 163.0 | 37.3 |
| Virgin *i*-PP | 0.00 | 0.00 | 166.0 | 32.4 |

*Acid number (mg KOH/g).

The degree of crystallinity of the sample was determined by measuring the enthalpic change. The crystallinity values depend on carboxylic acid contents of the grafted *i*-PPs. The observed differences between crystallinity values for all the grafted copolymers are related to carbonyl group contents of monomers on *i*-PP backbone. It is known that the functionalization caused an increase in the polarity of the medium, intensifying the interaction forces between the grafted *i*-PP molecules, and then increasing the polymer crystallinity [157,158]. The crystallization temperature was recorded from cooling scans of the samples. Maximum crystallization temperature of 122.3°C was obtained with MA grafted *i*-PP. Higher crystallinity and higher intensity in the cooling crystallization peak was observed for poly (*i*-PP-*g*-CA) in comparison with *i*-PP.

The results obtained by TGA analysis indicate that the degradation reaction does not occur under the reaction conditions. The thermal stability was slightly increased by the presence of grafted CA and MA units. The initial decomposition temperature and overall degradation values of all the grafted copolymers changed very little compared to *i*-PP [156].

## 2.2. Grafting in melt by reactive extrusion

The modification of themoplastics in twin-screw extruder systems is achieved to produce new materials is an inexpensive and rapid way to obtain new commercially valuable polymers. Twin-extruders act as continuous flow reactors for polymers playing an increasingly important role in producing high performance thermoplastics [127,130]. It was noted that this reactive extrusion technology is an increasingly important method of producing sizable quantities of modified polymers for various industrial applications. Twin-screw extruders as continuous flow reactors for polymer processing are playing an increasingly important role in producing high performance engineering thermolastics.

The most investigations were related with MA grafting of PP in the melt using various types of extruders and mixers. Thus, the MA functionalization of PP in the melt phase was studied by Gaylord and Mishra [132] and Ho et al. [146] and mechanism for the grafting was proposed. Their reactions were carried out in Brabender Plasticorder and their conclutions were that degradation of the polymer chain is, to a certain extent, a result of the disproportionation of polymer radicals generated by propagating MA homopolymer or graft copolymer. The effect of MA and peroxide initiator concentrations on the amount of grafts and on the degradation was determined using a twin screw extruder. It was showed that chain scission of PP is suppressed at high initial concentration of MA. Heinen et al. [159] used $^{13}C$ NMR in combination with specific labels in the MA monomer to determine the locations of the grafts formed from the radically induced grafting of MA onto high- and low-density PEs, isotactic PP and E-P copolymers. Obtained results showed grafting does not occur along the polymer chain. On the high- and low-density PEs and isotactic PP the MA was attached in the form of single succinic anhydride rings as well as short oligomers [160,161]. They suggested that formation of oligomeric graft structures in maleated PP, as described by Gaylord and Mishra [132], is of minor importance for high-temperature PP grafting and in these conditions, β-scission of PP chain does not occur [159].

The influence of residence time on degree of MA grafting onto PP in an internal mixer and a twin-screw extruder was studied by Cha and White [127] through measuring reaction yieds with respect to reaction time in the internal mixer as well as along the screw axis in the extruder using various peroxide type initiators. The degree of MA grafting in both systems increases with an increase of peroxide concentration. On the other hand, the degree of grafting with respect to initial monomer concentration exhibits a maximum reaction yield value at approximately 4 % of intial MA concentration in both used systems. Comparison of the degree of MA grafting in these sytems shows that the former is slightly higher than that of the twin-screw extruder because of air contact of the reaction medium. The grafting reaction in the twin-screw extruder is largely accomplished in the first half of the total extruder. It was also observed that the complex viscosity was slightly increased with an increase of MA grafting concentration while dramatically decreased by addition of peroxide, which plays a greater role in decreasing viscosity than shear effect.

Bettini and Agnelli [136,137] have investigated the effects of the monomer and initiator concentrations, rotor speed and reaction time on the grafting reaction of MA onto PP. Reactive processing was performed in a Haake torque rheometer at 180 $^oC$ under nitrogen atmosphere. They obtained the level of reacted MA by FTIR spectroscopy using 1790 $cm^{-1}$ (C=O anhydride unit) and 1167 $cm^{-1}$ ($CH_3$ in propylene unit) as the analytical bands; the Carbonyl Index (*CI*) was calculated using the ratio of absorbance of these bands:

$$CI = [A_{1790}(C{=}O)]/[A_{1167}(CH_3)] \qquad (1)$$

Authors also studied the effect of the rotor speed and reaction time on the extent of degradation by means of melt-flow index (MFI) measurements. It was shown that effect of these operational parameters on the reacted MA content and on the MFI depends on the levels of MA and peroxide initiator [46.5 % concentration of 2,5-dimethyl-2,5-di(*t*-butylperoxy) hexane in $CaCO_3$] concentration, which could not be analyzed separately. According to authors, the increase in rotor speed leads to better mixing of MA in the reaction mass, an increase in the production of macroradicals and sublimation of MA monomer, due to the increase in reaction temperature [136,137].

A reactive extrusion process for the functionalization (free radical grafting) of PP with MA in the presence of supercritical carbon dioxide was studied by Dorscht and Tzoganakis [161]. Carbon dioxide was used in this system to reduce the viscosity of the PP melt phase by forming a polymer-gas solution in order to promote better mixing of the reactants. Subsequently, the effect of supercritical carbon dioxide on the level of grafting, product homogeneity, and molecular weight was evaluated. It was observed that the use of carbon dioxide led to improved grafting when high levels of MA were used. Chang and White [162] used a range of continuous mixing machines (such as nonintermeshing modular counterrotating, intermeshing modular corotating and intermeshing modular counterrotating twin-screw extruders, and Kobelco Nex-T continuous mixer) as continuous reactors for grafting MA onto PP. They investigated the grafting of MA onto PP and degradation of PP during grafting reaction, as well as the influence of the extrusion parameters on the polymer characteristics as means for comparing these different machines for reactive extrusion. It was demonstrated that the Kobelco Nex-T continuous mixer had highest level of grafted MA (0.32 g/100g sample) as compared to other used extruder systems. The level of grafted MA was determined by both the alkali titration and infrared method. The intensity of the carbonyl absorption peak at 1785 $cm^{-1}$ divided by that of the $CH_3$ absorption peak at 1165 $cm^{-1}$, was defined as the carbonyl index, and used as a measure of the amount of grafted MA.

Dong and Liu [163] also studied the styrene-assisted free-radical graft copolymerization of MA onto PP in supercritical $CO_2$. It was shown that the addition of styrene drastically increased the MA functionality degree, which reached a maximum when the molar ratio of MA and styrene was 1:1. Styrene, an electron-donor monomer, could interact with MA through charge-transfer complex (CTC) to form the styrene–MA copolymer, which could then react with PP macroradicals to produce branches by termination between radicals. The highest MA functionality degree was obtained when the concentration of AIBN initiator was 0.6 wt % based on PP. An increase in the temperature increased the diffusion of monomers and radicals in the disperse reaction system of supercritical $CO_2$. The highest degree of grafting was observed at 80$^oC$. Degree of grafting decreased with an increase in the pressure of supercritical carbon dioxide within the experimental range. The morphologies of pure PP and grafted PP were significantly different under polarizing optical microscope. The PP spherulites were about 38 μm in size, and the grafted spherulites were significantly reduced because of heterogeneous nucleation.

Li et al. [164] found that the addition of styrene as a second monomer in the melt grafting system assisted in increasing the grafting degree of MA on PP. The styrene comonomer serves as a medium to bridge the gap between the PP macroradicals and the MA monomer. They have proposed that, in the aforementioned system, styrene preferentially reacts with the PP macroradicals to form more stable styrene macroradicals, which then copolymerize with MA to form branches. As noted by the authors, supercritical graft copolymerization is an advantageous process compared with known conventional methods (melt-grafting technology, solution graft copolymerization, etc.): (1) used $CO_2$ as a medium is nontoxic, nonflammable, and inexpensive; (2) reaction temperature can be controlled (60-90 $^oC$) and is much lower than that of melt-grafting technology; (3) supercritical $CO_2$ causes considerable polymer swelling and significant depression of the polymer glass-transition temperature; (4) mass transfer in a supercritical medium is very high; (5) the solvency of the medium can be controlled through changes in the density of the reaction medium through temperature and pressure profiling; (6) application of supercritical $CO_2$ fluid provides significant economic and social benefits for environmental protection and energy conservation.

The grafting reaction of MA with commercial polymers such as polyethylene (PE) [165-170], ethylene-propylene (E–P) copolymer [159], polypropylene PP [166,167-169] and polystyrene (PS) [150] in the presence of peroxide initiators were investigated extensively. MA has been grafted onto polyolefins especially in order to improve their compatibility with polar thermoplastic polymers such as polyamides and polyesters, and to promote their adhesion to glass fibers in polymer composites [169]. Melt functionalization of polyolefins (PO) [166,171] through MA grafting [166,171-173] was the topic of recent publications, too.

In several publications [160,174-176], the melt grafting of dibutyl maleate onto E–P copolymer has been investigated in detail. In most of these studies, the possibility of the addition of anhydride units on the polymer backbone, accompanied mainly by crosslinking and degradation of macromolecules has also been determined. The peroxide-initiated grafting of MA is accompanied by crosslinking for PEs [173-175], while degradation through chain scission is the dominant effect in the case of molten PP [168,175,176]. To the extent of saturation of E–P copolymers, both crosslinking and degradation reactions should occur in maleation in the presence of peroxide initiators [168,177].

The molecular characterization of MA melt-functionalized PP, poly(PP-*g*-MA), the grafting mechanism, the nature, the concentration, and location of grafted anhydride species onto the PP chain were described by Roover et al. in detail [167,168]. The PP functionalization was performed using a pre-heated Brabender plastograph (194$^{o}$C, 4 min of mixing time). It was demonstrated that the organic peroxide used in these experimental conditions undergoes an homolytic rupture and carries out a PP ternary hydrogen abstraction.. Resulting macroradical undergoes a β-scission leading to a radical chain end which reacts with MA. When a termination reaction occurs at this first step a succinic type anhydride chain end is obtained. However, oligomerization of MA was found to occur more frequently leading to poly(MA) chain end. Concentration of both anhydride units and minimal length of the grafted poly(MA) were determined. It was found that a fraction of MA does not react with PP, thus, homopolymerization leading to non-grafted poly(MA). However, it can be noted that at high temperature conditions free or grafted oligo(MA) fragments can easily undergo the decarboxylation reaction with formation of conjugated system [167].

The functionalization of PE with MA through a free radical reaction in both solution and melt processes have been widely studied [177-180]. It has been generally accepted that primary free radicals generated by initiators first abstract hydrogen atoms from the PE backbone to form macroradicals, after which the macroradicals undergo grafting with MA or cross-linking between themselves [181]. Heinen et al. [159] synthesized [2,3-$^{13}C_2$] MA-*g*-PE both in solution and in melt systems, including high-density PE (HDPE) and low-density PE (LDPE). From noise-docoupled and $^{1}$D inadequate $^{13}$C NMR spectroscopy, they found that MA groups grafted onto HDPE and LDPE are in the form of single succinic anhydride (SA) and sort of oligomeric SA, the average chain length of which is about 1-2 when the grafting is carried out in MA with 10 wt. % PE. It was found that the sites of attachment and the structures of the grafting fragments depend on (co)polymer composition. In E–P random copolymer, MA attaches to methylene and methine carbons in the backbone. In alternating E–P copolymer, MA attaches solely to polymer chain methine groups, indicating that $(CH_2)_n$ sequences with $n > 3$ are needed for MA attachment to backbone methylene carbons. In the copolymers and in isotactic PP, grafts are single succinic anhydride rings; in HDPE and LDPE short MA oligomers are also present. It was assumed that the chain scission can yield structures in which the anhydride ring is attached to the unsaturated end chain group through a fully substituted double bond.

Yang et al. [182] described the molecular structure of poly(MA-*g*-PE), which was synthesized in solution, as resolved with solution state NMR and FTIR spectroscopy. They demonstrated for the first time the formation of the fewer SA oligomeric grafts with a terminal unsaturated MA ring (oligo-MA) in addition to the more SA oligomeric grafts terminating in a saturated SA ring (oligo-SA). The only structure observed prior to this study was [2,3-$^{13}C_2$] MA grafted onto low molecular weight PE ($M_w$ = 4000 and $M_n$ = 1700). A grafting mechanism with two different termination processes, (a) dispropotionation between a grafting radical and a macroradical of a secondary carbon in low-molecular weight PE or another grafting radical as well as (2) the hydrogen abstraction of a grafting radical from a secondary carbon in this PE, is described to explain the formation of oligo-MA and oligo-SA.

Periodical and patent publications on the synthesis of polyolefin graft copolymers including polyolefins with grafted MA, maleates, fumarates, maleimides, etc. by reactive extrusion or other forms of melt phase processing were reviewed by Moad [150]. Primary concerns are the structures of the graft copolymers formed, the mechanism of grafting reactions and their relationship to processing conditions, as well as methods for characterizing modified polyolefins. Shi et al. [183] studied chemical structure and molecular parameters of grafted materials of poly(PP-*g*-MA) prepared by melt reactive extrusion by using electrospray ionization-mass spectrometer (EI-MS) and gel permeation chromatogrphy (GPC). It was found that the initial radicals, due to homolytic scission of dicumyl peroxide could be combined with MA monomer as well as isotactic PP molecular chains. Authors believed that the homopolymerization of MA cannot occur and the MA radicals undergo dismutational reaction under the processing condition (180-190 $^{o}$C). Authors tentatively proposed the mechanism of melt grafting MA onto PP as follows: (1) the initial radicals that were formed by initiator homolytic scission can combine with MA monomer as well as with isotactic PP molecular chain; (2) the homopolymerization of MA does not take place under the conditions used in this reactive extrusion process; (3) the MA radicals could be dismutationally terminated or acted as a chain transfer agent; (4) not all the ternary macroradicals can be involved in the β-scission that leads to the degradation of the polymer, and not all the secondary macroradicals can be involved in combination that leads to the increase of the molecular weight; (5) different structures were formed during melt reactive grafting of PP with MA. According to authors, the oxidation effect could be neglected during the processing.

Zhu et al. [184] investigated the grafting of MA onto isotactic PP initiated by dicumyl peroxide at higher temperature (190 °C) by means of Monte Carlo method. The ceiling temperature theory, i.e., no possibility for the homopolymerization of MA to occur at higher temperatures, was used in this study. The simulation results show that most MA monomers were grafted onto the radical ends arising from β-scission at a lower concentration of MA, whereas the amount of MA monomers attached to the ternary carbons was much larger than that of the grafted onto the radical chain ends at a higher MA concentration for various initiator concentrations. The variations of the grafting degree of MA and number-average molecular weight of PP with reaction time for 2 wt. % initiator concentration, respectively, have shown that the grafting degree of MA increases considerably with increasing reaction time up to about 100 s; thereafter, it almost remains unchanged with increasing reaction time. Obtained results are in agreement with the experimental results of Jaehyug and James [185], in which the grafting degree of MA reached a plateau value after 2 min for melt grafting of MA onto PP. On the other hand, the $M_n$ of PP decreases markedly with increasing reaction time up to about 100 s. Atactic PP grafted with 4.5 % MA exhibits a tensile strength of 1.91 MPa, elongation of 400 %, and hardness of 50-55 % (against for ungrafted PP of 0.53 MPa, 110 % and 10-15, respectively) [186,187].

Zhang et al. [188] assumed that the blend compatibilizing, oil dispersion, and rheological properties of poly[(E-*co*-P)-*g*-MA] can be expected to depend on the graft microstructure. A detailed knowledge of the nature of the grafted units whether isolated anhydride units or oligoanhydride fragments of clustered anhydride units (Figure 3) is quite important for improving the properties of this MA grafted copolymer. Grafting of single saturated succinic anhydride (SA) (micro-structure I) [188] is usually reported by many researchers [150,189-191]. However, it has also been shown that oligomeric or polymeric MA grafts can be formed as well (microstructure II) [136,178,192,193]. Microstructure III has been proposed, where MA is attached to the backbone as both individual and clustered units [190].

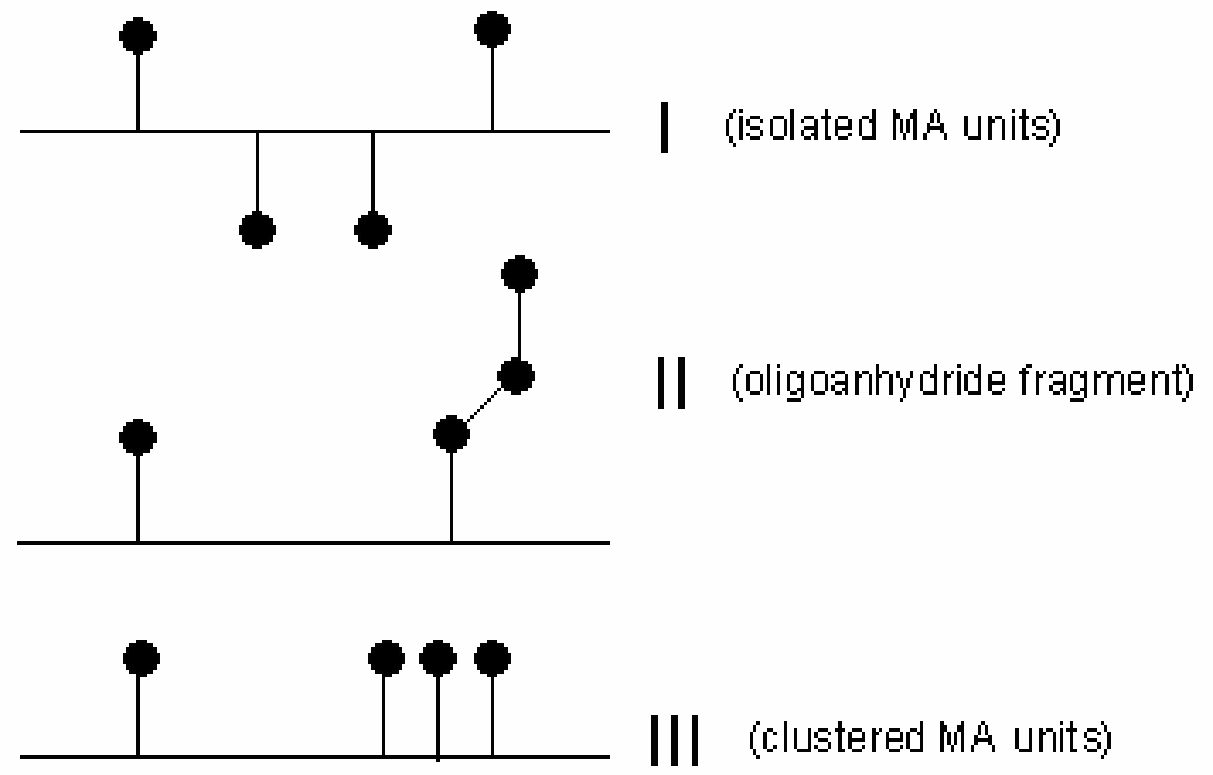

**Figure 3.** Schematic representation of graft microstructures in poly[(E-co-P)-g-MA]. Adapted from [188].

Vangani et al. [194,195] have shown that fluorescence experiments have the potential to yield the level of SA clustering of poly[(E-*co*-P)-*g*-MA]. This study requires attaching the pyrene dye onto the SA moieties of a poly[(E-*co*-P)-*g*-MA]. Upon absorption of a photon, the encounter of an excited pyrene with a ground-state pyrene leads to the formation of an excited complex called an excimer. The time scale of excimer formation depends on whether the two pyrenes are attached onto SA grafted units which are clustered or isolated.

An ethylene (E)–propylene (P) random copolymer was maleated according to two different procedures which were expected to yield different distributions of SA unit along the backbone [188]. The two maleated EP copolymers, referred to as EP-I and EP-II, were labeled with fluorescent dyes such as pyrenes and/or naphthalene by reacting 1-pyrenemethylamine and/or 1-pyrenebutanoic acid hydrazine (or 1-naphthalenemethyl amine) with the SA moieties randomly attached along the EP copolymer Authors detected the presence of MA attached onto the EP copolymer by monitoring the absorption band at 1778 $cm^{-1}$ characteristic of the C=O anhydride unit in the FTIR spectra. After EP-*g*-MA copolymer had reacted with 1-pyrenemethylamine (or other fluorescents) the 1778 $cm^{-1}$ band disappeared and a new peak appeared at 1710 $cm^{-1}$ characteristic of the presence of succinimide carbonyls. Hexane and THF were chosen as solvents. Pyrene excimer formation and fluorescence resonance energy transfer measurements were performed to demonstrate that the SA units are attached to EP-I in a less clustered manner than to EP-II (Figure 4).

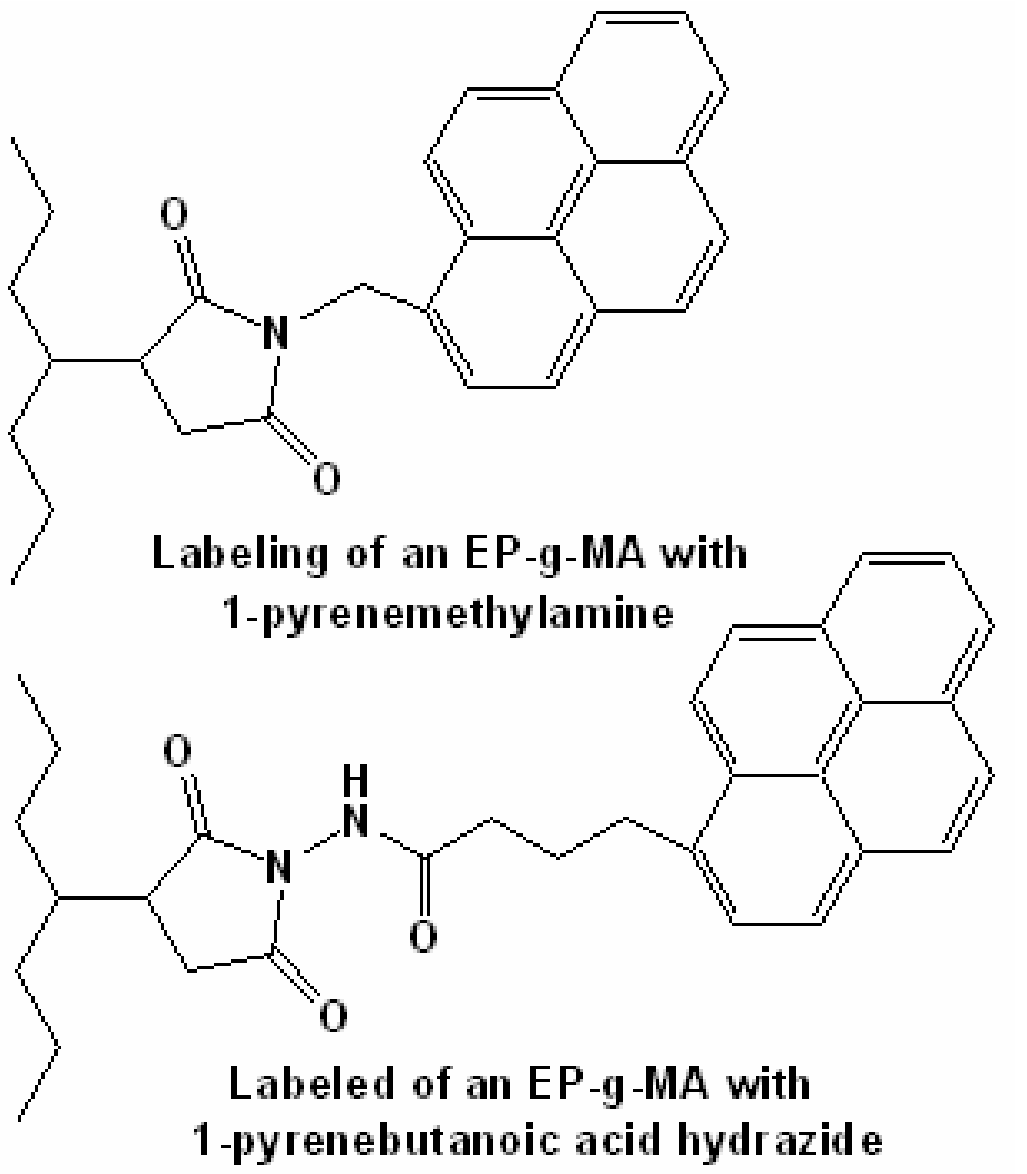

**Figure 4.** Chemical structure of pyrene-labeled derivatives of EP-g-MA graft copolymers. Adapted from [188].

Viscosity measurements were performed on both pyrene-labeled EP copolymers to evaluate the effect of SA clustering on the solution properties of the polymers in hexane. The solution viscosity of the pyrene-labeled EP-I increases less steeply with polymer concentration than that of the pyrene-labeled EP-II. Authors concluded that fluorescence can be applied to investigate the microstructure of maleated EP copolymers and that SA clustering of a maleated EP copolymer can affect its properties in nonpolar solvents [188].

Hong and Chen [196] surface grafted a thermoplastic olefin elastomer with MA/radical initiator (benzophenone, benzoyl peroxide or AIBN) mixture under ozone and studied adhesion and surface properties of the grafted polymer. Ozone was used to initiate the grafting reaction, and a procedure different from the conventional ozone treating procedures was applied. The surface of elastomer was first dip-coated with a MA/radical initiator mixture and then exposed to ozone to initiate the surface grafting reaction of MA. In the conventional ozone treating method, the functionalized surface grafting agents are applied after the polymer surface are exposed to ozone. Hence the grafting efficiency could be easily affected because the active radicals on the ozone treated surface are liable to be contaminated and then deactivated before the surface grafting process. The surfaces of modified polymer samples were characterized by FTIR-ART surface analysis, contact angle measurement, and lap shear strength measurement.

The modification of PP through free-radical grafting of itaconic acid onto LDPE by reactive extrusion [197-202] have also been reported. Many aspects of the grafting mechanism of itaconic acid (IA) onto polyolefin macromolecules have been investigated by Jurkowski et al. [198-202]. The most important formulation and technological factors have been determined and optimized so as to control the course of the grafting process, the grafting efficiency, the course of the by-process including crosslinking of PE macromolecules and the poly(PE-*g*-IA) physical structure. It is known that the chemical neutralization of carboxylic groups grafted onto polyolefin macromolecules greatly influences their compatibilizing activity in blends of thermoplastics and increases thermal stability and, as a result, the properties of the modified polyolefins [150,203]. Therefore, grafting of IA onto LDPE was performed by reactive extrusion in the presence of DCP initiator and metal oxide and hydroxides [ZnO, $Zn(OH)_2$, MgO and $Mg(OH)_2$] as the neutralizing agents (NAs) [202]. Free carboxylic groups of IA units were neutralized in molten LDPE directly in the course of acid grafting, and in prefabricated functionalized PE, i.e., poly(PE-*g*-IA) while the –COOH groups were neutralized partially or totally through chemical reactions. Authors have demonstrated that NAs, added into the initial reaction mixture improved the grafting efficiency of IA onto LDPE, and prepared poly(PE-*g*-$IA^-$ $M^+$)s exhibit higher thermal stability and resistance to thermal oxidation compared with that of poly(PE-*g*-IA).

Reactive blends of poly(PE-*g*-IA) with polyamide 6 (PA6) were chosen to show that IA grafted onto PE improves adhesion of interphases in the blends, increases the impact strength of the materials, and improves their processability [200]. A comparative study of the structure and properties of two-phase blends of PA6 and LDPE modified in the course of reactive extrusion, by grafting of İA without neutralization of carboxylic groups [poly(PE-*g*-IA)] and with neutralized carboxylic groups [poly(PE-*g*-$IA^-$ $M^+$)] was carried out. It was shown that 30 wt.% of poly(PE-*g*-$IA^-$ $M^+$) introduced to PA6 resulted in blends of higher Charpy impact strength compared with that of PA6/poly(PE-*g*-IA) blends. The maximum increase was achieved when $Mg(OH)_2$ was used as a neutralizing agent [201]. The structure and properties of PP/LDPE blends grafted with IA in the melt by reactive extrusion have also been studied [202]. The data of DSC and relaxation spectrometry indicated the incompatibility of PP and LDPE in the poly[PP/LDPE)-*g*-IA] systems on the level of crystalline phases; however, favorable interactions were observed within the amorphous phases of the polymers. Variations in the ratios of the polymers in the poly[PP/LDPE)-*g*-IA] systems led to both nonadditive and complex changes in the viscoelastic properties as well as mechanical characteristics for the composites.

MA grafted PE is of considerable importance for application as a copolymer precursor in polymer blends, and has received intense attention in the past decade. Monte Carlo simulation was used to study the graft of MA onto linear PE poly(PE-*g*-MA) initiated by dicumyl peroxide [204]. Simulation results revealed that major MA monomers were attracted onto PE chains as branched graft at higher MA content. However, at extremely low MA content, the fraction of bridged graft was very close to that of branched graft. This conclusion was somewhat different from the well known viewpoint, namely, the fraction of bridged graft was always much lower than that of branched graft under any conditions. Moreover, the results indicated that the grafting degree increased almost linearly to MA and dicumyl peroxide concentrations. On the other hand, it was found that the amount of grafted MA dropped sharply with increasing the length of grafted MA, indicating that MA monomers were mainly attracted onto the PE chain as single MA units or very short oligomers. Furthermore, the results suggested that the fraction of PE-$(MA)_n$-PE crosslink increased continuously with increasing MA content, whereas the fraction of PE-PE crosslink decreased correspondingly with increasing MA content.

The functionalization of ethylene polymers with diethylmaleate (DEM) initiated by dicumyl peroxide has been reported by Passaglis et al. [205]. The degree of functionalization has been determined by FTIR and NMR spectroscopy and through comparison with the mixtures of the same polyethylene and poly(diethyl- fumarate). A good linear correlation was obtained between the ratio $A_1/A_2$ (where $A_1$ and $A_2$ are the integrated areas of the vibrational bands at 1738 and 1460 $cm^{-1}$ for C=O and $CH_2$ groups, respectively) and the content of grafted 2-(diethylsuccinate) groups. The data obtained show that in all cases, the functionalization degree is appreciably affected by the amount and ratio of monomer and initiator used, thus allowing one to devise experiments aimed at controlled chemical modifications and to propose a reasonable reaction mechanism. In general, the grafting reaction occurs by the addition of the terminal macroradical, formed through β-scission, to the monomer double bond. PE [121], diethyl maleate (DEM) or a mixture of DEM/MA (2:1 molar ratio) with dicumyl peroxide (1:1) yielded PP with good functionalization degree, ranging from 0.2 to 1.0 mol %, but with a remarkable decrement of PP molecular weight [200]. The competition of the projected grafting reaction against scission can be improved via the formation of stabilized macroradicals. Romani et al. [206] have studied the chemical cross-linking of PP performed under dynamic conditions using a peroxide and a furan or bis(maleimide based coagent, such as bis(2-furanyl) aldazine or *N,N-p*-phenylene-bis(maleimide), as the crosslinking promoter [173,207]. These coagents are able to react properly with the macroradicals, produced by the primary radicals, thus spacing the free radical from the backbone of PP and promoting chain extension and/or preventing the β-scission.

Ciardelli et al. [208] studied the control of degradation reactions during radical functionalization of isotactic PP with MA in the melt by using furan derivative, butyl-3-(2-furanyl)propenoate (BFA). This coagent, successfully used for *i*-PP cross-linking without any remarkable increase of MFR, was used as a free radical remover and MA as a functionalizing monomer. The results indicate a detectable improvement with respect to the use of MA and peroxide only, allowing a significant grafting of functional groups and only partial degradation. Authors reperesented the proposed general mechanism of free radical functionalization of PP in the melt as follows (Schemes 5):

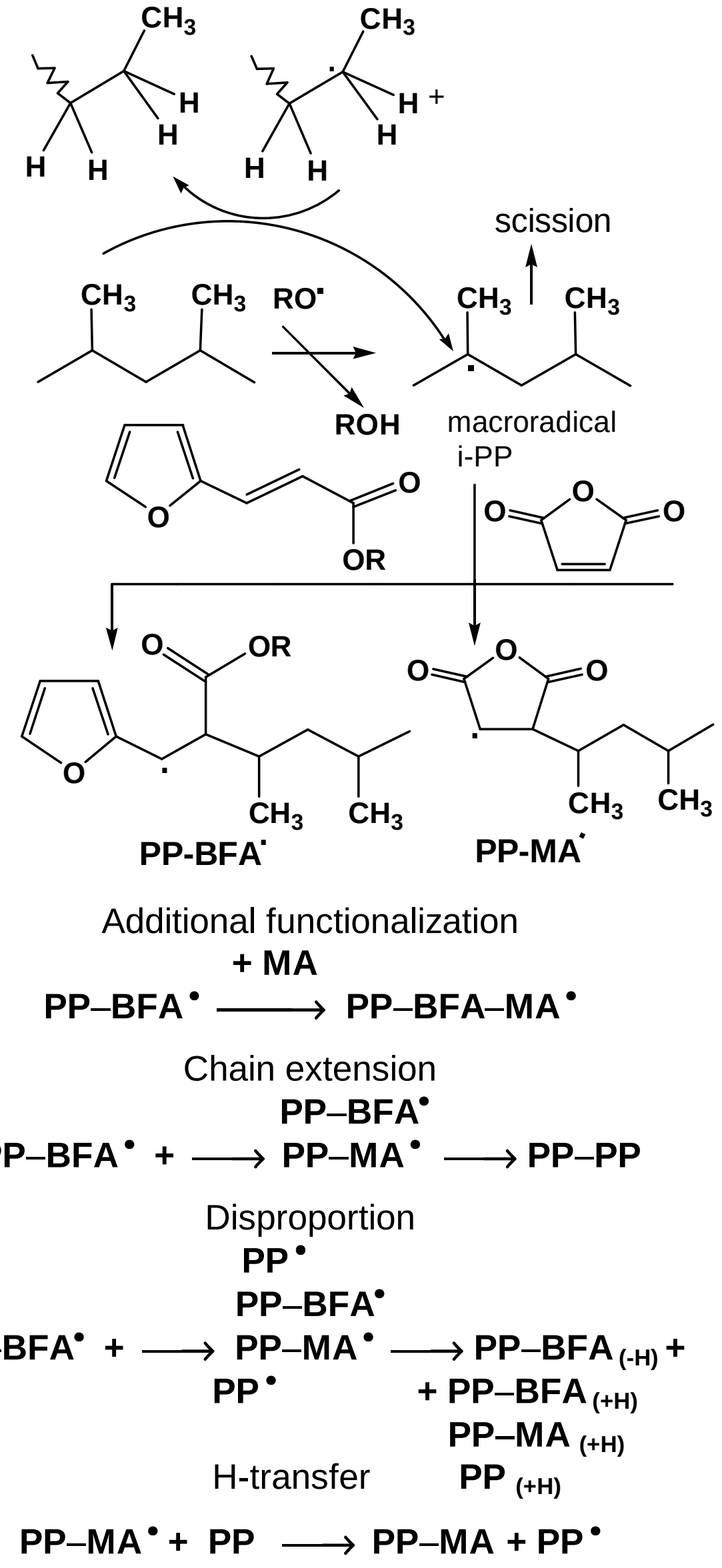

**Figure 5.** General mechanism of free radical functionalization of PP with furan derivatives in the melt. Adapted from [208].

The grafting of functional antioxidants in polymer melts, particularly polyolefins, is one of the most important approaches [209,210]. Thus, melt free-radical grafting of maleimide and newly designed monomeric antioxidants, maleimides with hindered phenol groups onto PE reported by Kim et al. [211,212]. These monomer-antioxidants with carbamate (1) and ester (2) linker functionality (Figure 6) were synthesized [212] and grafted onto PE via melt-processing with free-radical initiators (dicumyl peroxide and AIBN or BP) in a minimax molder at 120-180°C for 4-20 min. Authors used IR spectroscopy for the quantitative determination of the extent of grafting with the monomeric antioxidants. The peaks at 2025 and 1720 $cm^{-1}$ from PE and grafted monomers (C=O), respectively, were chosen as analytical bands, the integral ratios of which were used as a convenient measure of the degree of grafting. They observed that the grafted PE possessed some stabilizing effect against thermal oxidation with oven aging in air at 120°C for 5 days, with no intensity change indicated in the IR peak at 1720 cm$^{-1}$. Grafting was accompanied by partial crosslinking, the extent of which was determined from the gel content. To optimize the reaction conditions, authors studied the effects of the monomer and initiator concentrations, reaction time, and temperature on the extent of grafting [213].

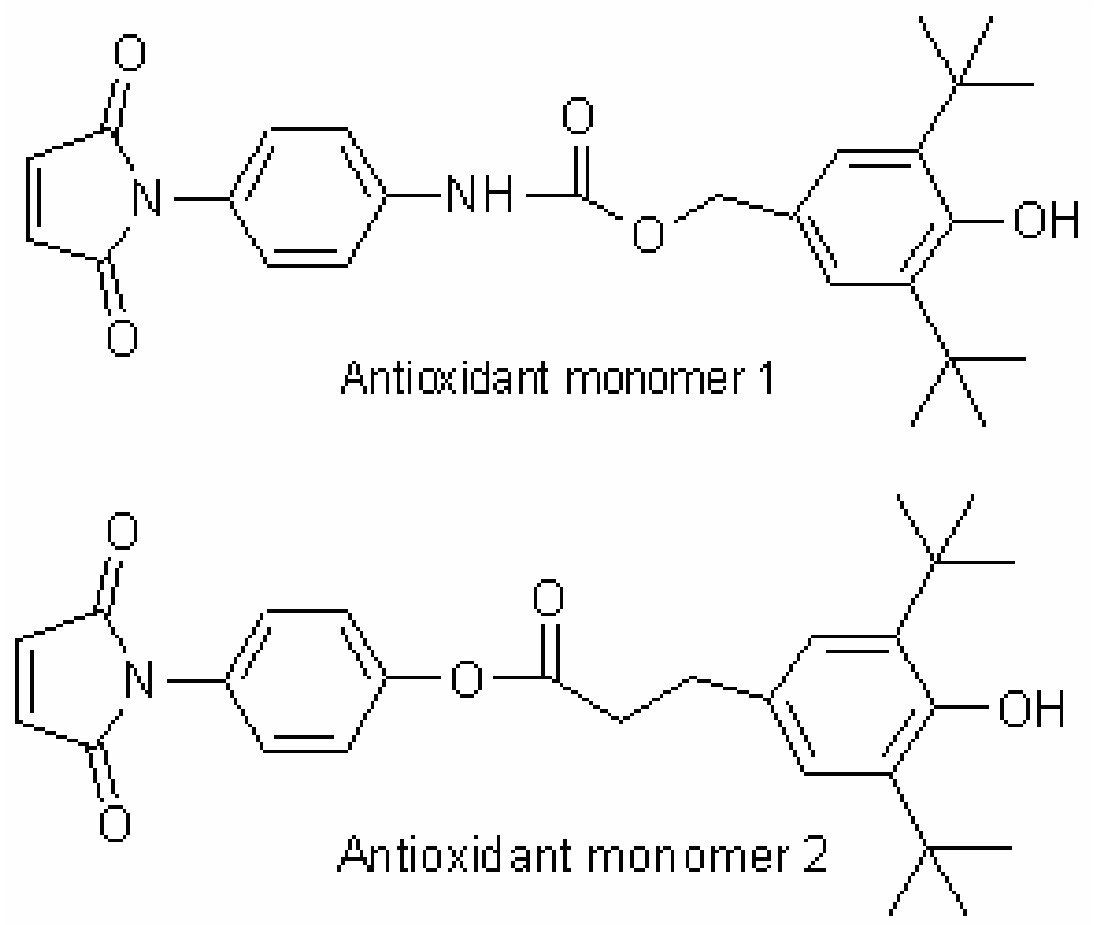

**Figure 6.** Chemical structure of monomer-antioxidants grafted onto PE by melt-processing. Adapted from [212].

The performance of adhesion and compatibilization of polyolefins functionalized with MA can be expected to depend on the microstructure of the graft copolymer and detailed knowledge of it is important for understanding structure-property relationships for these graft copolymers [150,209]. In polyolefins several structurally different radical sites can be produced by the action of radical initiators. Predominantly, secondary and tertiary macroradicals have different reactivities toward MA. It is generally accepted that cross-linking in PE and chain scission in PP may occur simultaneously with the graft reaction. E–P copolymers may undergo both side reactions. Besides the exact site of attachment, an additional item that has been speculated upon is the structure of the MA graft. A single, saturated succinic anhydride graft is usually suggested [1,214,215], but unsaturated [178], oligomeric [178,216], polymeric grafts [169], and combination of single and oligomeric grafts [132] have also been proposed.

Poly(MA-*g*-PP) was usually synthesized by chemical modification of preformed PP under free radical conditions [192,217]. Due to the inert nature of the PP structure and poor control of the free radical reaction, this MA grafting reaction includes many undesirable side reactions [218], such as β-scission, chain transfer, and coupling. The MA incorporation is usually inversely proportional to the resulting polymer molecular weight. In addition to the MA units incorporated along the polymer side chain, it has been generally suggested that a significant portion of poly(MA-*g*-PP) have a succinic anhydride group located at the polymer chain end [158,180] resulting from β-scission. In general, the inherent complexity of poly(MA-*g*-PP) molecular structure has significantly limited the understanding of its structure-property relationship, especially the ability of poly(MA-*g*-PP) as an interfacial agent in PP blends and composites.

A new synthetic route for preparing poly(MA-*g*-PP), which involves the borane terminated PP and a free radical graft-form reaction with MA reported by Priola et al. [219] and Ruggeri et al. [220]. It was shown that the resulting poly(MA-*g*-PP) containing a chain end terminated MA group was an effective compatibilizer in a PP/polyamide blend. Synthesis a new class of MA modified PP, having a relatively well-defined molecular structure was also described by Russell [217]. The chemistry involves the borane-terminated PP inter- mediate that prepared by hydroboration of the chain-end unsaturated PP. According to thje authors [220], the borane group at the chain end can be selectively oxidized by oxygen to form a "stable" polymer radical which then in situ react with MA to produce MA terminated PP (PP-*t*-MA) with a single MA unit. It was found that the polymeric radical initiated an alternating copolymerization of styrene and MA to produce PP-*b*-SMA diblock copolymers, and the incorporated MA units are proportional to the concentration of styrene and the reaction time. The resulting relatively well-defined PP–MA copolymers, with controllable PP molecular weight and MA concentration, evaluated systematically in the reactive PP/PP–MA/polyamide blends. It was established that the compatibility of these blend is dependent not only on the microstructure of the compatibilizer, formed in situ by reacting PP–MA with polyamide chains, but also on the composition of the PP/polyamide blend. According to authors, in polyamide-rich blends, all PP–MA copolymers show limited compatibility. Higher MA concentrations result in poor blend morphologies. Most of the compatibilizers formed fail to position themselves at the interfaces between the PP and polyamide domains. On the other hand, in PP-rich blends the in situ formed compatibilizers are generally very effective, especially those involving the PP-*b*-SMA with high PP molecular weight and multiple MA units. The MA concentration in the polymer estimated by FTIR spectroscopy with the following equation [220,221]:

$$\text{MA wt\%} = (k_1 A_{1780} + k_2 A_{1710})/d \qquad (2)$$

where $d$ is the film thickness, $A$ is the corresponding peak absorbances and $k_1$ and $k_2$ are the absorption constants for anhydride (1780 cm$^{-1}$) and acid (1710 cm$^{-1}$), respectively.

The method of synthesizing funtionalized polymers including MA grafting polyolefins by using reactive extrusion system increasingly finding favor as a procedure of new functional polymer synthesis [222,223]. Authors assumed that this method compared with alternative processes have the advantages: (1) no or little use solvents; (2) simple product isolation; (3) short reaction (grafting) time; (4) continuous process. The potential environmental and economic benefits of reactive extrusion process are such as to justify significant efforts aimed as resolving some difficulties relating to high reaction temperatures necessary to form the polymer melt and possibility of polymer degradation or polymer cross-linking accompanying processing. The patent and periodical publications in this area through 1990 were summarized and described in book edited by Xanthos [222].

Chemical modification of polyolefins with MA (so-called "maleation") is one of the most studied polyolefin modification processes yet many aspects of the process remain controversial. As a consequence, the process still attracts much attention in the periodical [167,168,223-228]. The maleation of conventional and metallocene linear low density polyethylenes (LLDPE) by reactive extrusion has been explored with a view to defining the conditions necessary for a robust process of both high grafting efficiencies (>80 %) and minimal degradation or cross-linking. The dependence of grafting efficiency on various operating parameters (MA level, MA/initiator ratio, throughput rate, direction of screw rotation, temperature) has been established [132]. A final graft level of 1.3-4.2 % MA was obtained at MA:initiator = 1:19 ratio and LLDPE throughput of 16 *kg/h*. A mechanism was suggested for the maleation of LLDPE which foresaw the following reaction directions: (1) formation of a macroradical from reaction initiator with polymer chain; (2) grafting of MA to macrochain with formation of $-MA^{\bullet}$ in the side chain; (3) transformation of $-MA^{\bullet}$ to succinic anhydride units via H-transfer (it should be noted that a general consensus implies that MA is grafted to polyolefins as single succinic anhydride units) [226]; (4) predominant oligomerization of $-MA^{\bullet}$ by the addition of free MA. It was shown that the structure of MA grafts may vary in terms of the fraction of grafts appearing as isolated succinic anhydride units vs. oligo(MA) chains, the length of any oligo(MA) grafts, and/or in the distribution of the grafts [217]. Applying FTIR analysis method for identification of both grafting fragments (1784 $cm^{-1}$ for the oligo-MA absorption band and 1792 $cm^{-1}$ for the single alkyl succinate band), authors suggested that the proportion of such grafts may be even higher, such that most grafts may be oligomeric. Analogous results were obtained by De Roover, *et al.* [168,224] in a systematic study of the IR spectra of various commercial and synthesized poly(MA-*g*-PP).

The MA grafted paraffin ($M_n$ = 785 g/mol, density 0.94 g/$cm^3$) was prepared by heating the wax-85/MA-10/benzoyl peroxide-5 mixture at 140 $^{o}$C for 10 min under nitrogen atmosphere [229]. The real concentration of grafted MA unit was 3.6 wt. %, which represented 36 % efficiency of grafting. For the maleated wax, the characteristic FTIR absorption bands were 1713 $cm^{-1}$ (C=O groups from carboxylic dimer acids) and 1790 $cm^{-1}$ (C=O of five-membered cyclic anhydride). Then authors studied the modification of the polarity and adhesive properties of LLDPE, LDPE and isotactic PP through MA grafted paraffin wax [230]. It was shown that MA grafted oligomer significantly increased the polar component of the total surface free energy of POs. Modified POs also had significantly higher adhesion to the polar substrate, a crosslinked, epoxy-based resin. The conservation of the good mechanical properties of the blends was observed up to 10 % grafted wax, except for PP blends, for which there was a reduction in the stress and strain at break at MA-*g*-wax concentrations higher than 5 %.

Hagiwara et al. [231] reported the original method of functionalization of the thelechelic oligopropylenes. They were synthesized from MA-containing oligopropylenes by the 'ene' reaction of MA with terminal vinyl or vinylidene groups on the polymer end (Figure 7).

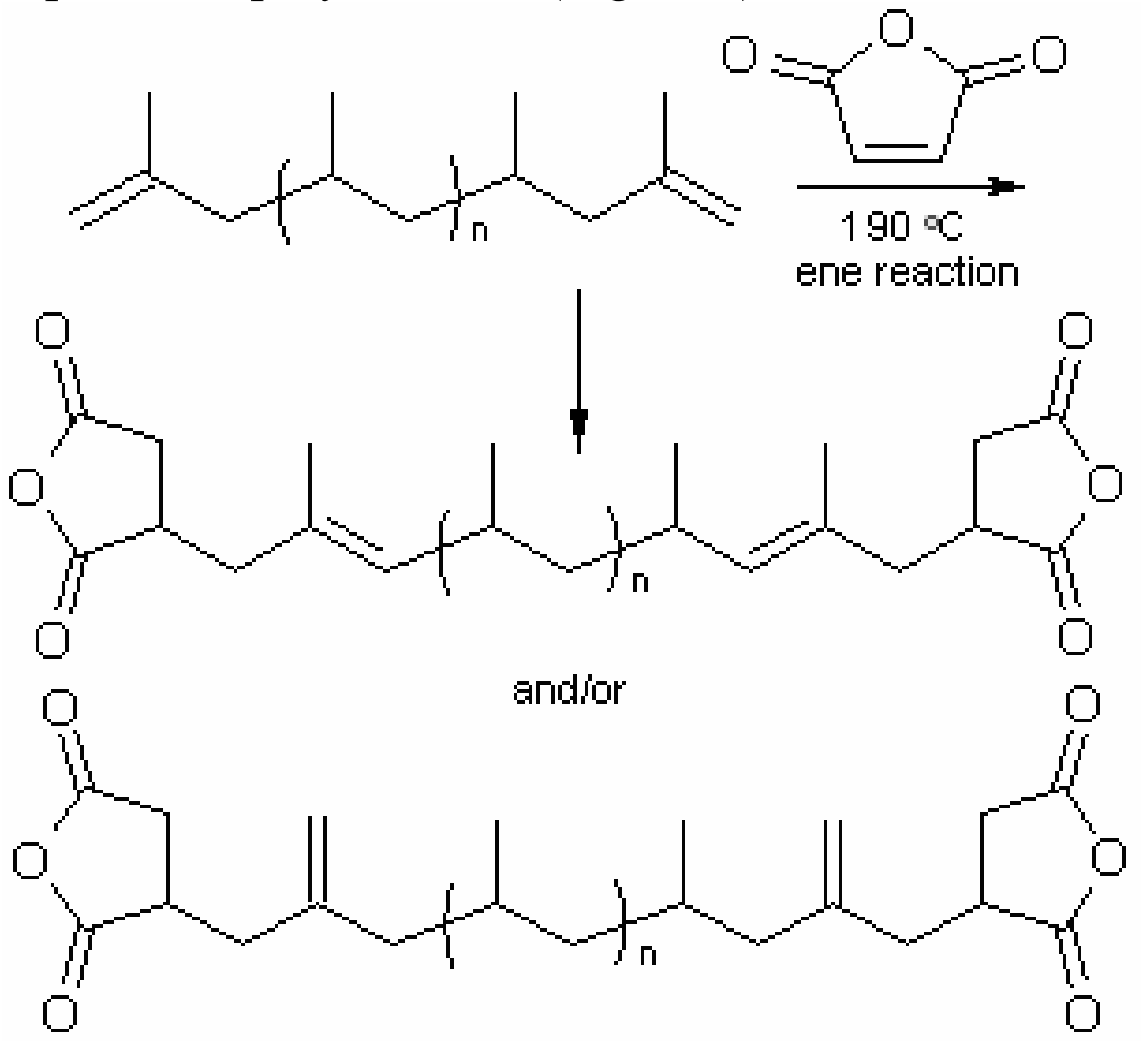

**Figure 7.** Schematic representation of MA functionalization of thelechelic oligopropylenes. Adapted from [231].

Multiblock copolyolefins, which have dissociable bonds in the main chain, prepared by the esterificastion polycondensation equimolar mixture of i-PP-MA ($M_n$ = 2400 g/mol, $M_w/M_n$ = 1.5) with telechelic oligomers such as i-PP-OH ($M_n$ = 2100 g/mol, $M_w/M_n$ = 1.8) in the presence of catalytic amount of p-toluenesulfonic acid [232].

Liu and Baker [229] also analyzed structure of poly(MA-*g*-LLDPE) by $^{1}$H-NMR and proposed that oligo(MA) and single succinic anhydride grafts may be distinguished by the chemical shift of the anhydride methine groups. They found that signals attributable to oligo(MA) could not be detected and concluded that poly(MA-*g*-LLDPE) contained predominantly single unit grafts. However, as noted in work [231], this determination is considered unreliable because of the relative broadness of the signal assigned to oligo(MA). Qualitative analysis of the FTIR spectra of poly(MA-*g*-LLDPE) and some model systems suggest that MA is most likely incorporated as a mixture of single unit and oligomeric grafts. This observed fact is supported by the study of the grafting of MA onto PE, PP and E–P copolymers [168,233] conducted as a study of the grafting of $^{13}$C-labeled MA onto polyolefins. Thus, the reaction products from the radically initiated grafting of specifically $^{13}$C-enriched MA ([2,3-$^{13}C_2$]MA) onto PE, isotactic PP and E−P random or alternating copolymers in the melt by using a small scale extruder in biphenyl solution at 170$^{o}$C, under nitrogen atmosphere were investigated by $^{13}$C-NMR spectroscopy [150,223]. It was shown that the sites of attachment and the structures of the grafts depend on polyolefin composition. In random E−P copolymer, MA attaches to methylene and methine carbons in the backbone. In alternating E−P copolymer, MA attaches solely to polymer methines, indicating that $(CH_2)_m$ sequences with $m$>3 are needed for MA attachment to backbone methylene carbons. In the copolymers and in PP, grafts are single succinic anhydride rings; chain scission can yield structures in which the anhydride ring is attached to the chain terminus via a fully substituted double bond. It was shown that a signal at $\delta$ = 44-55 *ppm* for the poly(MA-*g*-PE) should indeed arise from oligo(MA) grafts, taking into account that the chemical shift of the methine groups in oligo(MA) is about 45 ppm. In alternating E−P copolymer and PP, MA grafts onto the polymer backbone chiefly in the form of single succinic anhydride rings [227].

Atactic PP grafted with 4.5 % MA exhibits a tensile strength of 1.91 MPa, elongation of 400 %, and hardness of 50-55 % (against for ungrafted PP of 0.53 MPa, 110 % and 10-15, respectively) [228].

Some studies developed to the grafting of polar monomers (MA and IA) onto polyolefin binary blends especially onto PP/LDPE blends [133,134]. It has been shown that PP introduced into LDPE can affect both the amount of MA grafted onto a blend and its rheological behavior [233]. Itaconic acid (IAc) was grafted onto PP/LDPE blends by Pesetskii et al. [234] in melt with 2,5-dimethyl-2,5-di(*tert*-butyl peroxy)-hexane as a initiator using a reactive extrusion. Authors described the probable chemical reactions occurring at free-radical grafting of IAc onto PP/LDPE blends. According to the authors, grafting efficiency increased by introducing LDPE into PP. Apart from the interaction with the IAc monomer, LDPE macroradicals were involved in recombination reactions leading to crosslinking, branching, or chain extension, which increased the apparent viscosity (and decreased the related MFI) of LDPE-*g*-IAc as well as (PP/LDPE)-*g*-IAc systems containing higher amounts of LDPE. It was shown that variations in the ratio of polymeric components in the (PP/LDPE)-*g*-IAc systems cause nonaadditive, complex changes in the swell index of a molten jet and in the strength of the molten blended materials [234].

We have been reported [235-241] the synthesis of poly(PP-*g*-MA)s with different compositions as the precursor for the preparation of nanocomposites by radical grafting reaction of powder and granular PP with MA in melt by reactive extrusion using dicumyl peroxide (DCP) as an initiator. It was demonstrated that the structure, macrotacticity, crystallinity, crystallization rate, and thermal behavior of PP changed with grafting and depended on the grafting degree. The MA grafting efficiency of powder PP was higher than that obtained for the granular form of PP. In general, the grafting degree first increased with the MA or DCP content in the feed, then reached a maximum value, and finally decreased because of several possible anternative reactions during the grafting. The grafting of powder PP was more successful because of better initial mixing and less diffusional resistance during the grafting. We also have been investigated the grafting of MA onto powder PP in the thermal oxidative conditions using reactive extrusion [236,241]. Obtained results show that chain scission reactions are significantly suppressed at 3.0 wt. % MA concentration, especially when more than 1.0 wt. % of DCP is used. The effect of DCP concentration on the melt flow index (MFI) of isotactic PP in the absence of MA was also studied. In these conditions, the degradation of PP is not a controllable process and essentially depends on many factors, such as temperature, screw speed, and other extrusion parameters. The introduction of MA into this system allows the process to reasonably control the chain degradation reactions [241].

Above mentioned numerous investigations of free-radical grafting of MA and its isostructural analogues onto different POs (LDPE, HDPE, PP, EPR, EPDM, etc.) indicated that side reactions (crosslinking, degradation, oligomerization of monomer, β-scission of C–C bonds in the main chain, etc.) occurred simultaneously with the grafting reactions which are caused some changes in the molecular structure of POs, depending on the type of used polymers. The use of various types of reactive extrusion systems in the grafting process provides the important advantages in the production of high performance commercial materials.

## 2.3. Solid state grafting

Graft copolymerization of thermoplastic polymers and elastomers with MA and its isostructural analogues can also be performed in the solid state or in an aqueous suspension. Solid-phase graft copolymerization is a relatively new method developed by many researchers [124,202,242-266]. Many investigations on solid-phase grafting of PP are mainly based on the MA monomer [193,242-245]. Since the self-polymerization of MA is very difficult, the MA grafted units only exist as a monomer or a short branch with low grafting degree (<5 %) on the PP macromolecules. Jia et al. [245] found that solid-phase graft copolymerization of PP with two acceptor-donor type monomers, MA and styrene, significantly increased the grafting percentage and grafting efficiency. Solid-phase graft copolymerization was carried out by using benzoyl peroxide as an initiator and xylene as an interfacial agent. The effect of monomer concentration, monomer ratio and initiator on grafting percentage (acid number:r) and grafting efficiency were studied by these authors. The structure of the graft copolymer was characterized by FTIR spectroscopy (the absorption band at 1789 and 709 cm$^{-1}$ for C=O and benzene ring in MA and styrene grafted units, respectively), pyrolysis gas

chromatography–mass spectroscopy (the characteristic peaks at t = 4.66 and 5.51 min for MA and styrene units, respectively, at 423°C), and dynamic mechanical analysis. Authors found that the grafting percentage of the graft copolymer of PP with these two monomers were considerably higher than those of the graft copolymers of PP with MA and styrene alone; the graft segments were shown to be the copolymers of MA and styrene with a substantial molecular weight.

Grafting reaction in solid state between plasma activated polyolefin films and MA–vinyltriethoxysilane (VTES) cooligomer and surface properties of the modified films obtained were investigated [246]. The oligo(MA-*alt*-VTES) was synthesized by radical copolymerization with benzoyl peroxide in bulk at 80°C under nitrogen atmosphere [247]. For the activation of polyolefin film surfaces in the conditions of cold plasma, discharge parameters selected and used were as follows: pressure 0.2 *Torr*, power 1200 *J/min*, and discharde duration 30 *min*. A inductively coupled glow discharge system working a fixed RF frequency of 13.56 MHz was used for this plasma surface treatment. Thin coatings with characteristic thicknesses of 10-15 $\mu m$ on plasma-activated surfaces of PP and PE films are formed from a 5 % solution of oligomer in MEK by centrifugation. The observed systematic increase of surface energies both by the increase of modification times and temperature allows one to suggest that the chemical modification expected is accomplished. Increases of thermotreatment time and temperature are expected to increase the probability of formation of hydrophobic organosiloxane fragments of a crosslinked character. The grafted coats obtained on PP and PE films were found to be durable. A similar, thin and yet durable coating was found to be impossible to obtain on virgin polyolefin films without use of plasma, even if the same procedure is applied afterwards. FTIR-ART spectroscopic studies of modified film surfaces at different conditions of thermotreatment show that the chemical reaction between plasma-activated polyolefin surfaces and MA–VTES cooligomer has proceeded. New bands at 1720, 1770 and 1835 $cm^{-1}$ consisting of C=O (ester) and C=O (anhydride), respectively, appear in the spectra of the surface grafted PP films. Moreover, it is observed that in case of an increase of both modification temperature and duration of surface treatment increases the intensity of ester band and also decreases that of hydroxyl band (3360 $cm^{-1}$ , which allows one to demonstrate the probable reaction steps for the plasma-treated PP with oligomer as follows: (1) intermolecular esterification of anhydride units with PP surface hydroxyl groups, (2) intramolecular reaction between carboxyl and ethoxysilyl groups, and (3) polycondensation of $-Si(C_2H_5)_3$ fragments, as initiated by free –COOH, via formation of crosslinked structure. These can be presented schematically as follows (Figure 8):

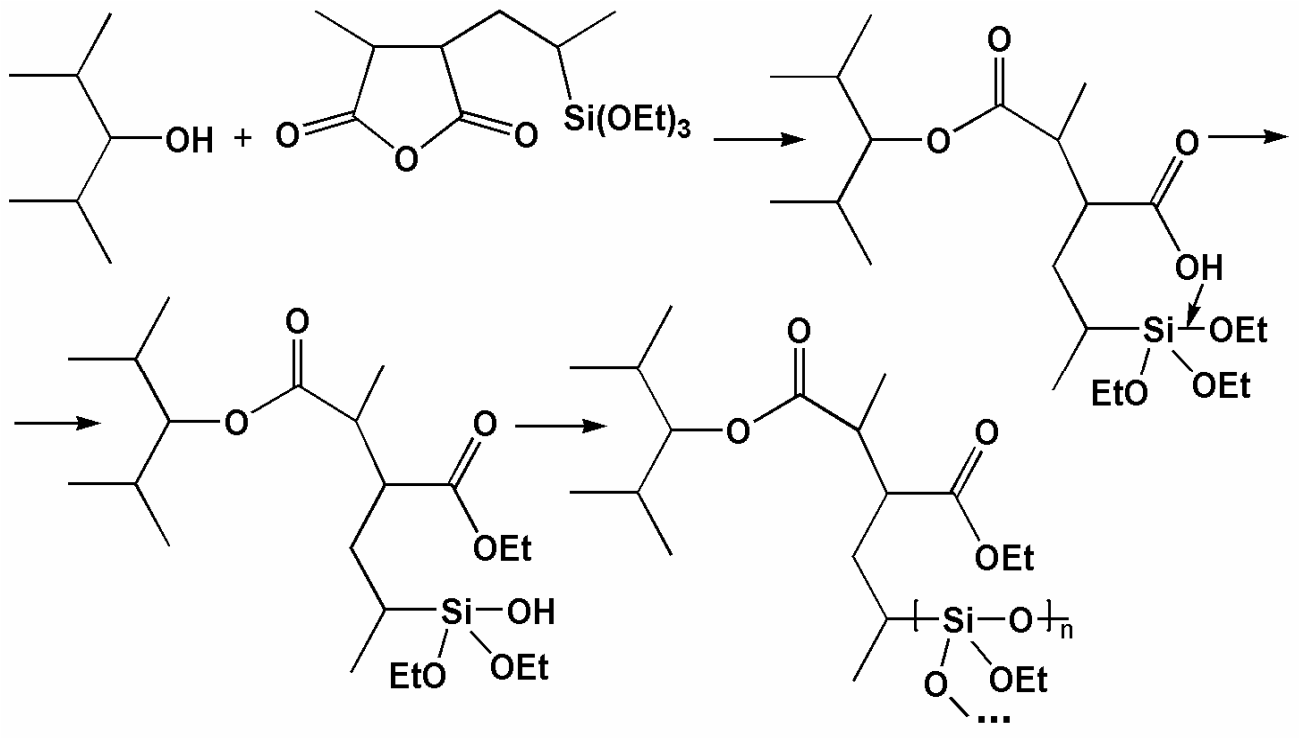

**Figure 8.** Proposed mechanism of formation of the PP/ organosilica structure via esterification (grafting)-hydrolysis-polycondensation reactions.

It was concluded that the modification of plasma-activated polyolefin surfaces by use of MA–VTES oligomer mainly proceeded via esterification and polycondensation reactions, as traced by the carboxyl and triethoxysilyl fragments, leading to crosslinked poly(organosilica) structures [246].

Zelenetskii et al. [248] have studied the effect of solid-phase MA grafting onto LDPE on the mechanical properties of LDPE and its wood-polymer composites, besides its adhesion to Al-foil. LDPE powder was mixed with MA and fed to a twin-screw continuous-action extruder. Treatment of the reaction mixtures was carried out in a solid-phase regime on an extruder at 80-90 °C, i,e., below the melting temperature of the polymer for 7-70 min. It was shown that when a small amount of MA groups were introduced into the chain of LDPE by means of solid-phase extrusion there was a considerable increase in the elastic modulus of the polymer and a reduction in the tensile strength. Solid phase grafting polyolefins possessed considerably higher adhesion properties compared to polyolefins modified in the melt.

## 2.4. Photo-induced surface grafting

Extensive attention has been paid to the surface chemical modification and functionalization of polyolefin substrates, especially starting from the 1990s, by which targeted products possessing various functional groups and different properties can be provided on the polyolefin surfaces using various developed methods [249,250]. Among these techniques, photografting (co)polymeriza-tion has drawn great interest, mostly because of its obvious advantages, such as low operation cost, mild reaction conditions, selectivity to absorb UV radiation, and relatively permanent modification effects, without the destruction of the bulk properties of substrates. For this purpose, acrylic acid and its derivatives and MA have been employed [251-254]. To make the technology of surface photografting polymeri- zation more practical and less costly but more versatile and environmentally friendly, anhydrides of unsaturated dicarboxylic acids, preferably MA has been selected. However, it is widely accepted that MA cannot homopolymerize under normal conditions but easily participates in donor-acceptor type copolymerization with various comonomers [1-4,27,255-257]. Surface photo-grafting polymerization of MA and its charge transfer complexes (CTCs) with vinyl acetate monomer and various solvents onto different polymeric substrates reported by many researchers [258-262]. Deng et al. [258-260] successfully photografted MA and vinyl acetate (VA) onto LDPE films, but for either single monomer, both the grafting rate and grafting efficiency were actually too low to be industrialized. Furthermore, authors identified the formation of grafted film with FTIR and ESCA surface analysis methods [262]. Accordint to authors [261,262], MA can be grafted onto polymer substrates under UV-irradiation rather easily. In this study, the photo-grafting polymerizatioın of MA was examined with LDPE film as a substrate in different solvents (THF, 2,4-dioxane, ethyl acetate and acetone) using benzophenone as a photoinitiator and benzoyl peroxide as a radical initiator. The effects of some principal factors, such as temperature, solvent, and UV intensity, on the grafting polymerization of MA also investigated. It was shown that solvents have a great influence on the grafting polymerization, and the grafting still occurs in the absence of photoinitiators. This fact authors explained by the abstraction of hydrogen atom from the LDPE chain with MA itself when UV-irradiated [260]. THF and dioxane solvents even seem to take part in the reaction. To explain this phenomena, authors should consider the known reaction between these solvents and MA. The ether oxygen atoms in both THF and dioxane are prone to give off ctrons. When they meet strong electron acceptor MA, the CTC between them and MA may be formed and, as shown in works [263], even participate in the reaction to a relatively high degree. For example, the equilibrium constant of the CTC formation ($K_c$) in the MA/THF system was determined to be 0.44 L/mol [264]; with irradiation from UV light, the following reactions can occur (Figure 9) [265]:

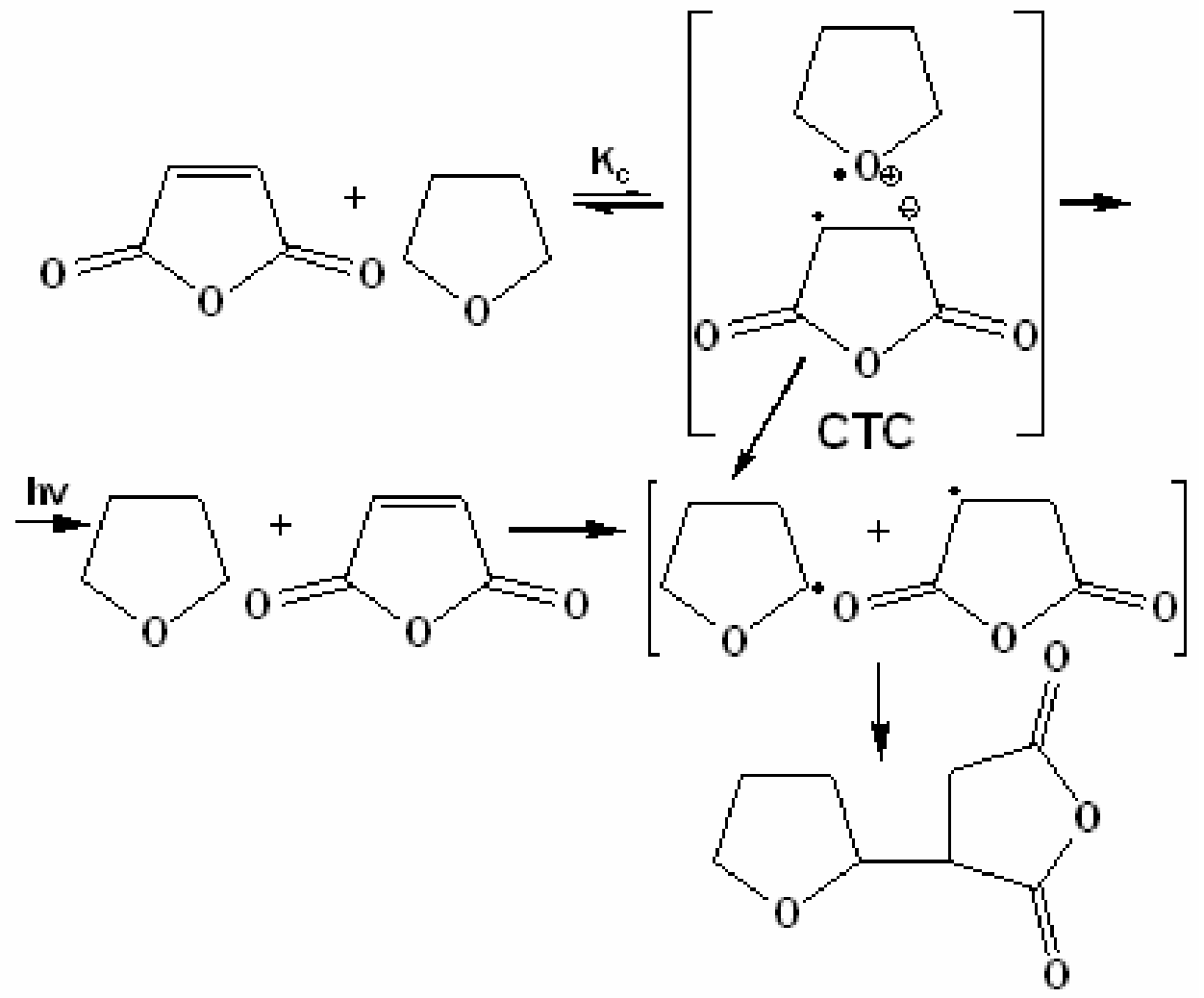

**Figure 9.** The formation of MA...THF CTC complex in the surface UV-grafting of LDPE with MA in THF solution. Adapted from [264].

Similarly, the irradiation of the MA/dioxane solution leads to the formation of an oligomer composed of MA and dioxane [261]. With respect to acetone (or ethyl acetate) , as noted by the authors [265], it may also be thought that the CTC can form between itself and MA [255], but its ability to give off electrons is lower than that of THF and dioxane; therefore the grafting polymerization of MA proceeds preferentially.

MA was photografted onto LDPE substrates at temperatures above the melting point of MA. The effects of some principal factors including irradiation temperature, photoinitiators, the intensity of UV radiation, and the far UV radiation on the grafting polymerization were studied in detail. It was shown that the photografting polymerization of of MA can proceed smoothly at temperatures higher than the melting point of MA ( >55 $^oC$); the far UV radiation and the intensity of the UV radiation affect the grafting polymerization greatly. According to FTIR spectra, it is confirmed that the grafted film samples contain anhydride groups (the newly appearing sharp absorption peaks at around 1870, 1850 and 1125 $cm^{-1}$). The contact angle measurements demonstrate that the wettability of the poly(LDPE-*g*-MA) films is enhanced obviously, especially for those grafted film samples through hydrolysis [263]. It is known that PP polymeric radical in situ reacts with MA to produce MA terminated PP with a single MA unit. In the presence of styrene, the polymeric radical initiates a "stable" copolymer of styrene and MA with an alternating manner [266]. The resulting PP-*b*-SMA diblock copolymer contains both PP and alternating styrene–MA segments.

In some studies, binary monomer systems were applied to photografting copolymerization, but most were carried out in the vapor phase [267-269]. That is, the substrate and MA and another monomer were placed in a reactor; by heating, the reactor was filled with monomer vapor, and some monomer precipated on the substrate, and with UV-light irradiation, photografting polymerization was started. According to these studies, adding MA definitely facilitated the photografting polymerization of some monomers, such as styrene and vinyl ethers, but for other monomers, this effect was not obvious. According to authors, one disadvantage of this technology lies in the fact that photografting polymerization in the vapor phase needs more time, as much as several hours, even though a higher grafting yield can be obtained.. The effects of several crucial factors, including the composition and total concentration of the monomer solution and different types of photoinitiators and solvents, on the grafting copolymerization were investigated in detail by Deng and Yang [268]. The conversion percentage (CP), grafting efficientcy (GE), and grafting percentage (GP) were measured by gravimetry using the following definitions:

$$CP\ (\%) = (W_P / W_M) \times 100 \qquad (3)$$

$$GE\ (\%) = (W_G / W_P) \times 100 \qquad (4)$$

$$GP\ (\%) = (W_G / W_F) \times 100 \qquad (5)$$

where $W_F$ is the weight of the LDPE flms before the grafting copolymerization; $W_M$ is the weight of the added monomers between the two films; $W_P$ is the weight of the polymer formed, including both the homopolymers of VA and MA and the copolymers grafted and not grafted onto LDPE films, obtained by weighing the films after the vaporization of the residual monomers; and $W_G$ is the weight of the grafted polymer, which was obtained after the extraction of the homopolymers and nongrafted copolymers with acetone.

The results of authors showed that the monomer composition played a great role in this binary system; appropriately increasing the MA content in the monomer feed was suited for grafting copolymerization. The three photoinitiators (2,2-dimethoxy-2-phenylaceto- phenone, benzoyl peroxide, and benzophenone) led to only slight differences in CP, but for GE, benzoyl peroxide was the most suitable. As for various other solvents (acetone, ethyl acetate, THF, and chloroform), the usage of those enabled the donation of electrons (acetone and THF) resulting in relatively higher CP values: on the contrary, the use of other solvents made GE obviously higher, and this should be attributed to the charge-transfer complex (CTC) formed in this system. It is known that in acceptor-donor MA(A)-VA(D) monomer system, when irradiated by UV light, an exciplex ( [D–A*] ) can be generated [268,269]:

$$D + A \xrightarrow{h\nu} [D\ldots A] \longrightarrow [D\ldots A]^* \qquad (6)$$

$$D \xrightarrow{h\nu} D^* + A \longrightarrow [D\ldots A]^* \qquad (7)$$

$$A \xrightarrow{h\nu} A^* + D \longrightarrow [D\ldots A]^* \qquad (8)$$

$$[D\ldots A]^* \longrightarrow [{}^{\bullet}D^{+}\ldots{}^{-}A^{\bullet}] \qquad (9)$$

Authors proved that [D…A]* is not stable; especially when irradiated by UV radiation, it can transform into $[{}^{\bullet}D^{+}\ldots{}^{-}A^{\bullet}]$, which contains two ions and two free radicals simultaneously. Deng and Yang [261] assumed that $[{}^{\bullet}D^{+}\ldots{}^{-}A^{\bullet}]$ can undergo hydrogen abstraction in the presence of active hydrogens in the system. According to authors, a chain-transfer reaction may make some contribution to the grafting copolymerization also. The photografting of such acceptor-donor monomer systems is a pathway different from that of general methods and will provide better insight into the mechanism of alternating copolymerization. Authors [268] studied the photo-grafting copolymerization of MA with VA from the perspective of dynamics. The resulting authors showed that in comparison with the photografting polymerization of the two single monomers, polymerization rate (PR) and RG noticeably increased for the MA–VA binary system. When the total monomer concentration was kept at 4*M* , the apparent activation energy ($E_a$) of the three photografting polymerization systems were as follows: $E_a$ = 41,0 and 43.9 kJ/mol (for the total polymerization and grafting polymerization of VA, respectively) and 39.65 and 43.23 kJ/mol (for MA, respectively), and 34.35 and 40.32 for the MA–VA binary monomer system, respectively. These results suggested that the polymerization of the binary system occured more readily than the other two. Authors inferred that in the binary monomer system, both the free monomers and CTC took part in the polymerization, which is very similar to the general principles of complex-radical alternating copolymerization [27]; to the termination of the propagating chains, two possible pathways, unimolecular termination and bimolecular termination, coexisted in this binary monomer system. According to the authors, the photografting copolymerization of MA and VA on the LDPE film surface and its mechanism can be presented as follows [268]:

CT-complex formation

$$VA + MA \overset{K_c}{\leftrightarrow} CTC \qquad (10)$$

Photodestruction of LDPE in the presence of BP

$$BP + PH \xrightarrow{h\nu} P^{\bullet} + \text{semibenzopinacol free radical} \qquad (11)$$

MA acts as a photoinitiator

$$MA + PH \xrightarrow{h\nu} P^{\bullet} + H\text{–}MA^{\bullet} \qquad (12)$$

Initiation of the grafting copolymerization by polymer radical

$$P^{\bullet} + VA \longrightarrow \sim P^{\bullet} \qquad (13)$$

$$P^{\bullet} + MA \longrightarrow \sim P^{\bullet} \qquad (14)$$

$$P^{\bullet} + CTC \longrightarrow \sim P^{\bullet} \qquad (15)$$

Bimolecular termination

$$2\sim P^{\bullet} \longrightarrow \text{Grafting polymer} \qquad (16)$$

Unimolecular termination

$$\sim P^{\bullet} + RH \longrightarrow \text{Grafting polymer} + R^{\bullet} \qquad (17)$$

The reaction orders of the copolymerization to the total monomer concentaration and concentration of BP were 1.34 and 0.81, respectively, reflecting that in the binary monomer system, both the free and complexed monomers participated in the polymerization [268].

## 2.5 Plasma-induced surface grafting

Various vinyl monomers successfully grafted onto PE and PP surfaces through plasma induced reaction [268-276]. The alkyl groups and tertiary radicals are shown to be created on the PE and PP surface after plasma exposure, respectively, which were proved to react with vinyl monomers such as styrene, 2-methoxyacrylate, methacrylic acid, acrylamide, allyl alcohol, glycidyl methacrylate, etc. The amount of monomer grafted depends on the plasma operating parameters. Hong and Ho [277] reported grafting of thermoplastic olefin elastomer such as E-P-norbornene rubber (EPDM) with MA under nitrogen plasma. The adhesion properties of EPDM rubber surface grafted with MA and different initiators such as benzoyl peroxide (BPO), azobisisobutyronitrile (AIBN) and benzophenone (BP) under nitrogen plasma condition were analyzed. The results of FTIR-ATR surface analyses indicated the the amounts of MA grafted depended on the plasma treating time and the type of initiator used. The total and the polar component of surface energies treated with EPDM rubber increased after plasma exposure, and followed the order of BP > BPO > AIBN after a 0.5-min plasma treatment. According to authors, the lap shear strength of all specimens singificantly decreased after prolonged exposure because of the possible deterioration of the grafted surface. The nature of the reacted surface layer in affecting the adhesion strength of the EPDM rubber was also confirmed.

Poly(MA) films can be formed in such a way that anhydride group functionality is retained, which can then be used for further chemical modification and bioconjugation [278-280]. Anhydride-containing surface films prepared by plasma polymerization method, and their functionality tuned by pulsing the input power [279]. Ryan et al. [278] reported the results of first detailed investigation of the radio frequency (RF 13.56 MHz) cold plasma-induced polymerization of MA. Authors demonstrated that variations in applied power and the duty cycle could be used to vary the chemical functionality of plasma polymer films of MA. By using XPS and FTIR spectroscopy, the retention of anhydride ring in the film was established , depending on both the equivalent power and the duty cycle. Schiller et al. [280] studied plasma-assisted polymerization of MA under different experimental conditions. They obtained the chemical structure of plasma deposited poly(MA) film using X-ray photoelectron and FTIR spectroscopy (FT-IR). Authors also carried out the surface reactions of anhydride groups with nucleophilic moieties such as decylamine and benzylamine. It was demonstrated a change the surface free energy of the plasma polymer films after surface derivatization reactions. Authors suggested that ontained results are of particular interest for future applications in the attachment of biological molecules and cells. They also developed a method of silicon substrate pretreatment to ensure reliable binding between the substrate and the plasma polymer film in aqueous solution. The hydrated films showed some resemblance to polyelectrolyte films and a clear correlation could be observed between the density of anhydride groups and the behavior of the films in solution.

Gaboury and Urban [281,282] developed a novel microwave plasma method and utilized chemically bonded solid monomers such as MA and acrylamide (AAm) to the surface of crosslinked poly(dimethyl- siloxane) (PDMS). By use of FTIR-ATR spectroscopy, they found that these monomers can be chemically bonded to the PDMS surface through the C=C bond opening. The advantage of this plasma energy source comes from its localized nature and short reaction times. For microwave reaction times of less than 15 s, a majority of the original anhydride or amide functionalities of the plasma deposited monomers was maintained. The microwave reaction time exceeding 15 s result in a cleavage of the surface anhydride or amide species and conversion of the monomer supply to maleic acid or polyacrylamide. The hydrolysis of the plasma reacted surfaces produces new acid salt groups which were also quantitatively evaluated. FTIR-ATR analysis reveals a 48 % conversion for MA. The amount of MA reacted onto the surface is of the order of 1 to 5 pmol/$cm^2$ [282].

Jenkins et al. [283] used a plasma polymerization method to deposit thin polymer films from MA monomer. Then anhydride functionalized surface on the silicone substrates was hydrolyzed and carboxylic acid functionalized film was used to support a negatively charged dimyristoyl phosphatidyl glycerol (DMPG) lipid bilayer using $Ca^{2+}$ as a chelating agent. This process was schematically represented in Figure 10.

The capacitance and resistance of adsorbed DMPG on three films prepared at different plasma duty cycles [$t_{on}/(t_{on} + t_{off})$], employed during the plasma treatment, could be correlated with the relative amount of anhydride groups present in the plasma polymer films.

Everson et al. [284] used pulsed plasma polymerization method for the deposition of well-defined MA functionalized films. Surface layeres of these films readily undergo reaction with amine-terminated nucleophiles to produce surface amide linkages that convert into cyclic imide groups upon heating.

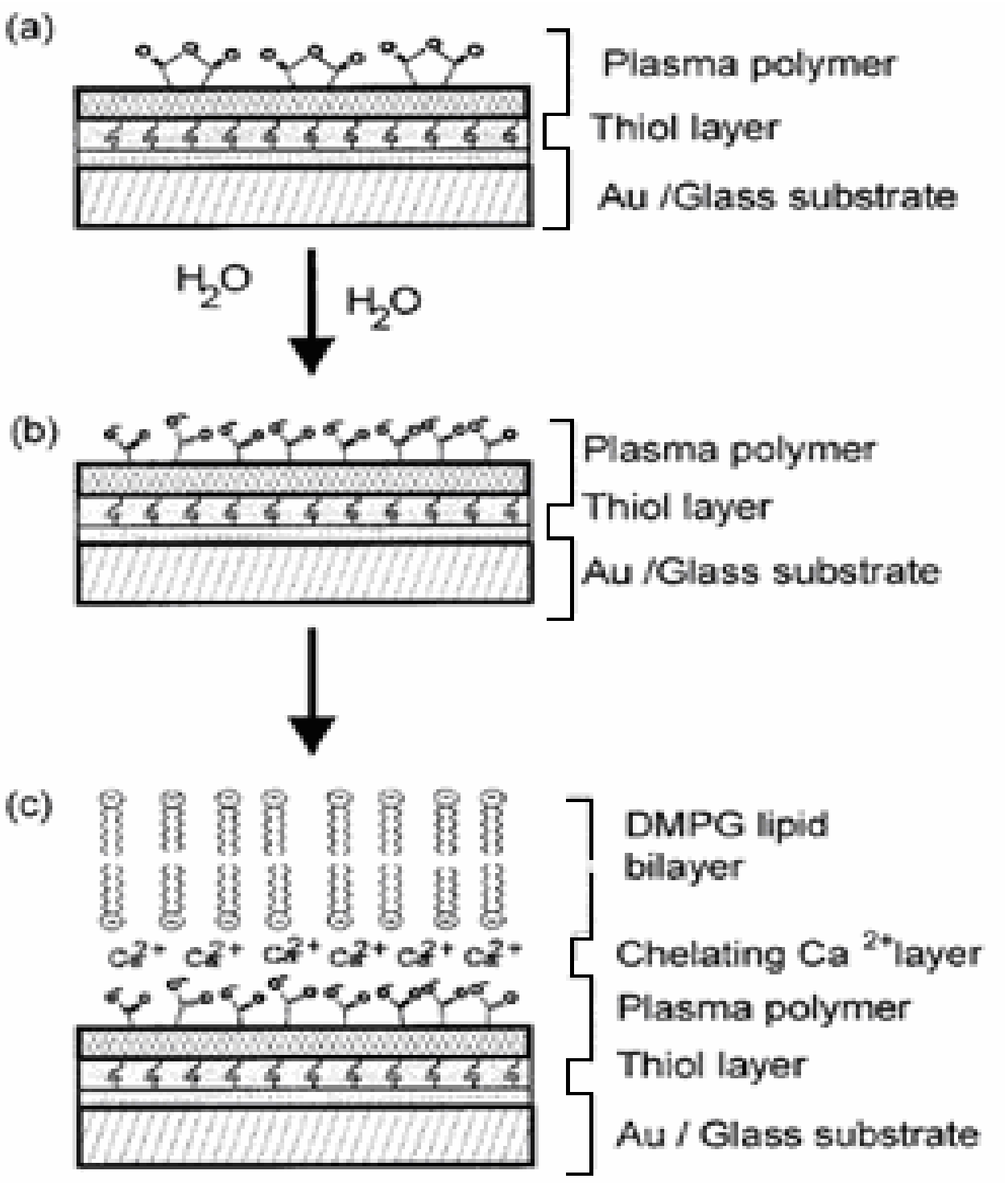

**Figure 10.** Plasma surface modification of the silicone substrate through plasma polymerization of MA, hydrolysis of MA units and chelating with biomolecules. Adapted from [283].

According to the authors, (1) amine-terminated nucleophiles can be chemically fixed onto any shaped solid substrate via reaction with a well-adhered MA pulsed plasma polymer layer; (2) the intermolecular spacing and concentration of amine-containing moieties attached to the surface can be finely tuned by controlling the level of anhydride group incorporation during pulsed plasma deposition; (3) tailoring solid surfaces using the combination of pulsed plasma deposition of anhydride groups and the subsequent aminolysis reaction can be attractived for a variety of applications such as the control of hydrophilicity and biocompatability (e.g., attachment of heparin), modification of polymer adsorption characteristics, microcontact printing, ion conducting membranes, patterned polymer multilayers, thin film electrodes, catalysis, and adhesion. In the case of propylamine- and allylamine-derivatized MA pulsed plasma polymer layers, XPS, FTIR, diphenylpicrylhydrazyl assay, water contact angle and reflectometry all indicate a marked change in the film. Such film expansion during aminolysis enables greater access to subsurface trapped radicals. Imidization of these amide films by heating in vacuum at 120$^o$C for 1 h [283,285,286] leads to the disappearance of acid-base interactions and a rise in hydrophobicity. This increase in hydrophobicity should enable apolar solvents to penetrate into the film more effectively, thus assisting access to the subsurface free radicals.

According to Teare et al. [286] pulsed plasma polymerization is reorganized as being a practical, substrate-independent route for functioalizing solid surface with specific chemical groups, especially with functional monomers. Other advantages include the fact that this approach is quick (single step), solvent free, energy efficient, and can be applied to a whole host of complex geometries (e.g. microspheres, fibers, tubes, etc.). Authors described a subsrate-independent method for the synthesis of polymer brushes on solid surface. This entails pulsed plasma deposition of MA to generate a free radical initiator containing thin film followed by styrene polymerization. In the case of the graft polymerization of styrene onto MA plasma polymer surfaces, authors found significantly greater extent of polymerization for pulsed compared to continuous wave deposition conditions. According to the authors, presence of "trapped radicals" in MA molecule, pulsed plasma polymer will be enhanced by resonance stabilization (Scheme 11).

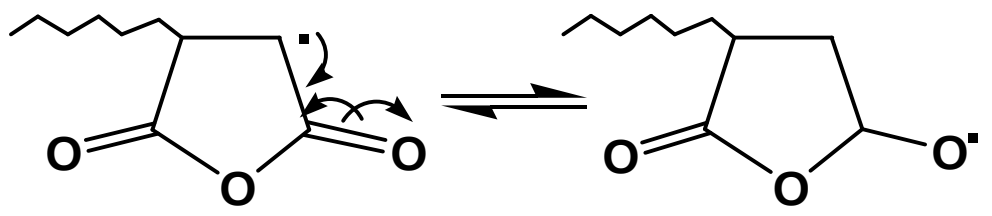

**Figure 11.** Resonance stabilization of "trapped radicals" in the pulsed plasma graft polymeriztion of styrene onto MA plasma polymer surface. Adapted from [286].

They demonstrated that a pulsed plasma-chemical deposition of MA polymer is a substrate-independent method for functionalizing solid surfaces with initiator sites for nitroxy-mediated controlled radical graft polymerization, and swelling of the initiator film via aminolysis can lead to grafted polymer brushes that are first order magnitude thicker than those obtained by existing methods on solid surfaces [286]. According to Yasuda [287], the graft polymerization of styrene is most likely to be concentrated at radical sites located at or near the surface, where resonance and polar effects will help to stabilize the intermediate radical. Authors schematically represented resonance stabilization of reaction intermediate between MA and styrene as follows (Figure 12) [286]:

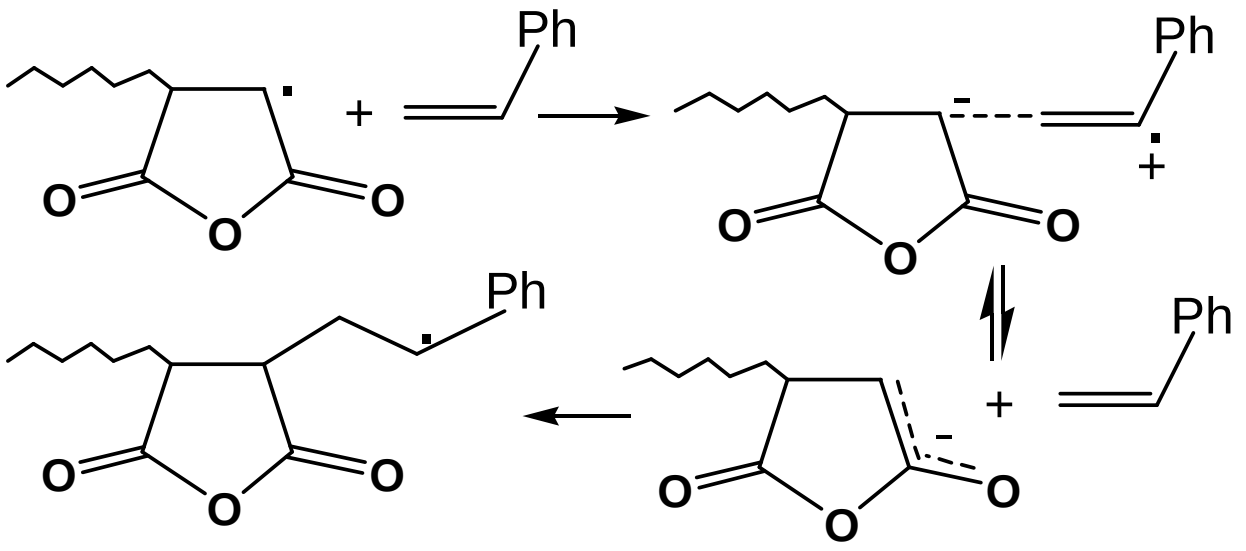

**Figure 12.** Plasma graft polymerization via resonance stabilization of intermediate reaction between MA unit and styrene. Adapted from [286].

Tarducci et al. [288] used furfuryl methacrylate as the precursor for pulsed plasma polymerization in order to provide furan ring functionalized surfaces, which are capable of participating in [4+2] cycloaddition reaction with MA as a dienophil component. Reaction of the furfuryl methacrylate pulsed plasma polymer layers with MA was represented as follows (Figure 13):

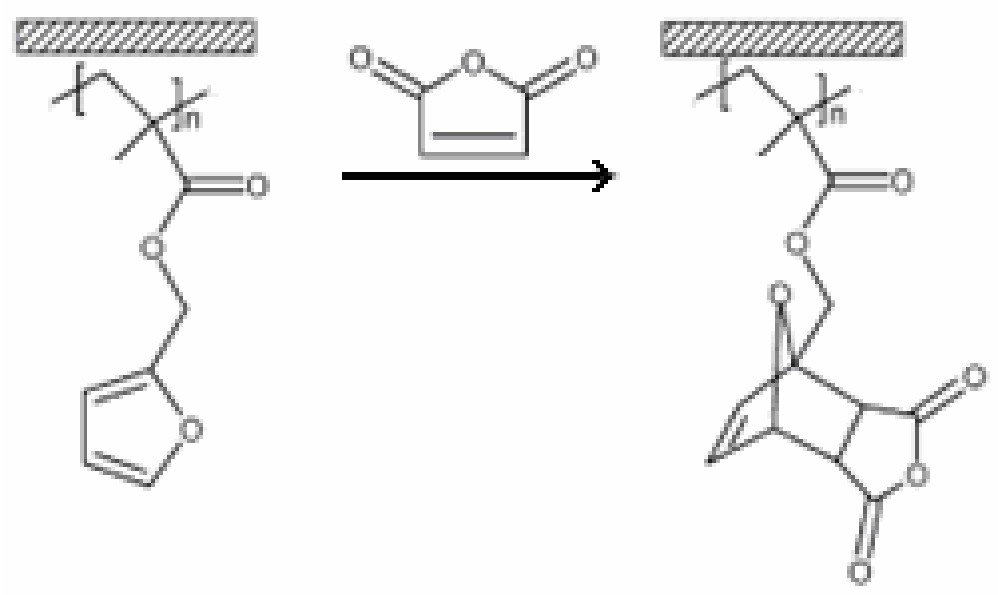

**Figure 13.** Reaction of the furfuryl methacrylate pulsed plasma polymer layers with MA monomer. Adapted from [288].

Grazing angle IR spectroscopy of the Diels-Alder derivatized pulsed plasma polymer film indicated the presence of new peaks attributable to: anhydride C=O stretching (1860 cm$^{-1}$), C=C stretching (1636 cm$^{-1}$), and an intense broad peak due to C-O-C stretching (1070 cm$^{-1}$).

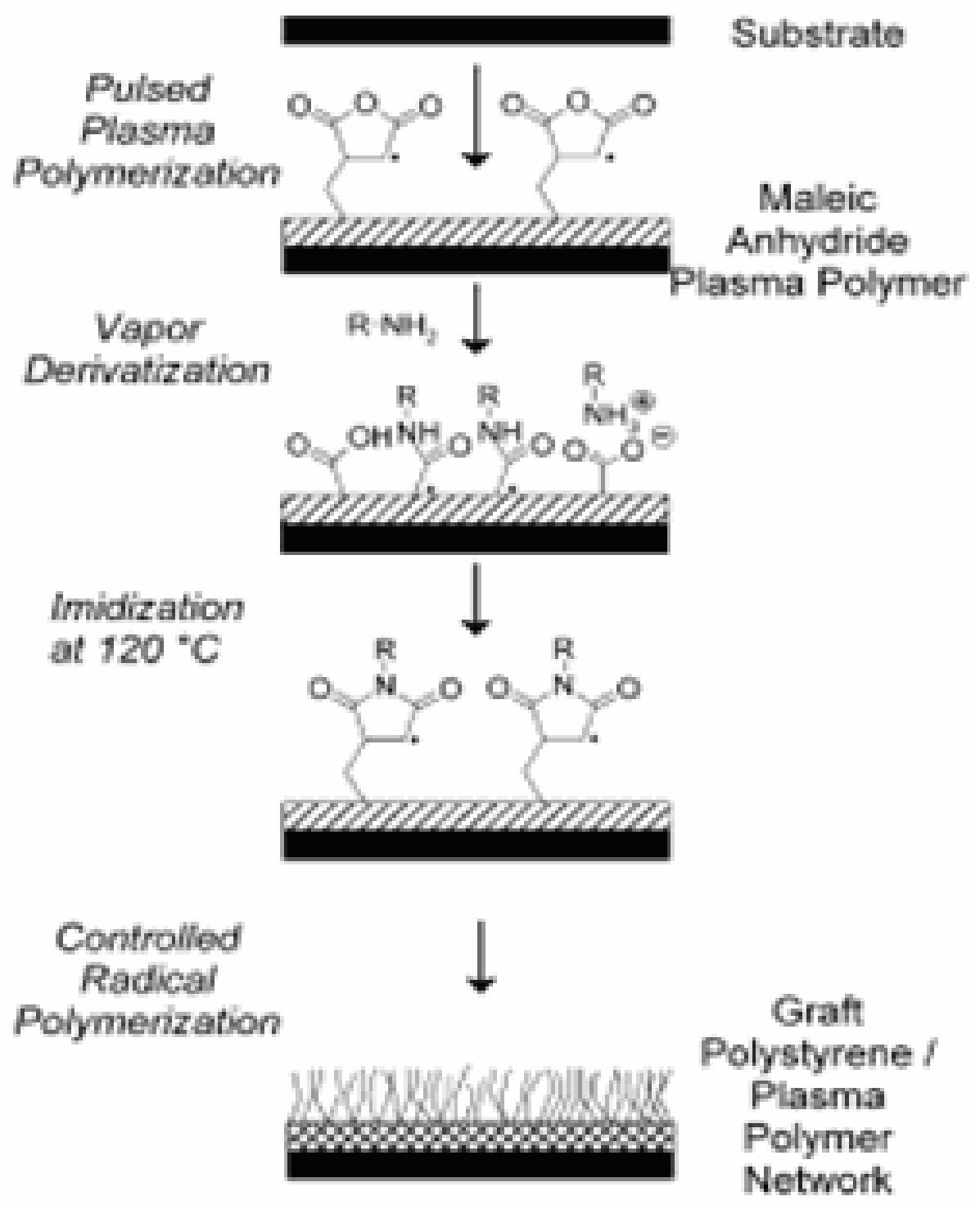

**Figure 14.** Nitroxide (TEMPO)-mediated controlled free-radical graft polymerization of styrene onto pulsed plasma deposited poly(MA) layers. Adapted from [290].

Authors conclude that the methacrylate C=C double bond in furfuryl methacrylate (together with the substrate) is activated during the short pulsed plasma duty cycle on-time followed by conventional polymerization proceeding within each associated extinction period. This yields furan funtionalized surface which readily undergoes the cycloaddition reaction with MA [288]. Authors found that heating the MA pulsed plasma polymer surface in the presence of diphenylpicrylhydrazyl (a free radical scavenger) gave rise to the capping of trapped radical sites at the surface, thereby inhibiting styrene graft polymerization. UV irradiation through photomask of such capped surfaces leads to localized reactivation of the initiator sites, which in turn provides a method for the microfabrication of patterned polystyrene brush surfaces [289]. In their other publication, authors [290] also described a substrate-independent approach entailing the use of trapped-free-radical containing MA pulsed plasma polymer films as surface-bound initiators for the nitroxide (TEMPO)-mediated controlled free-radical graft polymerization of styrene. This controlled radical graft polymerization on pulsed plasma deposited poly(MA) layers was schematically represented in Figure 14.

The benefits of swelling the MA plasma polymer initiator films via aminolysis or imidization (through reaction of surface MA unit with propylamine or allylamine) prior to nitroxide-mediated controlled radical graft polymerization is highlighted [290]. Of particular importance is the fact that derivatization of the plasma polymer with a spacer molecule yields grafted polymer brush film at 1 order of magnitude thicker than previously reported for TEMPO-mediated surface-controlled free-radical graft polymerization [291].

It is known that the Diels-Alder reaction is a well-established solution-phase method for marking complex ring molecules in a single step [292]. Recently, it was demonstrated that this type of cycloaddition reaction can also take place at a solid surface, thereby marking in of potential interest for a number of technological applications. The examples of surface Diels-Alder chemistry include self-assembled monolayers, the immobilization of dyes [293], and direct reaction with polymer substrates containing polymer repeat units which participate in the Diels-Alder reaction [294-296]. Roucoules et al. [297] reported a simple, multisubstrate approach based on pulsed plasma deposited MA thin films. Pulsed plasmachemical surface functionalization comprises two distinct reaction regimes corresponding to monomer activation and the generation of surface sites during the duty cycle on-period (via UV irradiation, ion or electron bombardment) followed by conventional polymerization during the subsequent off-time (in the absence of any irradiation-induced damage). The overall approach in the study of authors entails allylamine undergoing aminolysis with MA pulsed plasma polymer films to yield alkene functionalized surfaces [277,278, 280]. The generated covalent amide linkages are subsequently converted into cyclic imide groups upon heating [298]. These surfaces are then shown to be suitable dienophiles for the Diels-Adler cycloaddition reaction with 1,3-cyclohexadiene to give mixtures of *endo*- and *exo*-bicyclo[2.2.2]octene-2 groups (Figure 15).

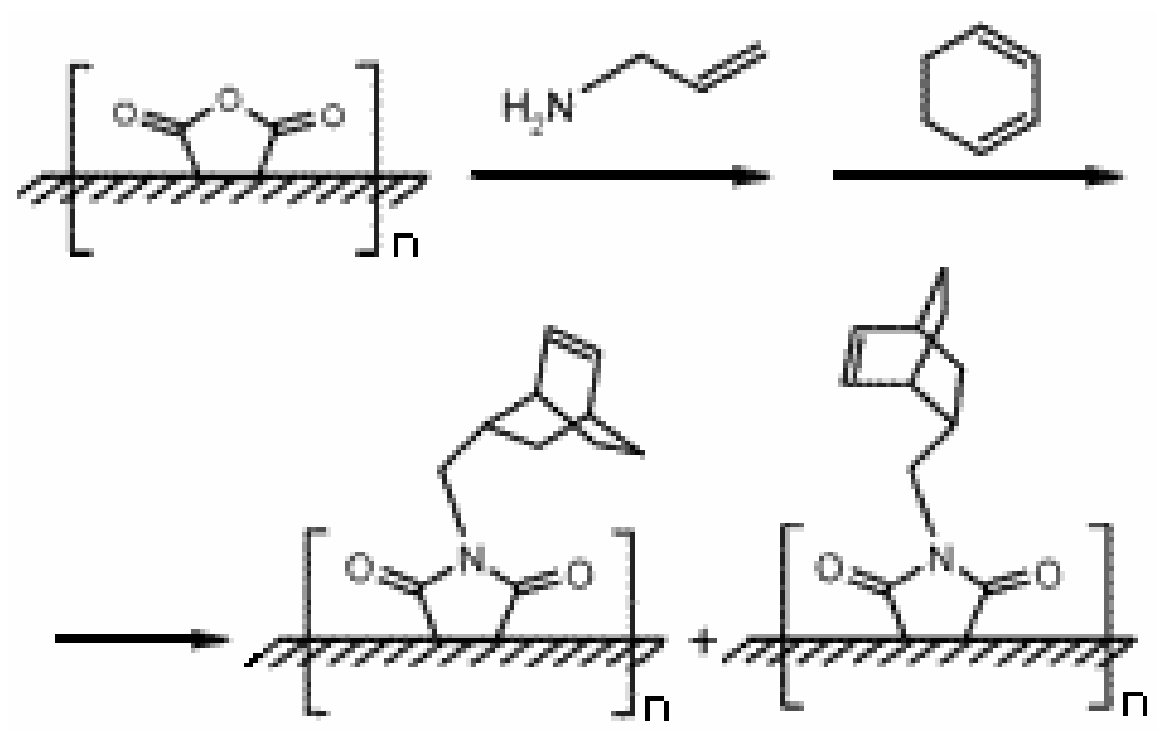

**Figure 15.** Pulsed plasmachemical surface functionalize-zation of poly(MA) plasma polymer films via amidization and Diels-Adler cycloaddition reactions. Adapted from [298].

Reflection-absorption IR analysis of MA pulsed plasma polymer films (95 nm thickness) confirmed a high degree of anhydride group incorporation [298]. The following characteristic infrared absorption features of cyclic anhydride groups were identified: asym. and sym. C=O stretching (1860 and 1796 $cm^{-1}$), cyclic conjugated anhydride group stretching (1241 and 1196 $cm^{-1}$), C–O–C stretching vibration (1097 and 1062 $cm^{-1}$), and cyclic unconjugated anhydride group stretching (964, 938, and 906 $cm^{-1}$) [187]. According to the authors, reaction of allylamine vapor with the deposited MA pulsed plasma polymer layer resulted in ring opening of the cyclic anhydride centers to yield amide (amide I at 1658 $cm^{-1}$ and amide II at 1550-1510 $cm^{-1}$), carboxylic acid stretching (1716 $cm^{-1}$ ), and NH wagging (820-750 $cm^{-1}$ ) bands. Heating of these surfaces to 120 $^{o}C$ gave rise to ring closure and the formation of cyclic imides, as seen by the drop in the intensity of strong amide, NH wagging, and C=O acid bands, while two imide stretching vibration bands appear at 1775 and 1710 $cm^{-1}$. Then, authors functionalized the aforementioned plasma polymer surface with cyclic imide groups was exposed to 1,3-cyclohexadiene solution in toluene at 90 $^{o}C$ for 15 h. This gave rise to a loss in intensity of the C=C stretching band (1638 $cm^{-1}$) belonging to the surface alkene in conjunction with the appearance of a new band at 1618 $cm^{-1}$. These reactions were then repeated on MA pulsed plasma polymer films deposited onto steel plates, PE films, PS microspheres, and paper in order to demonstrate the general applicability of this approach.

Adhesion of thermoplastic olefin elastomer (TPO), consisting dispersions of poly(E-*co*-P-*co*-diene) rubber particles (EPDM) in a PP matrix, is low compared to other materials because of its low surface energy due to hydrocarbon constituents. To facilitate adhesion, the apolar aliphatic polymer surface can be modified into a high-energy polar surface by a number of methods such as bulk grafting, compounding, and variuos surface grafting methods [299-305]. As a surface modification technique, ozone has also been used successfully in treating polymers such as PP, PE, PET and PU to improve adhesion [298-308]. Polar carbonyl, carboxylic and hydroxyl groups formed on the ozone treated polymer surfaces were identified [308,309]. It was also recognized that hydroxyperoxides formed on the ozonied surface can induce radical polymerization of monomers containing C=C bonds, resulting in surface grafting of functional monomers onto the base polymer surface including PE and PP film surfaces [301,305,308]. It is well known that MA has been used extensively in producing compatibilizers for polymer alloying and in functionalizing polyolefins because of its activite vinylidene double bonds and anhydride functional groups.

Hong et al. [196,309] reported surface grafting of TPO with MA under ozone exposure using one of the initiators benzophenone (BP), benzoyl peroxide (BPO) or 2,2'-azobisisobutyronitrile (AIBN). The modified TPO surfaces were characterized by ATR-FTIR spectroscopy, contact angle measurement, and lap shear strength mesurement (LSS). They showed that TPO oxidizes to form –COOH (1746 $cm^{-1}$) and C=O (1714 $cm^{-1}$ ) groups under ozone exposure [309]. Two forms of MA, grafted MA (MA reacted with TPO) and homo- polymerized MA, can be formed during ozone treatment [290]. The radicals generated from BP then react with TPO to form MA grafted TPO or with each other to produce poly(MA). It is believed that MA is formed either from grafted MA or tightly entangled MA homopolymer. The grafted MA greatly improved the LSS of the TPO in TPO/epoxy adhesive/steel laminates. The LSS also depends on the used initiator type. The total and the polar component of the surface energy of the MA-grafted TPO increased after ozone exposure, following the order AIBN < BPO < BP, while the LSS increased in the order of BP < AIBN < BPO after long time ozone treatment. According to authors, this method of surface modification through grafting reaction is considered to be a good alternative to the conventional ozone treatment [196]. Surface plasmon resonance was used to follow the binding of the antibody, anti-bovine serum albumin (BSA), to protein-modified plasma polymerized MA films [310]. Authors found that BSA could be irreversibly bound to plasma polymer of MA (PPMA) under pulse plasma conditions. Moreover, the degree of antibody binding, which is directly related to the quantity of BSA on the PPMA, correlated with the plasma duty cycle. Authors speculated that BSA is being covalently bound to the PPMA via the reaction of amine groups on lysine residues in BSA with the retained anhydride group functionality in the polymer.

## 3. Grafting of Styrene (Co)polymers

Jo and Park [311] reported the fuctionalization of atactic polystyrene (*a*-PS) with MA in melt by reactive extrusion method. Passaglia et al. [322] studied the functionalization of PS block in styrene-*b*-(ethylene-*co*-1-butene)-*b*-styrene (SEBS) triblock copolymer with MA and diethyl maleate DEM). According to the authors, grafting of MA onto *a*-PS can be accomplished through a radical process, although the reaction mechanism is not clarity. The bulk functionalization of SEBS triblock copolymer with DEM or MA and dicumyl peroxide as initiator was carried out in a Brabender mixer. The functionalization degree (FD) of SEBS was determined by NMR analysis, the results of which indicated that the FD values depend on the feed composition and in particular on the DEM/initiator ratio. All obtained products were fractionated by solvent extraction and characterized by IR, NMR and GPC. They established that the functionalization takes place with a very large preference at the aliphatic carbons of the polyolefin block. Moreover occurrence of degradation and chain extension reactions gives a functionalized product with a molecular weight distribution larger than 1.

Preparation and characterization of MA-functionalized syndiotactic PS (*s*-PS) reported by Li et al. [313]. They accomplished the free radical-induced grafting of MA onto *s*-PS in the solution process at 110$^{o}$C by using 1,1,2-trichloroethane as solvent and dicumyl peroxide as free radical initiator. According to the authors, functionalization reaction in a melt state was proved to be very difficult, since *s*-PS is a semicrystalline polymer and with a high melting poit of about 270 $^{o}C$. FTIR and $^{1}H$ NMR spectra used by authors to confirm that maleated s-PS in the form of single succinic anhydride rings as well as short oligomers. The results obtained by GPC analysis indicate that the degradation and chain extension reaction do not occur under the reaction conditions. In addition, it was found that the crystallization behavior of the poly(PS-*g*-MA) exhibited somewhat differences in comparison to the neat *s*-PS. The MA grafted PS crystallizes at higher rate than the unmodified polymer, while the values of the degree of crystallinity were lowered by the presence of the MA grafts. It was observed that the melting polint of the maleated *s*-PS samples changed little or did not change. Obtained glass transiton temperature ($T_g$ ) data helps in undestanding the effects of MA grafts on the movement of copolymer chains. The $T_g$ values of the modified polymer alter slightly in comparison to the neat PS.

Chemical modification of the phenyl ring in the PS macromolecule is also possible and has been frequently employed. For example, introduction of carboxyl groups in phenyl rings of PS [314-316] can be easily carried out by (1) cleavage of benzoylated or chlorobenzoylated PS [316] and (2) oxidation of commercial formyl resin, chloromethylated PS [315] and acylation of PS with acetic anhydride in the presence of Lewis acids [317]. It was shown that new coating system with improved heat resistance, anticorrosive, and impact properties can be obtained by using these modified PSs (317,318]. According to the authors, functionalization of PS by grafting reaction can be carried out either via the polymer backbone in the case of radical initiators or through the side phenyl rings using cationic catalysts. They chemically modified PSs ($M_n$ around 5000-100000) with MA by use of certain cationic catalysts of Lewis acid type ($BF_3$.$OEt_2$, $AlCl_3$, $TiCl_4$, $ZnCl_2$, $FeCl_3$, and $SnCl_4$) in chloroform at 0-30$^o$C. The results obtained and the known fact that benzene is easily acylated with MA in the presence of $AlCl_3$ allow one to present the general scheme for the side chain modification of PS in the presence of Lewis acid in the following way (Figure 16) [318]:

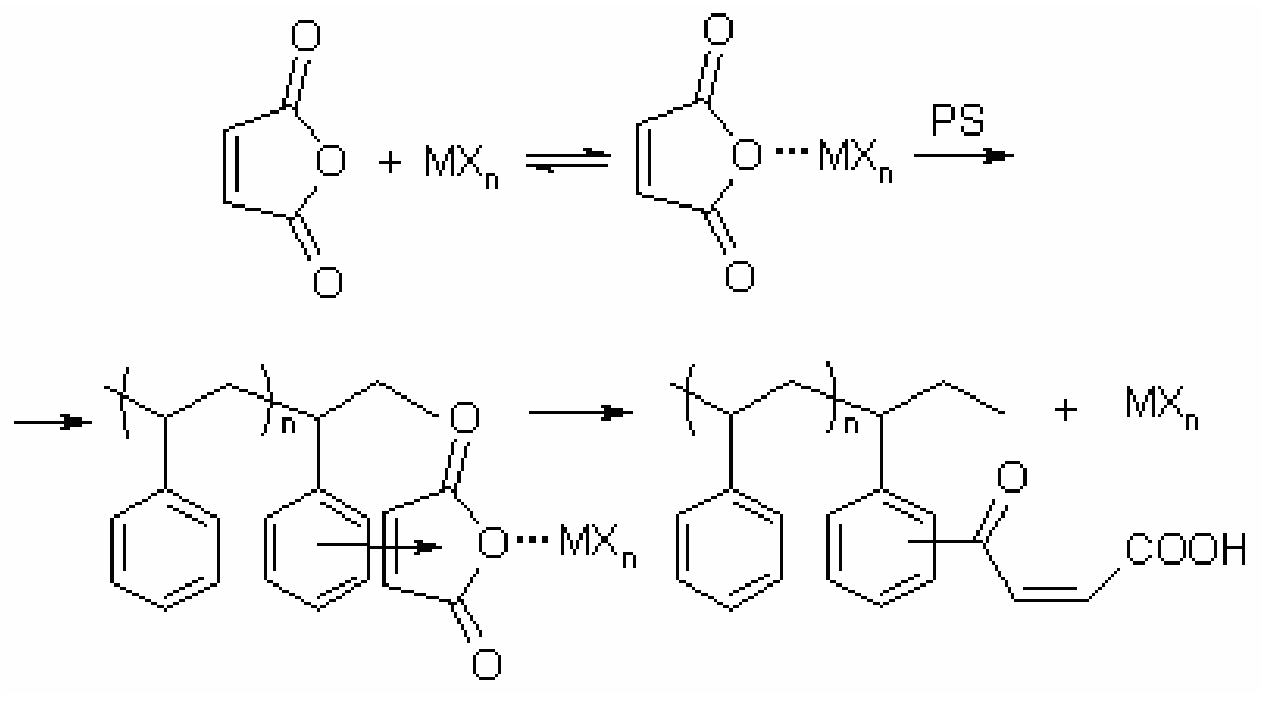

**Figure 16.** Side-chain acylation of PS with MA in the presence of Lewis acid.

As one can see, the scheme consists of several stages: (1) complex formation of Lewis acid with MA ($MX_n$), (2) addition of MA…$MX_n$ complex to the phenyl ring of PS, and (3) break of the hydrogen atom from the *p*- or *o*-position of phenyl ring and its addition to maleate fragment. In this scheme, complex formation plays a very significant role in the acylation reaction. On the other hand, it can also by suggested that in the acylation reaction the CTC between MA and phenyl ring of PS or between complexed MA and PS also can take place, which helps facilitate the acylation reaction [208].

Complex formation between MA and aromatic compounds such as benzene and toluene was noticed in earlier study [264]. It was established that modification of PS by MA occurred in all cationic catalyst media used-however, at different levels. The highest carboxyl group concentration (17.8 and 14.3 %) is found to occur with $BF_3$.$OEt_2$ and $TiCl_4$. But in the case of $TiCl_4$, acetylated PS has a relatively low value of viscosity, which indicates that acylation is accompanied by degradation of the main chain. The highest $M_n$ value was obtained when the $BF_3$.$OEt_2$ catalyst was used.

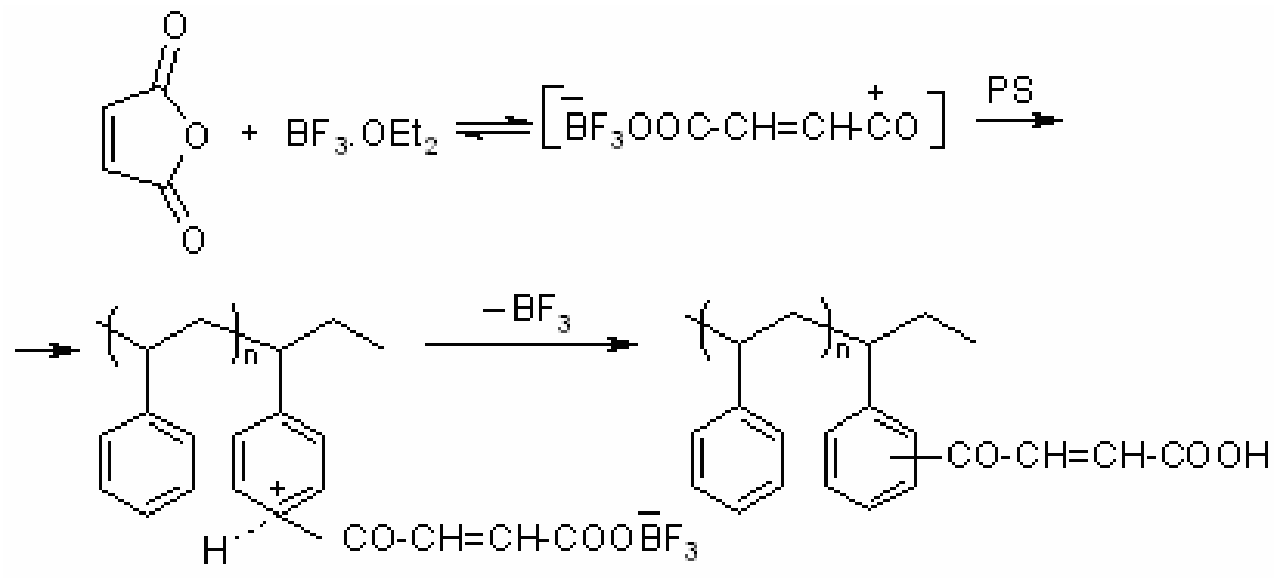

**Figure 17.** Mechanism of side-chain acylation of PS with MA in the presence of $BF_3$.$OEt_2$ catalyst.

In the case of $BF_3$.$OEt_2$ as the catalyst, the reaction scheme can be suggested as follows (Figure 17) [318].

The activities of Lewis acids in the reaction studied are as follow: $BF_3$.$OEt_2$ ≥ $TiCl_4$ > $AlCl_3$ > $SnCl_4$ > $ZnCl_2$ > $FeCl_3$. Functionalized PSs containing –CO–CH=CH–COOH fragments in side chains are characterized by their high thermostability, adhesion, and photosensitivity. Thermal analysis (DTA and TGA) of functionalized PS, especially the observed weight loss and exo-peaks on the curves TGA and DTA, rspectively, allow proposed that the decarboxylation reaction of unsaturated acyl groups with the formation of vinyl ketone fragments in the side chain proceeds in the used isothermal conditions. Then these vinyl ketone groups took place easily in the crosslinking reaction as shown in Figure 18 [318]:

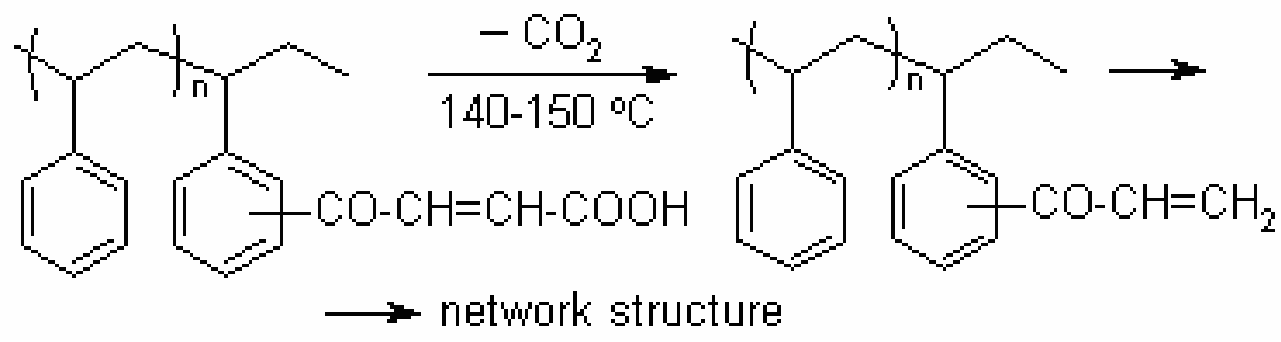

**Figure 18.** Proposed mechanism of the side-chain fragmentation through decarboxylation and the formation of crosslinkable vinyl ketone group in the isothermal conditions.

Similar acylation reaction of benzene ring in bisphenol A was observed Can et al. [319]. They were studied reaction of MA with the bisfenol A(BPA)/soybean oil monoglyceride in the presence of triphenyl antimony as a catalyst at 125 °C during 9 h. From the $^1H$ NMR analysis data, which were indicated that the integral areas of BPA's *o*- and *m*-position protons signals at 7.0 and 6.7 pmm, respectively, were not equal, authors proposed that Friedel-Grafts acylation may be taken place as a side reaction with the maleate half ester formation, i.e., esterification of phenolic hydroxyl groups with MA (Figure 19):

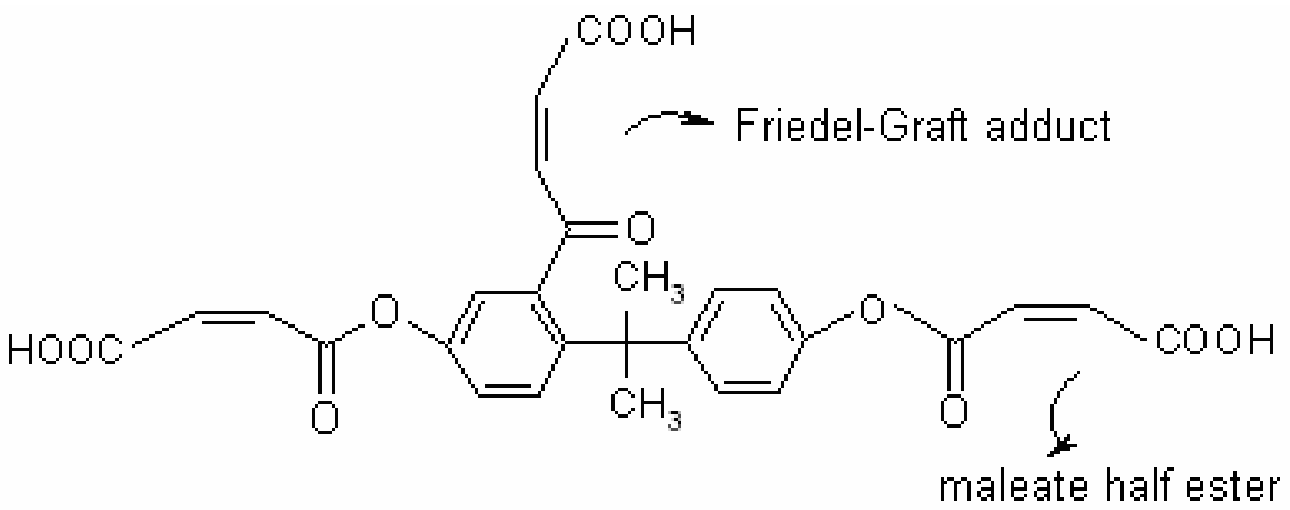

**Figure 19.** Friedel-Grafts acylation of bisphenol A with MA via esterification of phenolic hydroxyl groups. Adapted from [319].

First the studies on the preparation of functionalized high impact PS (HIPS) via melt grafting of MA onto NIPS and its effects on the compatibility of NIPS and PA-1010 reported by Chen et al. [320]. They attempt to graft MA on HIPS through reactive processing and effect of *in situ* compatibilizer of poly(HIPS-*g*-MA) on the final morphology and mechanical property of the immiscible polymer blends of HIPS and PA-1010. Authors noted that commercial high-impact polystyrene (HIPS) with MFI 3.1 g/10 min and polyamide 1010 (PA-1010) with MFI 10 g/10 min are commondity polymers that possess unique properties individually; HIPS is a low price, but tough thermoplastic with relatively pooer solvent resistance; in practice, however, it is difficult to obtain good performance because HIPS and PA-1010 are immiscible. Many studies on the compatibility of immiscible PS/PA-6 blends also reported [128,321-323]. The HIPS grafted with MA (NIPS-*g*-MA) was prepared with melt mixing in the presence of dicumyl peroxide as a free-radical initiator [320]. The grafting reaction was confirmed by IR analysis by the new absorption bands at 1218 (C-O), 1780 and 1857 $cm^{-1}$ (C=O)]. The amount of the MA (1-5 %) grafted on HIPS was evaluated by a titration method. According to authors, the anhydride group in grafted copolymer reacts easily with the amine group at the chain in PA-1010 (I) (Figure 20), propably through both intermediate amide-anhydride and/or secondary amine-anhydride (preferably) complexation. However, poly(HIPA-*g*-MA) has good compatibility with HIPS due to the presence of structurally similar HIPS units in the polymer backbone; thus, authors assumed that the physical and chemical interactions across the phase boundaries will control the overall performance of the polymer blends.

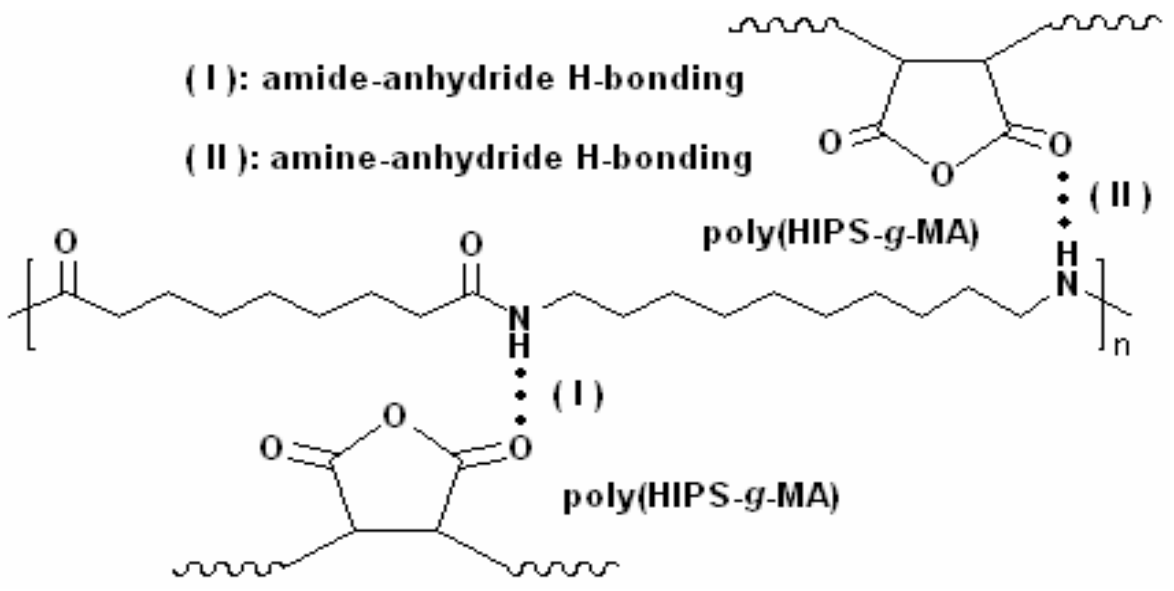

**Figure 20.** Interaction of (NIPS-g-MA) with PA via anhydride-amine complex formation (Effect of compatibilization). Adapted from [320].

Evidence of reactions in the blends of poly(NIPS-*g*-MA)/HIPS/PA-1010 confirmed by authors in the morphology and mechanical behavior of the blends. The improved adhesion in a NIPS-*g*-MA(10%)/HIPS(15%)/polyamide(75%) blend was also detected [320]. Park et al. [322] synthesized several oligostyrenes with terminal anhydride groups using two different methods of functionalization. These maleated oligostryrenes and high molecular weight poly(styrene-*rand*-MA)s were examined as the reactive compatibilizers for the immicible Nylon-6/PS blends. The effect of molecular weight on particle size reduction depended on the basis of comparison, mass of MA functionalized additive, or moles of MA units. It was shown that a high molecular weight random copolymer is most effective when compared on a mass basis. The increase in adhesion between the Nylon-6 and the styrenic phases caused by the *in situ* resction was evaluated by a lap shear technique. The free PS, Nylon-6, and MA functionalized oligostyrene or poly(styrene-*rand*-MA) formed were separated by solvent extraction technique using formic acid and toluene. It was founded that the extent of coupling reaction between the functionalized PSs and Nylon-6 ranged from 25 to 43 %.

Bartus et al. [324] studied the copolymerization of maleimide type monomers such as *N*-phenylmaleimide, *N*-(2,6-dimethylphenyl)maleimide and [2-(2-hydroxy-3-methylenemaleimido-5-methylphenyl)-2*H*-benzotriazole] with acrylonitrile and styrene in the presence of acrylonitrile– butadiene or styrene–butadiene latexes. A new type of polymerizable 2-(2-hydroxyphenyl)-2*H*-benzotriazole UV stabilizer with a reactive maleimide group prepared and utilized for modification of acrylonitrile–butadiene–styrene (ABS) resins to improve the thermomechanical properties and increased resistance to UV radiation. Authors showed that the use of these maleimides as patial substitute for styrene in the ABS composition increases the glass transition temperature of ABS resin. Styrene/acrylonitrile graft-polymer in the presence of the SBR latex containing 51 % of butadiene, 46 % of styrene and 3 % of acrylic monomer ($T_g$ = 110°C) as well as styrene/acrylonitrile graft-polymers with the replecement of 5 mol.% of *N*-phenylmaleimide ($T_g$ = 115 °C), 10 mol. % ($T_g$ = 120 °C) and 20 mol.% ($T_g$ = 130 °C) in the presence of the SBR latex gave high yields of final polymer. Similar results also obtained in graft polymerizations of *N*-(2,6-dimethylphenyl) maleimide of 5 mol.% ($T_g$ = 140 °C), 10 mol. % ($T_g$ = 165 °C) and 20 mol. % ($T_g$ = 225 °C).

Nguyen et al. [325] investigated the possibility of using styrene, a weak donor forming donor/acceptor pairs with electron-poor (EP) vinyl monomers, such as MA, methyl methacrylate (MMA), methyl acrylate (MAC), dimethyl maleate (DMMA), acrylonitrile (AN) and acrylic acid (AA) for initiating spontaneous photopolymerization and photografting of the copolymers onto polypropylene. The possibilities of CT complex formation between styrene, a weak donor, and the EP vinyl monomers MA, MMA, MAC, DMMA, AN and AA were also explored. Among these EP vinyl monomers, MA was found to have initiated spontaneous polymerization with styrene under UV irradiation. Authors used grafting yields together with FTIR analyses to confirm the presence of grafting. Styrene/MMA and styrene/AN systems achieved significant grafting, but such levels of grafting were not observed in the styrene/MAC and styrene/DMMA systems. No grafting was observed for styrene/AA or styrene/MA systems, but the latter system underwent photopolymerization. They evaluated the effect of solvents on grafting in the styrene/MMA and styrene/AN systems, and found that dimethylformamide inhibites the grafting of both donor/acceptor systems. In contrast, chloroform and methanol enhanced grafting of the styrene/AN system although these two solvents had no significant effect on the grafting of the styrene/MMA system.

## 4. Grafting of Synthetic and Natural Rubbers

MA is one of the most widely used functional monomers for the graft modification of PE, PP, PS, ethylene-propylene elastomer and ethylene-propylene-diene monomer (EPDM) rubbers [151,162,164, 313,326-329]. Introduction of MA on the non-polar backbone of polymers has overcome the disadvantage of low surface energy of these polymers. Th,s not only improves the hydrophilicity of the surface of the polymers for the benefit of printing and coating applications but also the adhesion of these polyolefins and rubbers with polar polymers such as polyamide, metal, and glass fibers. These maleated polymers are also used as impact modifiers and compatibilizer in polymer blends [330,331]. Natural rubber (NR) is widely used in various applications particularly for tires because of its excellent elastic properties over other synthetic counterparts. However, its non-polar character limits its application due to poor oil resistance and high air permeability [332]. Grafting of MA or other polar monomers onto NR backbone improves the compatibility between NR and other polar elastomers and some engineering plastics such as polyamide. According to several authors [333-336], the grafting process of MA onto polydienes, such as NR, styrene-isoprene, and styrene-butadiene-styrene block copolymer, involver two different mechanisms, i.e., radical-induced grafting and thermal grafting via an ene mechanism. For the ene mechanism, high temperature (160-240 $^oC$) is required for the fixation of MA to polydiene backbone. In the case of radical mechanism, the reaction usually occurs quickly and at a lower temperature than the ene reaction [336].

Currently, grafting of MA on NR has mostly been carried out in a solid phase by mixing the NR with MA in a kneader, a roll mill, or an internal mixer [332,337]. In these cases, a small percentage of MA grafting achieved and the crosslinking reaction was predominated. Nakason et al. [338] used the MA grafted NR as a compatibilizer in polyamide/NR blends [339] and maleated NR/cassava starch blends. They estimated in these two cases that the improvement of compatibility and mechanical properties of the blends are due to the intermolecular interaction between the anhydride polar groups grafted on NR and the amine end-group of polyamide (amidization reaction) or hydroxyl group of cassava srarch macromolecules (esterification reaction), respectively.

For the characterization of MA-modified polymers, FTIR spectroscopy [263,326-329,337,338,340] and titration [133,215,262,313,339,341] techniques were used to determine the amount of MA grafted on the polymer backbone. According to the authors, it would be possible to widen the area of application of NR by chemical modification of NR with MA, as it is a way to introduce not only the polar moiety on the backbone of NR bur also to introduce another type modification if a required amount of the MA can be prepared and quantitatively determined [338]. Grafting of MA onto NR and influence of styrene on the grafting efficiency investigated by Saelao and Phinyocheep [332] using a radical initiated grafting process. Grafting was carried out with 5-25 mol % of MA in toluene solution in the presence of benzoyl peroxide (0.5-1.5 mol %) at 60-80$^oC$ under nitrogen atmosphere. It was found that the grafting degree is higher at relatively high temperature (80$^oC$). Compared to the reaction without styrene, the addition of 0.1 mol % of styrene monomer, in the grafting process increased % MA grafting up to two times. Authors postulated that the styrene may act as a charge transfer complex for the reaction of NR with MA.

It is known that styrene as a electron-donor monomer was used to activate the MA monomer for grafting onto PE and PP [164,329]. The styrene comonomer may serve as a medium to bridge the gap between the NR macroradical and the MA. In the case of styrene-assisted MA grafting on to PP, increasing the amount of styrene resulted in increasing the amount of MA as well as styrene grafted on the polymer backbone [164]. It was found that a very small amount of styrene (0.01 mol %) also affected an increase of MA grafting onto NR at an early stage of reaction time (up to 2 h) but at longer reaction times there is not much difference in content of MA with or without the styrene. Authors assumed that at early stage of rection time, when styrene is added, styrene and MA can interact with each other one-to-one to form a charge transfer complex, which increases the electron density of MA double bond, resulting in enhancing the MA’s reactivity, hence increasing the grafting degree [332]. Authors used NR with highly *cis*-1,4-polyisoprenic structure. It possesses labile allylic protons in every repeating unit. These protons are prone to hydrogen abstraction by radical active species generated from radical initiators similar to the grafting process of MA on styrene–isoprene and styrene–

butadiene–styrene (SBS) block copolymer [333,334]. Anhydride functionalized NR easily hydrolyzed with formation free carboxylic groups in side-chain macromolecules (Figure 21).

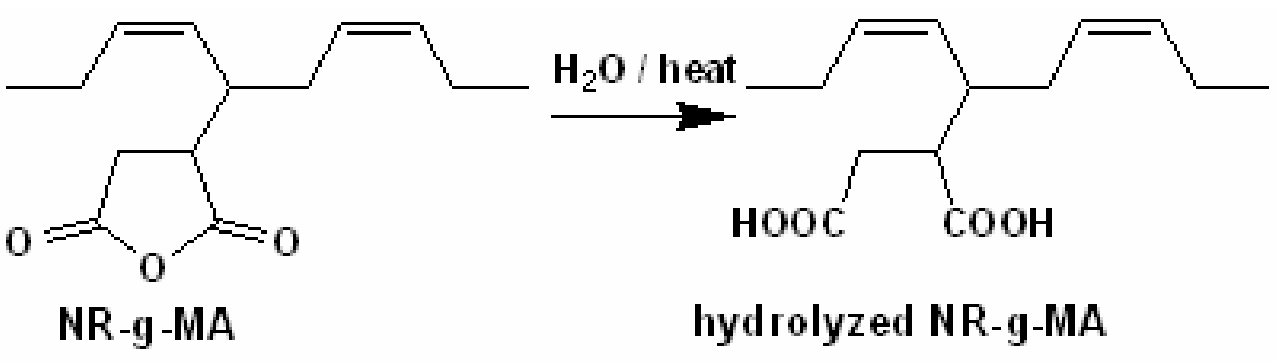

**Figure 21.** Hydrolysis of MA functionalized NR. Adapted from [333].

Authors detected the gel formation in same conditions of MA grafting onto NR. According to the authors, this was not surprising as the grafting reaction that occurred through the radical mechanism (*a* and *b*), as well as via acid-anhydride interaction (*c*) as shown in the following scheme (Figure 22) [332]:

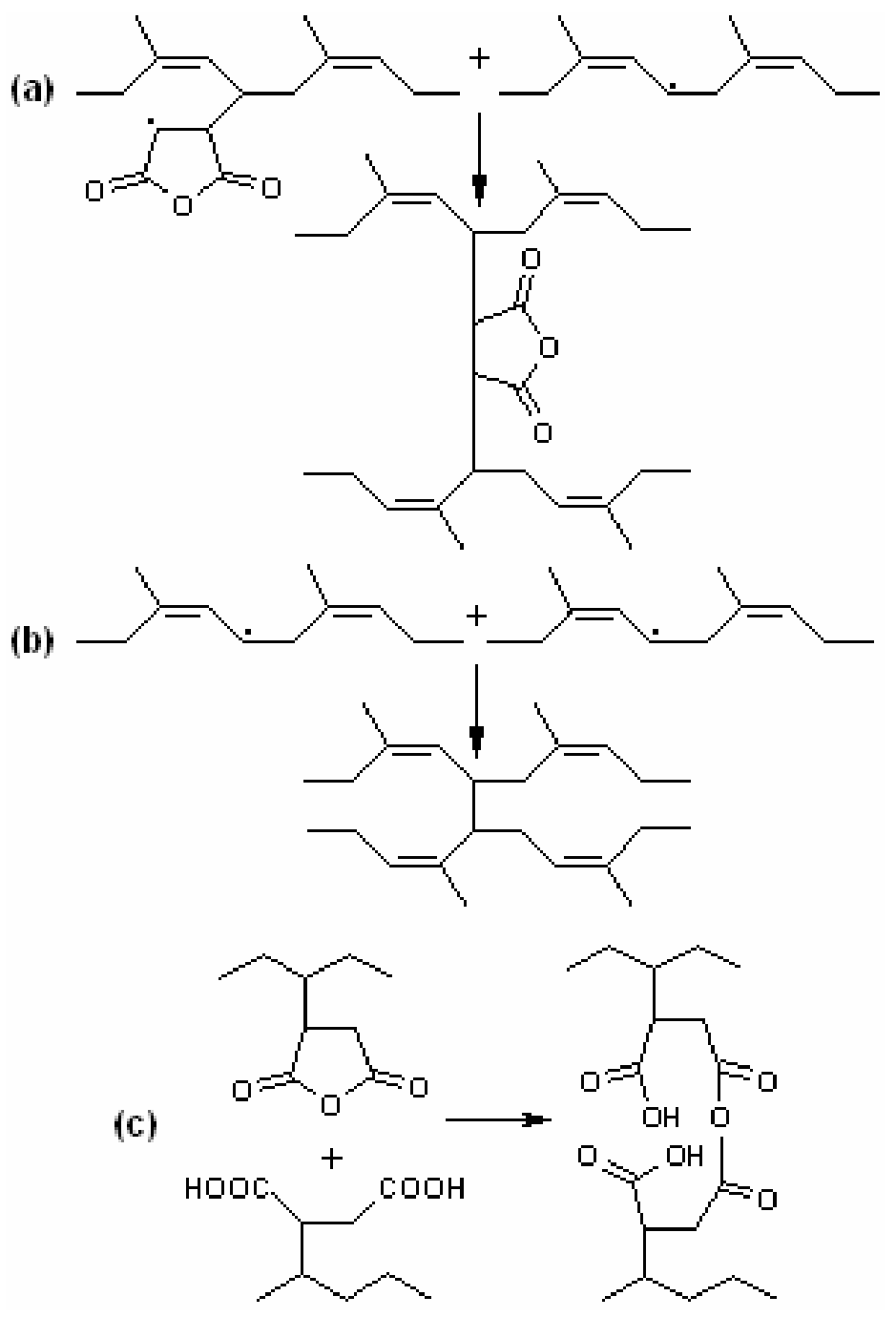

**Figure 22.** Radical grafting of MA onto NR via acid-anhydride interaction. Adapted from [332].

The MA modification of different kinds of rubbers is a useful way of compatibilizing immiscible polymer blends as well as improving interfacial adhesion in poliymeric composites. Thus, the modification of EPDM rubber with MA in a twin screw extrusion process has been described by Oostenbrink and Gaymans [341], in the presence or absence of bis(*t*-butyl peroxyisopropyl)benzene as an initiator. They proposed that the knowledge of MA content in the rubber by covalent linking is important not only to evaluate its application but also to choose the best manner of grafting. For the determination of MA unit in the grafted polymers, different method, such as gravimetry, titration of acid group, IR and NMR spectroscopy, and so on were used.

Couitinho et al. [342] used Multiple Reflectance (MR) FTIR spectroscopy in order to simplify the method of quantitative determination of the percentage of grafting in the polymers, especially in the EPDM chain. The EPDM elastomers (1.7 % of norbornene unit) was chemically modified with MA in chlorobenzene solution with benzoyl peroxide as an initiator. The presence of MA grafting on elastomer was made evident by MR-FTIR analysis using 1856 and 1780 $cm^{-1}$ bands of anhydride carbonyl groups.

MA grafted to PE, PP, polyisobutylene, poly(vinyl chloride) (PVC), and PS using mechanochemical (with and without free-radical) and free-radical, ionic, and radiation-initiation techniques [140,141]. Grafting MA onto PE, PP, PS and polyisoprene (natural and synthetic rubber) provides copolymers with a typical structural residue (**I**). According to the authors, two structures (**II** and **III**) are possible in the case of MA grafted polyisoprene (Figure 23), depending on the initiation mechanism.

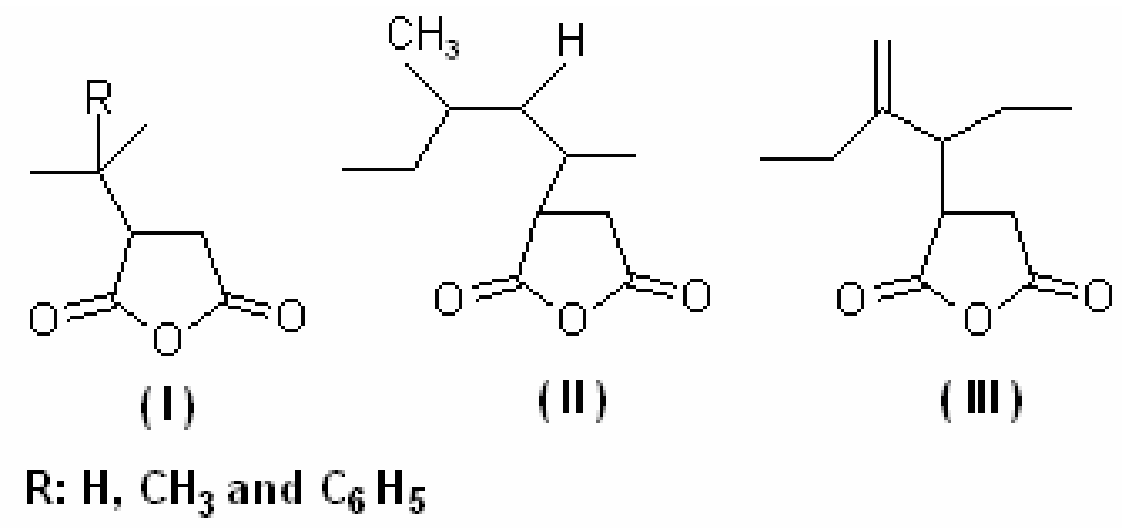

**Figure 23.** The different graft copolymer structures as results of MA grafting onto PO and PS (I) and natural rubbers (II and III). Adapted from [141].

Maleinization of liquid polybutadienes with molecular weight around 1000-5000 is industrial process [233]. This process is similar to that unsaturated natural oil and mechanism of "ene reaction" was suggested [343-345]. Maleinated products contain succinic anhydride units randomly distributed along the polymer chain. These graft copolymers, after hydrolysis of half-esterification by alcohols of the anhydride ring followed by neutralization, become water soluble or dispersible and can yield coatings by crosslinking with polyamines, polyols, epoxy resins, etc. [346-348]. Obtained coatings being water-based, are environment-friendly and show good adhesion on various substrates including steel and synthetic fabrics. Moreover, they have good water resistance, electrical insulation properties and low temperature stability. Therefore, typical applications are in the field of electrodeposition paints, electrical potting and encapsulation, sealing and adhesive compounds [349-351]. On the other hand, the residual unsaturations enable its use as water soluble and oxidative binder. They are compatible with rubbers and resins so that the grafting of acrylonitrile–butadiene rubber with MA has been reported to improve the compatibility with natural rubbers [351].

Ferrero et al. [352] studied the kinetics of polybutadiene bulk maleinization in working conditions of an industrial process (temperature in the rangr from 180 to 220$^oC$ and time from 2 to 4 h). It was found that the reaction rates are affected by the polybutadiene microstructure, since higher reaction rates were shown by the 1,4-oligomers. Then this bulk reaction was investigated by calorimetric analysis, which has confirmed that crosslinking reactions also take place. They also studied solvent effect in grafting of liquid polybutadienes with MA using *o*-xylene and decahydro-naphtalene as solvents. It was observed that a final constant yield depends on microstructure and MA concentration. In decahydronaphtalene at 130°C, the reaction was very fast with very high yields. Nevertheless, in the range from 150 to 180°C, amount of linked MA in the final products did not exceed 11 %. This fact author explained by probably decarboxylation of the maleated polymers which favoured by decahydronaphtalene solvent. The effective grafting yields were higher below 150°C.

Several researchers [353-356] grafted MA–styrene copolymer onto a variety of conventional polymers using free-radical initiated graft copolymerization in melt state. They reported spontaneous bulk polymerization of MA/styrene mixture in the presence of PE, PP, PS, PVC, poly(ethylene-*co*-propylene), poly(styrene-*co*-acrylonitrile), poly(*cis*-1,4-polybutadiene), poly(acrylonitrile-*co*-butadiene-*co*-styrene) (ABS) and poly(butadiene-*co*-acrylonitrile) by using various type mixing apparatus and extruders.

Triblock copolymer of styrene-*b*-(ethylene-*co*-butene-1)-*b*-styrene (SEBS) was functionalized with MA by several researchers [357-361]. These grafted copolymers were utilized as compatibilizers for the immiscible polymer blends. According to the authors, SEBS-*g*-MA copolymer seems to be a better compatibilizer than similarly functionalized polyolefin such as (ethylene-*co*- propylene)-*g*-MA.

## 5. Grafting onto Biodegradable Polymers

### 5.1. Polysaccharides

In recent years, there has been a great interest in the development of biodegradable plastics from renewable resources. Earlier publications Vaidya et al. [362], Bhattacharya et al. [363-366] and other researchers [367,369] indicates that blends of anhydride functional polymers and starch could lead to products with useful end properties. Wu et al. [370] described the researchs and application areas of modified plant cellulose, preparing by radiation grafting of functional monomers onto celullose macromolecules. Research on blending between polysaccharides (preferably cellulose and starch) and polyolefins has a long history, but their significant dfference in character results in poor compatibility of starch and polyolefins. In order to improve this poor compatibility, many researchers have done much work, such as the chemical modification of both starch [367-370] and polyolefins [371,372] and/or the inroduction of compatibilizer [373-377] into the blends of starch and polyethylene (PE). These compatibilizers contain poly(E-*co*-acrylic acid), poly(E-*co*-MA) or poly(E-*co*-vinyl alcohol) copolymers. Of all the modifying approaches, the compatibilizer intraduced into the blends is considered the most effective method. However, most of the compatiblizers are either expensive or difficult to produce.

Recently, *in situ* grafting methods were developed for the compatibilization of many immiscible polymer blends. One method is the “two-step” process, in which polymers are functionalized selectively in the first step, and then blended in an extruder in the second step. In this case, *in situ* grafting reactions should occur between the functional groups of both polymers in melt. The other approach is a “one-step” process, including the addition of low molecular weight compunds into the melted blends to initiate graft/coupling reactions at the interface and form graft/block copolymers during extrision. This method proved to be successful and low-cost. Because of the formation of the graft and block copolymers, authors [378] used as compatibilization during one-step reactive extrusion. Thus, Wang et al. [379] developed an effective method of compatibilization of the thermoplastic starch/PE blends by one-step reactive extrusion, in which the formation of poly(PE-g-MA) played a role of compatibilizer between PE and starch. The blending samples preparing in the presence of dicumyl peroxide, MA, starch and PE in a single-screw extruder were characterized by means of TGA, SEM and FTIR analysis. According to authors, a uniform viewpoint that poly(PE-g-MA) using as compatibilizer is based on two factors: (a) the ester-forming ability of anhydride groups with hysdroxyl groups on starch, and the hydrogen-bond-forming ability between carboxyl groups of hydrolyzed MA and hydroxyl groups on strach; (b) the good compatibility between grafted PE chains and the PE phase.

Cellulose is themost abundant natural polymer, with as much as 300 billion tons a year formed on each. Because of its renewability, biodegradability, and non-toxicity, increasing attention has been paid to the use of cellulsic materials for novel composites [380-386]. On the other hand, it is know fact that the inherent incompatibility of hydrophlic cellulose fibers with hydrophilic thermoplastic polyolefins inhibits good adhesion between cellulose and the matrix, making the composites impractical. Many researchers performed to enhance the interfacial adhesion [376,379,387-390]. It was found that MA grafted PP, poly(PP-*g*-MA) is very efficient for preparing cellulosic PP composites with improved mechanical properties. The improvement imparted by the use of poly(PP-*g*-MA) is considered to be attributed to esterification reaction between anhydride and hydroxyl groups of used polymers in solution in the presence of catalyst. The formation of ester groups was confirmed by the appearance of a new IR band at 1729 and 1750 $cm^{-1}$ [391-395]. However, no direct evidence of esterification demonstrated by authors from the melt-mixing of cellulosic materials with poly(PP-*g*-MA), although the relevent composites are generally prepared by a melt-mixing method. According to the authors, this implies that highly crystalline cellulose is too stable to form a number of ester bonds with poly(PP-*g*-MA) in a melted state, because it processess very few free OH groups because of extensive formation of hydrogen bonds [396,397].

For the melt-mixed composites of crystalline cellulose and PP compatibilized with poly(PP-*g*-MA), even an amount of grafted MA as small as 0.2 wt.% is sufficient to impact maximum mechanical properties to the resulting composites [173,398,399]. This result is in agreement with the limited esterification between crystalline cellulose and poly(PP-*g*-MA). However, Hirotsu et al. [398,400] found that ball-milling induces intensive esterification between the OH groups on microparticles of cellulose and the MA units of poly(PE-*g*-MA) or poly(PP-*g*-MA), in marked contrast to the melt-mixing of the original cellulose and these graft copolymers, in which the number of ester bonds is too few to be observed by FTIR spectroscopy [400].

Blends and composites of biopolymers and synthetic polymers also constitute a promising area of material science and engineering. Recently, growing interect has been directed to the development of novel composites of polyolefins/natural polymers, which are promising as new environmental propection plastics. According to Bumbu et al. [401], the properties of pure synthetic polymers and pure biological polymers are often inadequate for producing materials with good chemical, mechanical, thermal, and biological performance charateristics, so researchers have been trying to prepare blends of synthetic polymers with biological macromolecules to obtain new materials with enhanced functional properties and biodegradability at a relatively low cost, so-called bioartificial polymeric materials. They are produced in different forms, including films, sponges, and hydrogels, and have been evaluated as biomaterials for dialysis membranes, wound dressings, drug delivery systems, and so foth. Thus, the compatibility of pullulan and dextran polysaccharides with poly[MA-*alt*-vinyl acetate (VA)] in the solid state in the form of thin film was studied with TGA, DSC, FTIR and electron microscopy. With respect to observed morphology by authors, blends with a content of pullulan greater than 85 wt % exhibited an even distribution of finely disperced particles. From comparative analysis of the obtained results for both dextran/copolymer and pullulan/copolymer blends, authors remark that the type of glycosidic linkage from the main chain of a polysaccharide with the same structural unit (but different number of primary OH groups) is decisive for the morphology and thermal behabior of their blends with poly(MA-*alt*-VA). The thermal properties were dependent on the mixing ratio, and the interactions between components were quite pronounced in the pullulan-rich blends. Both pullulan-containing and dextran-containing films can undergo a crosslinking esterification teaction with MA copolymer via heating. New potential biomaterials based on mixtures of PVC and pullulan with different MA copolymers as compatibilizing agents have also been reported [402].

In the last years, increased interest has developed with respect to reactive blends on the base of MA-containing polymers, such as poly(MA-*alt*-styrene), poly(MA-*alt*-ethylene) and poly[(E-*co*-P)-*g*-MA], and polysaccharides (preferably starch) for the preparation of the compatible and biodegadable polymer materials [374,403-405]. As evidenced from these studies, in the reactive poly(PE-*g*-MA)/strach blends [375,376,406], MA grafted PE as a compatibilizer plays three important roles: (1) the anhydride units easily esterifies with hydroxyl groups of strach, (2) hyrolysed MA unit forms hydrogen-bonding with free OH groups of strach, and (3) provides the substantial compability between grafted PE macromolecules and pristine PE phase. With the similar purpose, Wang et al. [407] developed a single-screw extrusion method of preparing the *in situ* modified polymer blend on the base of LLDPE, MA as a chemical active modifier, dicumyl proxide as a initiator, strach, glycerol as a plasticizer. The morphology of these blend with various composition, studiying by authors using SEM, indicated that with the addition of MA {with formation of poly[(LLDPE-g-MA(0.1 or 0.2 wt.%)]}, the blends have good interfacial adhesion and finely dispersed strach and LLDPE phases, which is reflected in the mechanical and thermal behavior of the blends. Authors illustrated the mechanism of interfacial chemical reaction as follows (Figure 24):

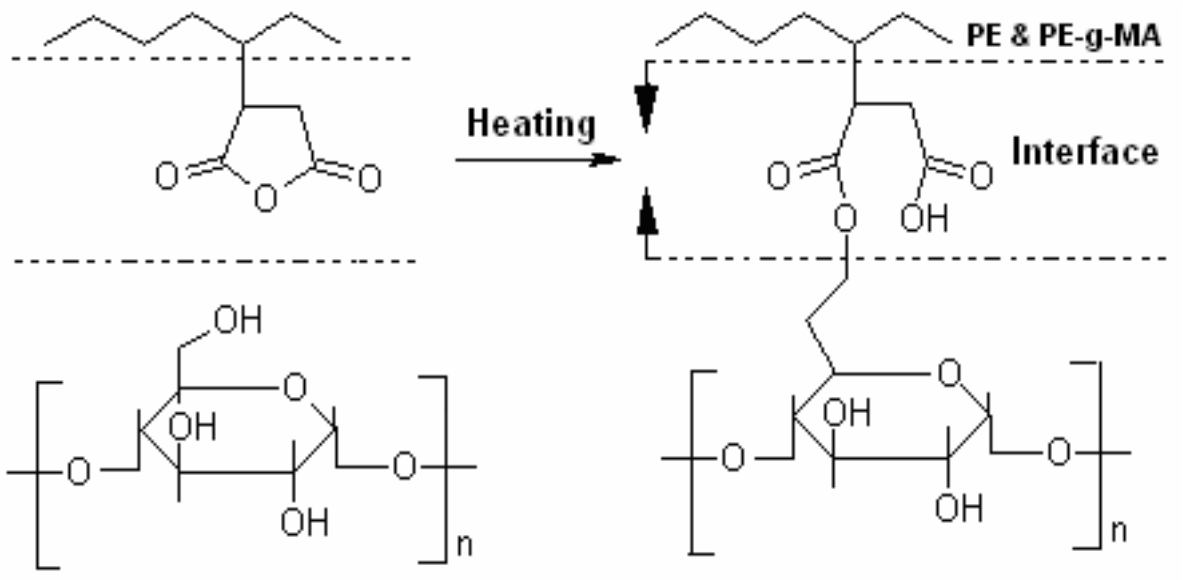

**Figure 24.** Combatibilizing effect of PLLDPE-*g*-MA via interfacial chemical reaction in starch/PE blends.
Adapted from [407].

These composites containing MA modified linkages showed higher tensile strength, elongation at break, and thermal stability than those of blends without MA. The rheologic properties of the blends demonstrated the existence of processing [407].

To clarify the nature and structure of the interface, Bumbu et al. [408] described the structure and thermal properties of PP-*g*-MA chains bound to microparticles of cellulose, which are derived from a composite prepared by ball-milling highly crystalline fibrous cellulose (CF11, 50-350 μm in length, about 20 μm in diameter, and 93 % in crystallinity) and poly(PP-*g*-MA) (MA unit content ~0.6 wt. %, MFI 115 g/10 min at 190 $^{o}$C, 2.16 kg) (Figure 25), and demonstrated that the nature of the bound PP-*g*-MA chains determines the interfacial structure and, accordingly, the tensile properety of the composite. The crystallinity of the PP-*g*-MA bound to the cellulose particles in the extraction residues is confirmed directly by wide-angle X-ray diffraction (WAXD). The XRS (X-ray photoelectron spectroscopy) analysis of C/O ratio for the residues reveals that the cellulose particles bind the MA grafts at the end of PP-*g*-MA chains, with the hydrophilic hydrocarbon sides stretching out from the core of cellulose. Authors observed interesting fact that the PP-*g*-MA chains bound to the cellulose particles crystallize with each other as well as in theneselves, the extent of crystallinity increasing with the number of bound PP-*g*-MA chains, even though the crystallinity is much smaller than that neat PP-*g*-MA [408].

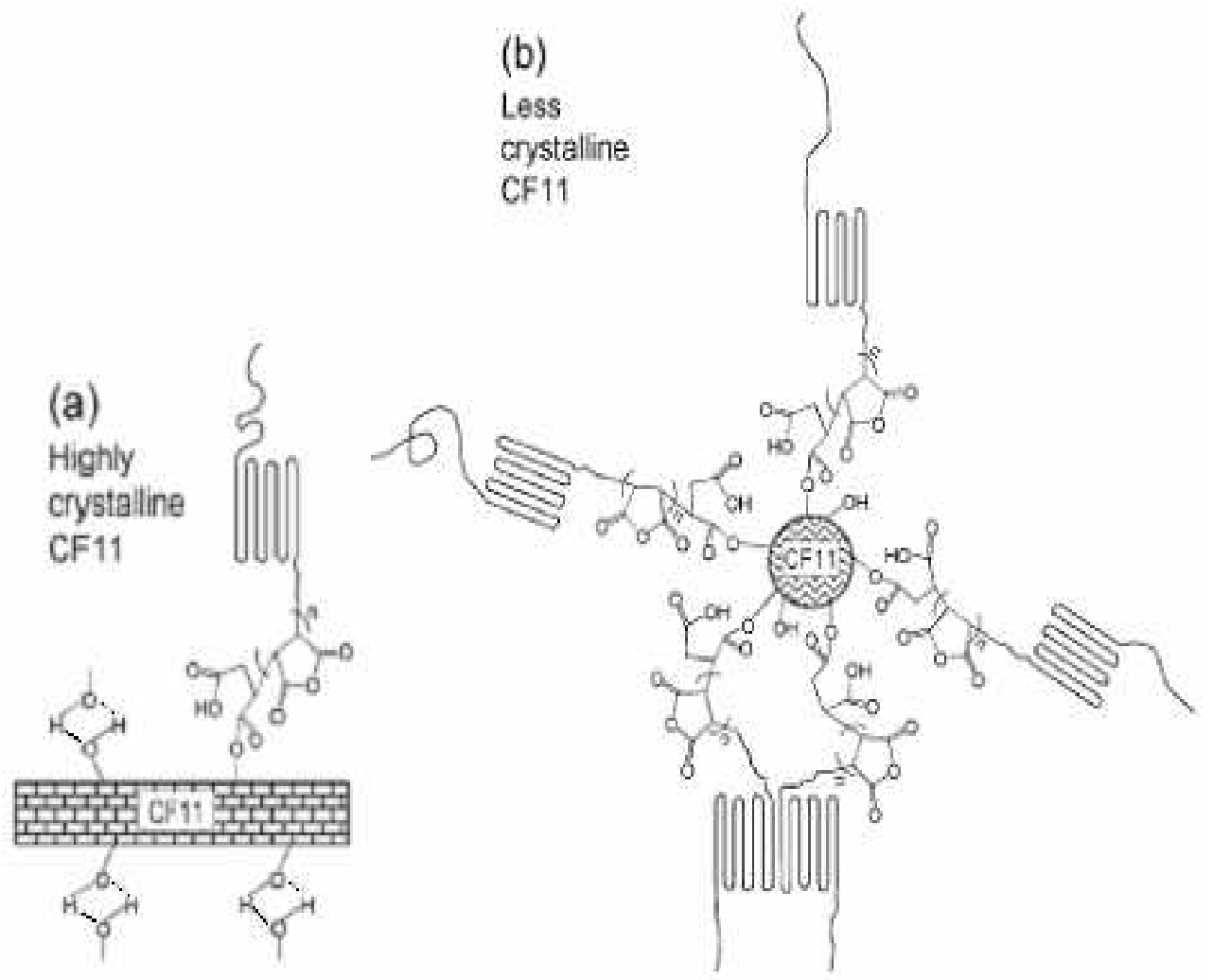

**Figure 25.** The nature and structural model of interface linkages in PP-*g*-MA chains bound to microparticles of (a) high and (b) less crystalline cellulose. Adapted from [408].

According to Zhang et al. [173] ball milling of crystalline cellulose with maleated polyethylene (PE-*g*-MA) yields a novel composite with ester bonds formed by reaction of the hydroxy groups of the cellulose with the MA groups of graft copolymer (MPE) (Figure 26).

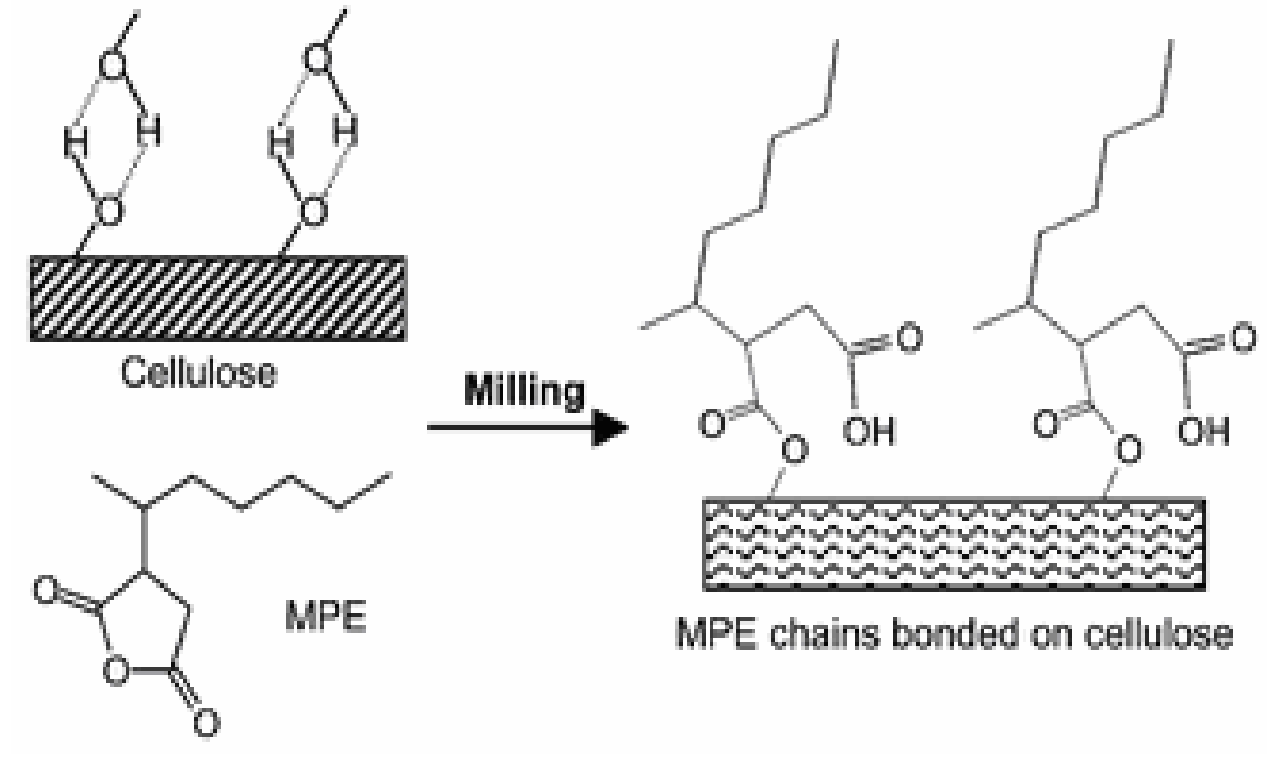

**Figure 26.** Model of reaction between crystalline cellulose and PE-*g*-MA via esterification for the preparation of cellulose particles with improved properties. Adapted from [173].

Authors found that the composite exhibits much improved toughness and ductility compared with the product formed by melt-mixing, probably because of the formation of an interphase of MPE chains bonded on cellulose particles; the mixtures compatibilized with poly(MA-*alt*-S) and poly(MA-*co*-MMA) or with a poly(E-*co*-P-*g*-MA) exhibited a higher hydrophilicity and biocompatibility that the pullulan/PVC blend.

Compatibile blend is also formed from mixture of hydroxypropyl cellulose and poly(MA-*alt*-VA) [408]. Authors established that the physical interactions are prevalent in blends with a high content of polysaccharide, whereas chemical interactions predominate in blends with a medium and low content of this polysaccharide. By increasing the temperature, the thermochemical reactions (esterification/ crosslinking) are favored. Poly[E-*co*-ethyl acrylate(32%)-*co*-MA(1.5-2.5%)] terpolymer, containing a reactive anhydride unit, can react with amino and hydroxyl groups of protein.

Some studies already showed that the synthetic polymers with the MA group had improved compatibility with soy protein and starch [409-412]. Liu et al. [412] prepared starch-LDPE blends with the addition poly(PE-*g*-MA) using a lab-scale twin-screw extruder. The effect of MA grafted PE on the thermal properties, morphology and tensile properties of blends were studied by DSC, SEM methods and mechanical test technique. They showed that the interfacial properties between starch and LDPE improved after poly[PE-*g*-MA (3 wt %)] addition, as evidenced by the structure morphology revealed by SEM. According to the authors, the mechanical properties (tensile strength and elongation ar break) of starch-LDPE-poly(PE-*g*-MA) blends were greater than those for LDPE-starch blends, and their differences became more pronounced at higher starch content.

The esterification reaction between different wood fibers and poly(PP-*g*-MA) has been a subject of investigation of several researchers [391,392,394]. X-ray photoelectron spectroscopy and FTIR spectroscopy are the methods most often used to confirm the this esterification reaction. Paunikallio et al. [413] studied the influence of maleated PP as a coupling agent on the mechanical properties of viscose fiber/PP composites. These composites were characterized by FTIR and mechanical testing. The most notable results was the effect of the poly(PP-*co*-MA) concentration on the tensile strength of the composites; the tensile strength increased from 40 to 69 Mpa when maleated PP was added in amounts to 6 wt.% of the fiber weight. The interaction between graft copolymer and fiber was confirmed by FTIR analysis. A new band at 1740 $cm^{-1}$ was observed in the FTIR spectra of the poly(PP-*co*-MA) treated fibers with different extrusion times. The anhydride groups react more easily with hydroxyl groups than to the corresponding carboxylic groups.

Sugama and Cook [414] showed that poly(itaconic acid) or poly(maleic acid) as a more effective reactant than other polycarboxylic acids containing a single carboxylic groups was used for the chemical modification of industrial chitosan (CS) biopolymer in water at 200°C. One the other hand, it is known that in an aqueous medium at pH > 6.5, CS, containing amino group, becomes a linear polybase electrolyte having a highly positive charge density [415].

Authors found that the grafting and crosslinking reactions proceed between poly(itaconic acid) and CS leading to the formation of amide bonds [414]. These amidizated SCs recommended by authors as effective anticorrosion coatings to prevent the corrosion process of aluminum substrates. The interaction of the free carcoxylic groups in the modified SC with the hydroxylated Al metal surfaces is formed an interfacial covalent oxane-bond structure, such as –COO–Al–linkages; according to the authors, this is one of important factors playing a key role in conferring resistance to Al surface corrosion.

Zhang et al. [416] used MA as a nontoxic reactive compabilizer to improve mechanical properties of poly(lactic acid) (PLA)/starch blends in extrusion. In this case, interfacial adhesion between PLA and starch was significantly improved. Mechanical properties increased markedly compared to the virgin composites of PLA/starch. The PLA/starch composites at a constant ratio of 55/45 compatibilized by 1% MA and initiated by 10% L101 (MA basis) resulted in the highest tensile strength and elongation. A PLA/starch blend (55/45) with 1 % MA and 10 % initiator (MA basis) had a tensile strength of 52.4 Mpa, significantly higher than the 30.0 Mpa of a virgin PLA/starch blend with similar composition. However, it was shown that elongation at break remained almost the same as the virgin PLA/starch blend. Plastizers improved the blend's elongation but also reduced tensile strength. The plastizer in the blend also suppressed compatibilizer efficiency [417]. Zhang and Sun [418] studied mechanical and thermal properties of PLA/starch/dioctyl maleate (DOM as the compatibilizer and plasticizer) blends, as well as the influence of a polymeric DOM, a derivative of MA, on the mechanical properties of these blends. It was observed that DO acted as a compatibilizer at low concentrations (below 5 %), and markedly improved tensile strength of the blend. However, DOM functioned as a plasticizer at concentrations over 5 %, significantly enhanced elongation. Compatibi-lization and plasticization took place simultaneously according to the analysis of mechanical and thermal properties of blends.

The investigations concerning interpolymer complex formation between a synthetic and a natural or semi-synthetic water-soluble polymer have become of great interest. Iwaka et al. [419] studied polymer/polymer complex formation in poly(ethylene oxide)/polyacid series systems including maleic acid (MAc)-styrene copolymer [poly(MAc-*co*-S)]. Recently, the ability of MAc-vinyl acetate copolymer to form interpolymer complexes with a series of polybases also reported by Vasile et al. [420]. Bumbu et al. [421,422] discussed the capability of hyroxypropyl-cellulose (HPC) to form an interpolymer complex with poly(MAc-*co*-S) in aqueous dilute solution and evaluated the thermodynamic function of the complexation process. The formation of interpolymer complexes between HPC and poly(MAc-*co*-VA), poly[MAc-*co*-acrylic acid (AA)] and poly(MAc-*co*-S) in aqueous solution also investigated by these authors using turbidimetry, viscometry and fluorescence measurements [422]. The results of viscometry studies of authors indicated a compact structure for the complexes between HPC and maleic acid copolymers. Besides H-bonding, strong hydrophobic forces materialize between HPC and poly(MAc-*co*-S), strengthening this this interpolymer complex. The strengh of the interpolymer interactions was estimated to increase in the order: HPC/ poly(MAc-*co*-VA) < HPC/poly(MAc-*co*-AA) < HPC/poly- (MAc-*co*-S). The fluorescence measurements demonstrated the contribution of hydrophobic interactions to the stabilization of interpolymer complexes formed between HPC and maleic acid copolymers. The strength of the hydrophobic interaction depended on the hydrophobic/ hydrophilic behavior of the maleic acid comonomer unit. According to the authors, the observed strongest hydrophobic interactions appeared between HPC and poly(MAc-*co*-S) due to the presence of the hydrophobic styrene unit in the polyacid. The interpolymer complexes of HPC with poly(MAc-*co*-VA) or poly(MAc-*co*-AA) were water-soluble at pH values higher than 3, while the HPC:poly(MAc-*co*-S) complex was soluble at pH values above 4.5 Dicarboxylic acid monomers such as itaconic acid (IAc) and others were used as crosslining agents for cellulose [423].

Naguib [314] studied grafting reaction of itaconic acid (IAc) onto sisal fiber, chemical compositon of which a multicellular nature containing 62 % of true cellulose, 16 % of pentosan, 10 % of hemicellulose, 8 % of lignin and 2 % of waxes etc. using potassium persulfate as initiator. It was demonstrated that this grafting method is an efficient way to improve the dyeability of sisal fibers. The ctystalline lattice of the fibers is not changed after the graft copolymerization. The thermal stability and tensile strength decrease by increasing the grafting degree. According to the author, this observed effect probably due to the ease of the decarboxy-lation reaction of the itaconic unit.

A great deal of attention has been paid to the application of biodegradable polymers, such as starch, liglin, cellulose, gluten, and chitin and its derivatives to replace the conventional thermolastic polymers, predomnantly polyolefins that cause significant environmental problem because of their non- biodegradability [425]. Because PE is one of the most extensively produced nondegradable polymer and various types of PEs are used extensively in many fields, including agricultural and food-packaging films [426], ther has been an increased interect in enhancing the biodegradability of PEs by blending them with a cheap natural biopolymer [427]. Starch that shows a high biodegradation rate is a blend of amylose and amylapectin, both of which are polysaccharides composed of α-D-glucopyranosyl units, $(C_6H_{10}O_5)x$ [428]. Even since Griffin [429-431] used granular starch as a filler in PE to enhance the biodegradability of the PE blend system, many studies have been focused on the increase of the mechanical properties and processibility of the starch/polyolefin blend system. According to Bikiaris et al. [432], compatibility between low-density PE and plasticized starch could be increased with a relatively small amount of poly(PE-*g*-

MA), and the resultant blend system still showed good biodegradability. Incorporation of MA and styrene onto cellulose produces a symmetrical amphiphilic graft copolymer. Membranes with hydrophobic and hydrophilic/ ionic moieties are expected to posssess a high degree of permselectivity and permeability combined with longer stability [433]. Such polymer chains also form complexes with low molecular weight surfactants both in water and media of low polarity [434]. MA is an unique monomer which, in its ionized or derivatized form, acts as a “molecular anchor” for aggregating additional molecules on a membrane surface. However, Wu et al. [140] assumed that independent graft yield of MA is both low and dependent on its feed concentration.

According to Chauhan et al. [435] copolymerization of MA with styrene, however, not only confers alternation and amphiphilic characteristics, but optimizes graft yield through the formation of charge transfer complexes. To develop amphiphilic alternating graft copolymers for use as cost effective membrane materials in separation and enrichment technologies and as potential biomaterials, authors studied graft copolymerization of MA and styrene onto *Pinus roxburghii* cellulose initiated by γ-irradiation. Total conversion, grafting efficiency, grafting degree, and rates of polymerization and grafting were determined as a function of MA concentration. Observed the high degree of kinetic regularity and the linear dependence of the rate of polymerization on the MA concentration, along with the low and nearly constant rate of homopolymerization allow authors to suggest that the monomers first form a complexomer which then polymerizes to form grafted chains with an alternating sequence. Grafting parameters and reaction rates achieve maximum values when the molar ratio of styrene to MA is 1:1. Further evidence for the alternating monomer sequence is obtained from quantitatively evaluating the composition of the grafted chains from the FTIR spectra, in which the ratio of anhydride absorbance (1733 and 1728 $cm^{-1}$ for C=O) to aromatic (1646 and 1613 $cm^{-1}$ for C=C) absorbance for the stretching bands assigned to the grafted monomers remained constant and independent of the feed ratio of MA to styrene. Thermal behavior of the graft copolymers revealed that all graft copolymers exhibit single stage decomposition with characteristic transitions at 161-165 $^{o}$C ($T_m$ for graft chains) and 290-300 $^{o}$C ($T_d$).

Raj et al. [436] reported lignocellulose-filled thermoplastic composites with improved mechanical and physical properties. Rozman et al. [437] used MA to chemically modify the lignocellulose fiber before its incorporation into PP. The anhydride group was expected to be sufficiently reactive with the hydroxyl groups of the lignocellulose fiber. MA chemically attached to the lingnocellulose surface, might serve as a bridge between the former and the PP matrix. According to the authors, the better compability between these two components provides high grafting degree, and therefore, subsequently enhance the mechanical and physical properties of the composite; MA treated fiber/PP composites showed lower water absorption and thickness swelling than those with untreated fiber.

## 5. 2. Polyesters

Polylactide (PLA) relating to class of natural polyesters is an important biodegradable polymer which has been used in such established applications as medical implants [438], sutures [439], and drug delivery systems [440]. Introducing new functional groups onto the PLA backbone paves the way to prepare composites, laminates, coated items, and blends/allows with improved proerties and cost effectiveness. Functionalizing matrix polymer and the fiber/filler with highly reactive groups is perhaps the most successful strategy, leading to a variety of commercial composites and alloys made by reactive processing [441].

An effective way to improve the compatibility between polyester matrix and starch is to functionalize the polyester matrix by grafting highly reactive functions. These grafted reactive functions can react with the hydroxyl groups of the starch to form covalent bonds; thus, they provide better control of the size of phase and strong interfacial adhesion. Among the various production methods of graft polymers such as melt grafting, solid state grafting, solution grafting and suspension grafting in aqueous or organic solvents, the reactive extrusion technique is more effective way to produce a variety functional groups onto the surface of natural polymers [442,443]. Functional groups such as isocyanate, amine, anhydride, carboxylic acid, epoxide, and oxazoline are ofen itroduced during reactive extrusion with a short residence time. Combination of amine/anhydride, amine/epoxide, anhydride/epoxide and amine/lactam [444,445] provides practical routes for reactive processing. Such coupling reactions provide interfacial bondings in composites, laminates, and coated items [446], and in immiscible polymer blends, they provide better control of the phase size and strong interfacial adhesion.

Triple-detector-size-exlusion chromatography, melt flow index and TGA analysis were used to characterize the maleated PLA polymers and their reactive blends with starch. Increasing the intiator concentration resulted in an increase in the grafting of MA, as well as a decrease in he molecular weight of the polymer. The MA functionalization of PLA proved to be very efficient in promoting strong interfacial adhesion with corn native starch in composites as obtained by melt blending. Thus, improved interfacial adhesion could be obtained in PLA/starch blends through chemical modificationm of PLA with low levels of MA monomer [447].

MA was first used as a monomer to graft onto biodegradable polymers such as poly(caprolactone), poly(butylene succinate-*co*-adipate) and poly(lactic acid) [439,447-451]. Carlson et al. [451] performed free-radical-initiated grafting of MA onto a PLA backbone at 180-200$^{o}$C with an 2,5-dimethyl-2,5-di-(*tert*-butylperoxy) hexane (Lupersol 101) concentration, ranging from 0 to 0.5 wt. % and 2 wt. % of MA concentration, by using twin-screw reactive extruder. Under these conditions, between 0.066 and 0.672 wt. % MA was grafted onto the PLA chains. They proposed maleation reaction mechanism as follows: The formation of a radical is the first step in the MA grafting of PLA. Once the radical is formed, hydrogen abstraction can occur, producing a PLA which may react with the MA.

Similar hydrogen abstraction on the PLA backbone recently reported by Hvella et al. [452] in the case of radical polymerization of butyl acrylate in the presence of performed PLA. According to the authors [447,452], the resulting polymer radical may then combine with another radical (MA, proxide, or polymer radicals or hydrogen) and further undergo a β-scission as shown in Figure 27.

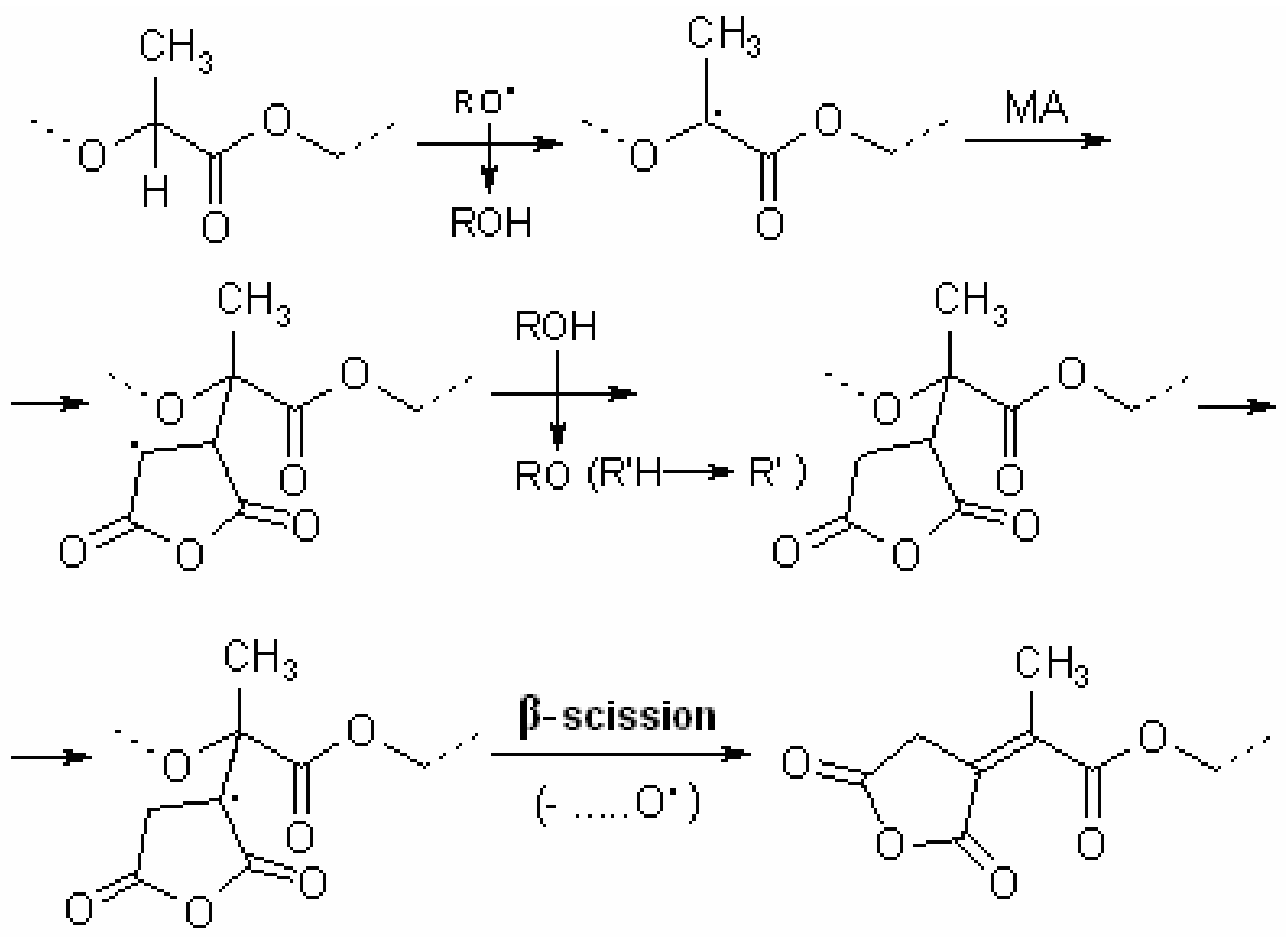

**Figure 27.** Chain β-Scission in grafting of MA onto polylactide acid (PLA). Adapted from [452].

Lanzilotta et al. [453] also studied the melt modification of PLA with MA. According to the authors, the property enhancement achieved in injection-molded flax-PLA composites, in which PLA had been modified with MA by reactive extrusion, did not justify the costs of the process. Plackett [454] grafted PLA with MA in the chlorobenzene solution with benzoyl peroxide at 130$^{o}$C under argon atmosphere for 4 h. The aim of this work was to synthesize and characterize a maleated PLA and to study whether use of this additive might enhance the properties of PLA biocomposites. Author investigated the effect of poly(PLA-*g*-MA) as an additive by three methods: (1) compounding with a commercially available L-PLA and wood fiber in a Brabender mixer, (2) compounding with a commercially available L-PLA and nanoclay in a Haake mini-extruder, and (3) solution treatment of jute fiber mats that were then used to prepare jute-PLA composites by a compression molding process. SEM photomicrographs of the wood-PLA compound indicated some improvement in adhesion might have occurred in the presence of poly(PLA-*g*-MA). Tensile testing of jute-PLA composites showed a reduction in composite tensile strength resulting from addition of poly(PLA-*g*-MA) to the fibers. Addition of grafted PLA changes in the X-ray diffraction pattern. According to the authors this effect may be related to changes in polymer crystallinity.

Polyhydroxyalkanoates (PHAs) are biodegradable and biocompatible thermoplastic polyesters with properties similar to those of classical polyolefins [456]. According to Mechanical properties of poly(3-hydroxybutyrate) (PHB) are similar to those of isotactic PP. PHAs have attracted much attention as environmentally degradable resins, which are useful for a wide range of applications. However, PHAs are highly hydrophobic and degrade thermally during processing. More particularly, polyhydroxybutyrate (PHB) one of the members of PHA family is a crystalline polymer and its main drawbacks are its brittleness, narrow processing window and thermal instability. However, PHB thermally degraded easily during processing, which is of great disadvantage to its widespread commercial use. Graft copolymerization method used to improve important properties of PHB [454,455]. A process and composition using anhydride grafted polyhydroxyalkanoate (PHA) polymer (grafted polymer) which has been extruded with a PHA polymer (non-grafted) and a dried cellulose fiber which reacts with the maleated PHA is described in a patent publication [454]. The composites formed have improved mechanical properties. The present invention also relates to the process for the fabrication of biocomposites of polyhydroxyalkanoates (PHAs) with fibers with the use of anhydride grafted PHAs (PHAs bearing anhydride groups) as compatibilizers. The fibers react with the grafted PHAs.

Bahari et al. [455] reported the radiation grafting of MMA, 2-hydroxyethyl methacrylate, acryilic acid and styrene onto PHB and its copolymers, and were found that the thermal stability or the biodegradability was obviously promoted. Lee et al. [456] studied graft copolymerization of acrylamide onto poly(HB-*co*-hydroxyvalerate) film to test the application of grafted film on permselectivity. Chen et al. [457] selected MA as the grafting monomer to be grafted onto PHB chains because of its good reactivity and controllability in free-radical polymerization. Authors supposed that introducing a certain monomer onto PHB chains, such as MA, could disturb the regularity of PHB chains, then control the morphological structures and improve its properties. Graft copolymerization of MA onto PHB was carried out in chlorobenzene solution in the presence of benzoyl peroxide as an initiator at 130$^{o}$C. It was shown that the monomer and initiator concentrations play an important role in grafting reaction, and graft degree initially increases in monomer and initiator concentrations, and then plateaus above a ceptain level. According to authors, by changing the reaction conditions, graft degree can be controlled in the range from 0.2 to 0.85 %. This grafting reaction was represented as follows (Figure 28):

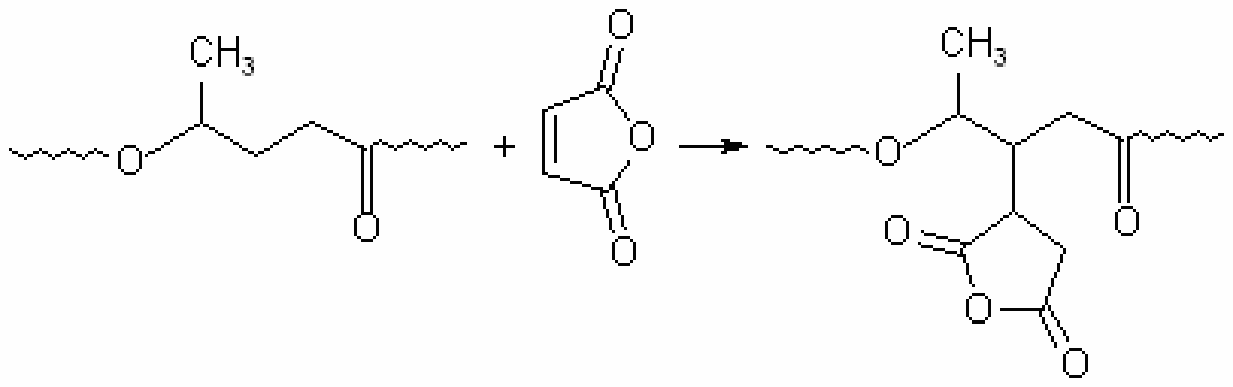

**Figure 28.** Grafting of MA onto biodegradable poly(3-hydroxybutyrate). Adapted from [457].

Authors found that the thermal stability poly(HB-g-MA)

obviously improved, compared with that of pure PHB; the crystallization behavior of PHB obviously changed after MA grafting; the cold crystallization temperature from the glass state increased, the crystallization temperature from the melted state decreased, and the growth rate of spherulite decreased. PHB is more stable in nitrogen than in air, and a small amount of MA can gratly retard the thermal decomposition of PHB. According to the authors, the biodegradability of PHB improved after grafting MA because of the improvement in the wettability of PHB with the enzyme (*Penicillin* sp.) solution [457].

The possibility of grafting MA onto polymer with ester functionality, such as poly(E-*co*-methyl acrylate), reported by Ranganathan et al. [158]. Authors used two low molecular weight esters, a linear ester methyl deconate (MD) and a branched ester methyl 2-ethylhexanoate as model compounds. Grafting reaction was carried out in a round-bottom flask at 180°C in the presence of commericial dialkyl proxide initiators (Lupersol 101 or 130). The grafted products were characterized extensively using several analytical techniques such as FTIR, NMR, MS (mass spectroscopy), and chromatography (SEC). The nature of the grafts was established using 2,3-$^{13}C_2$ labeled MA. The FTIR spectroscopy provides qualitative evidence that the grafting of MA to ester has occur. The spectra of grafted product show bands at 1864 and 1788 $cm^{-1}$ (antisymm. and symm. C=O, respectively), and 952-909 $cm^{-1}$ (ring stretching of a saturated cyclic five membered anhydride). The percent MA unit was determined to be 1.25 wt. %. These results are in agreement with those of several researchers who have used FTIR, both qualitatively to identify the presence of succinic anhydride residues [134,151,192,458,459] and quantitatively to determine the MA percent in grafted polymer [168,224] and model systems [140,191]. Proton NMR spectra of grafted MA exhibits resonances centered around 3,3, 3,2 and 2.83 ppm, which have been assigned to the CH and $CH_2$ protons of the grafted MA ring. The carbon atoms of these groups in the $^{13}C$ NMR spectra are characterized by 30-35 ppm ($CH_2$) and 40-47 ppm (CH) peaks of the grafted MA residues. Authors represented proposed mechanism of grafting as follows (Figure 29) [158]:

Free-radical-initiated grafting of MA onto poly(butylene adipate-*co*-terephthalate) (PBAT) a biodegradable aliphatic–aromatic copolyester, was performed in a twin-screw extrusion system at 185°C in the presence of 0.5 wt. % organic peroxide (Lupersol 101) by Nabar et al. [460].

The MA concentration varied between 1.0 and 5.0 wt. %. In these conditions, authors prepared poly(PBAT-*g*-MA)s containing MA grafted unit between 0.194 % and 0.691 %. Authors proposed the following mechanism of grafting reactions which are accompanied by the β-scission along the polyester backbone (Figure 30).

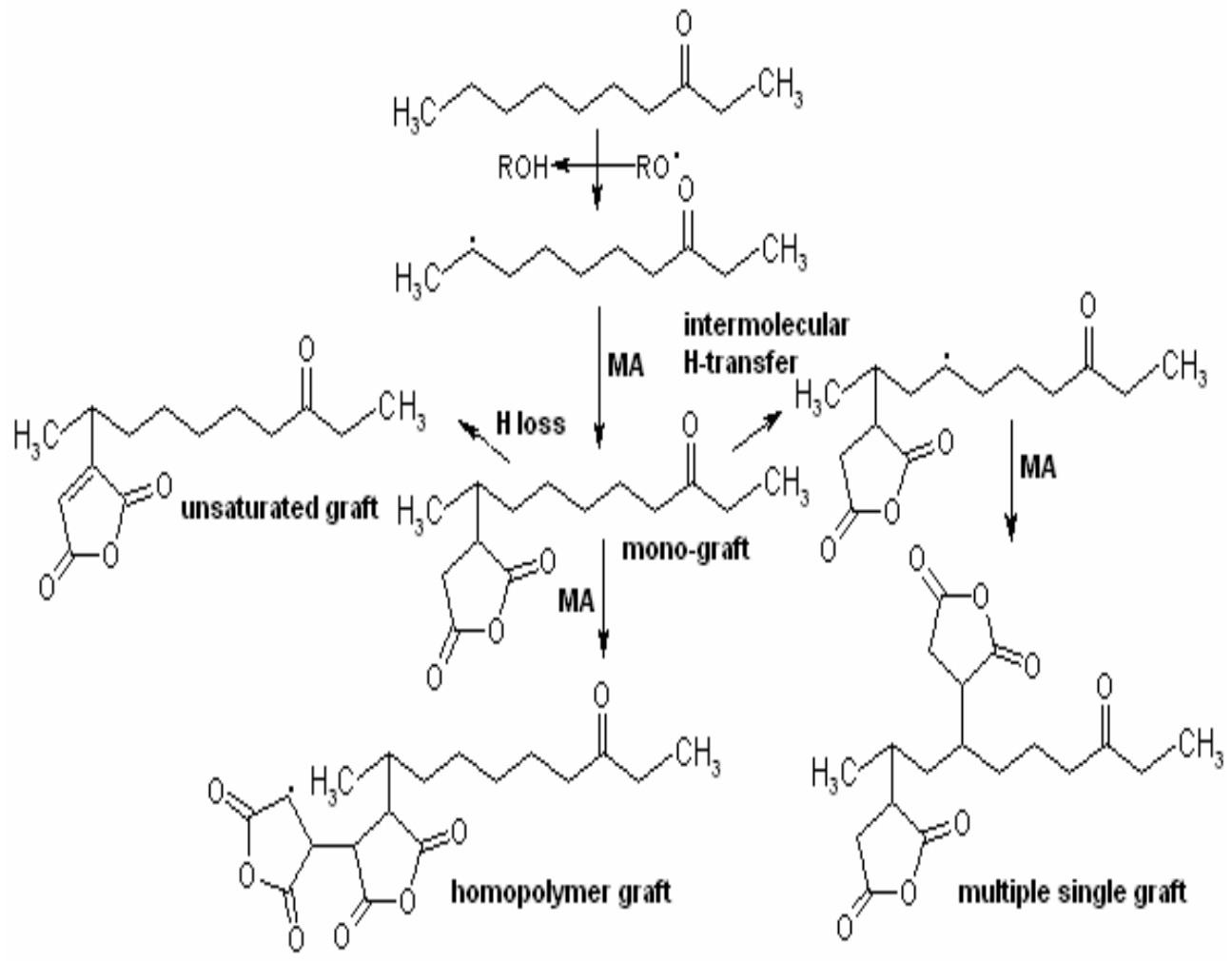

**Figure 29.** Grafting of MA onto a linear ester methyl deconate as a model biodegradable compound. Adapted from [158].

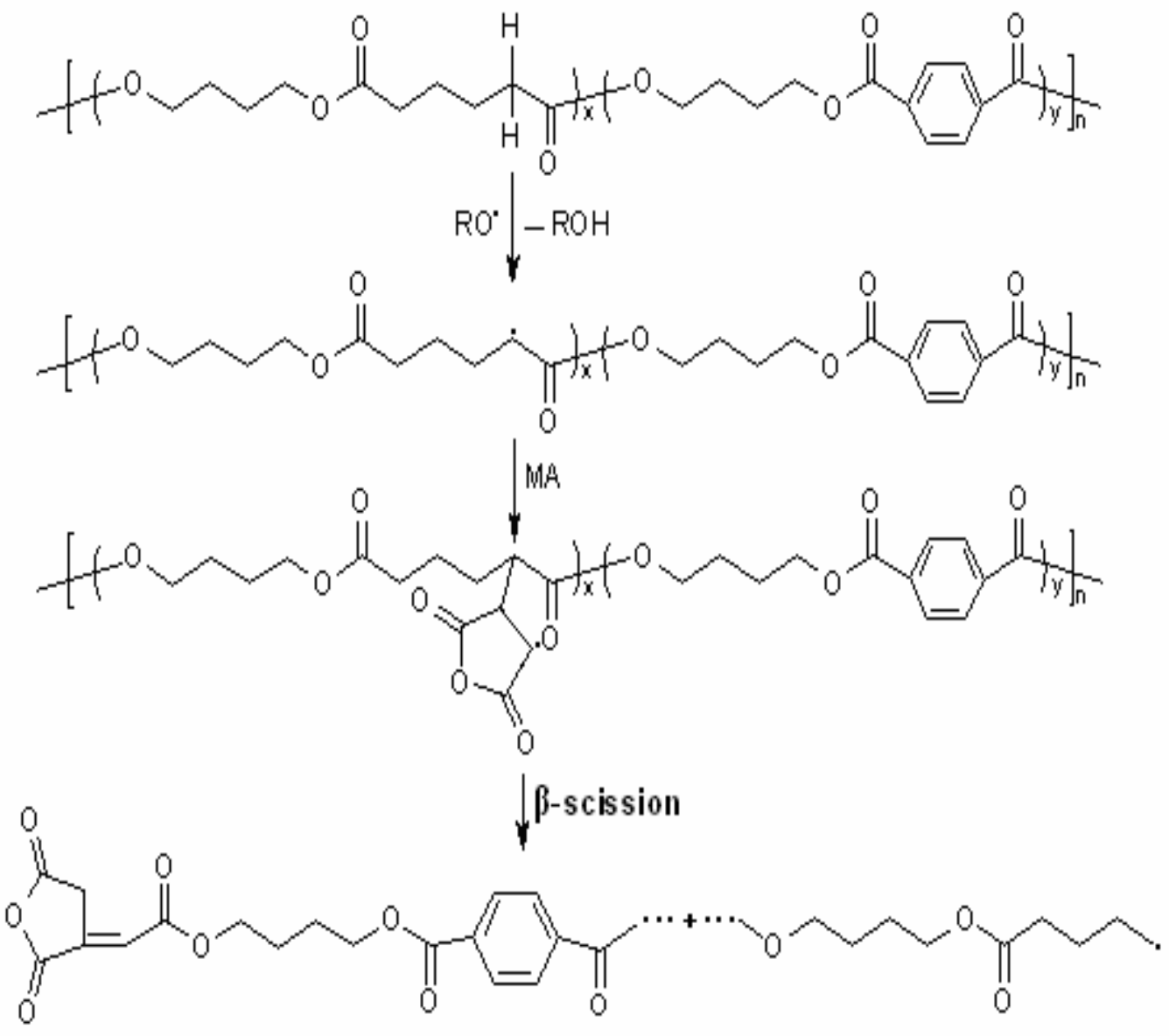

**Figure 30.** Grafting of MA onto biodegradable poly(butylene adipate-*co*-terephthalate). Adapted from [460].

But in this system, oligomerization of MA was not observed. According to the authors, the β-scission is responsible for the drop in molecular weight and melt viscosity would indeed be the dominant mechanism over the hydrogen abstraction rections. Observed unusual relationship between molecular weight and initiator or MA concentration, authors explaned by the presence of other reactions, such as branching, favorable radical recombination, radical termination before the grafting and thermal hydrolysis during the maleation of PBAT in the extrusion process. They also showed that the maleation of the polyester proved to be very efficient in promoting strong interfacial adhesion with high amylose cornstarch in starch foams as prepared by melt blending.

Bhattacharya et al. [121] studied grafting reactions of MA with aliphatic and aromatic/aliphatic copolyesters in melt by reactive extrusion at 130-180 °C in the presence of dicumyl peroxide as an initiator, as well as in toluene solution. The following polymers were used in this study: polybutylene succinate adipinate (PBS), polybutylene succinate adipinate copolyesters (PBSA), copolymer of tetraphthalic and adipic acid and butanediol (Eastar) and 2- and/or 4- and 4-carbon diacids and glycols (Bionolle). The percent grafting was found to vary with the experimental conditions. Thus, the graft content increased
from o.46 to 1.15 % as the temperature increased from 140 to 180 °C. FTIR studies confirm the presence of siccinic anhydride groups; 2D $^{1}$H NMR spectra (COSY) suggest that the grating reaction takes place at aliphatic dicarboxylic acid units of the copolyesters; minimal degradation of the polyester is evident as observed from intrinsic viscosity measurements. The grafting behavior of different copolyesters studied in this work varied with the [peroxide]/[MA] ratio. In PLA, low grafting efficiency was observed due to its limited availability of free radical sites on the polymer backbone for grafting.

### 5.3. Polyethers

Poly(propylene oxide) (PPO) is a widely applied type of aliphatic polyether for the preparation of thermoplastic elastomers as soft segment component, surfactants and poltmeric additives. For the development of advanced thermoplastic elastomers and other high performance materials, functionalization of PPO was carried out by radical grafting of maleic and fumaric acids onto PPO macromolecules [461,462]. Authors of these works suggested single succinic acid units linked to PPO methine carbons as the structure of graft products of maleic acid (MAc) and PPO. The proposed structure shows similarities to the structure of the graft products of MA and PP. However, studies concerning the reactivity of ethers in radical reactions [463,464] and the grafting of poly(tetrahydrofurane) (as a cycloaliphatic polyether) with MA [465] indicated that the ether oxygen has a high influence on the radical hydrogen abstraction. Taking into account the higher reactivity of MA in the radical reactions compared to maleic and fumaric acids, as well as possibility more efficient grafting expection by using MA as graft monomer, Rische et al. [459,466] studied radical-initiated grafting reactions and structure of grafted products of MA/poly(tetrahydrofuran) (PTHF) and MA/PPO systems [459]. The grafting of MA onto PTHF mainly occurred through carbon atoms in α-positon. It was found that approximately 10 % of all graftings took place onto PTHF backbone carbons in β-position. Indications for the formation of poly(MA) graft units were not found. A protection of hydroxyl end-groups of PPO realized by acetylation (structure D) to prevent side-reactions of these groups with MA. It was shown that in contrast to grafting reactions onto polyolefins, unusual high grafting degree (up to 20 wt.%) were obtained by the grafting of MA onto poly(tetrahydrofuran) and PPO. According to the authors, by extraction with hexane highly MA grafted PPO chains could be isolated easily from the reaction products. The procentage of grafting of the isolated products was up to 24.3 wt.%. The synthesized graft products characterized by authors using FTIR and $^{13}$C NMR and GPC techniques, results of which are confirmed the formation of the following graft structures (A, B and C) (Figure 31) [466]:

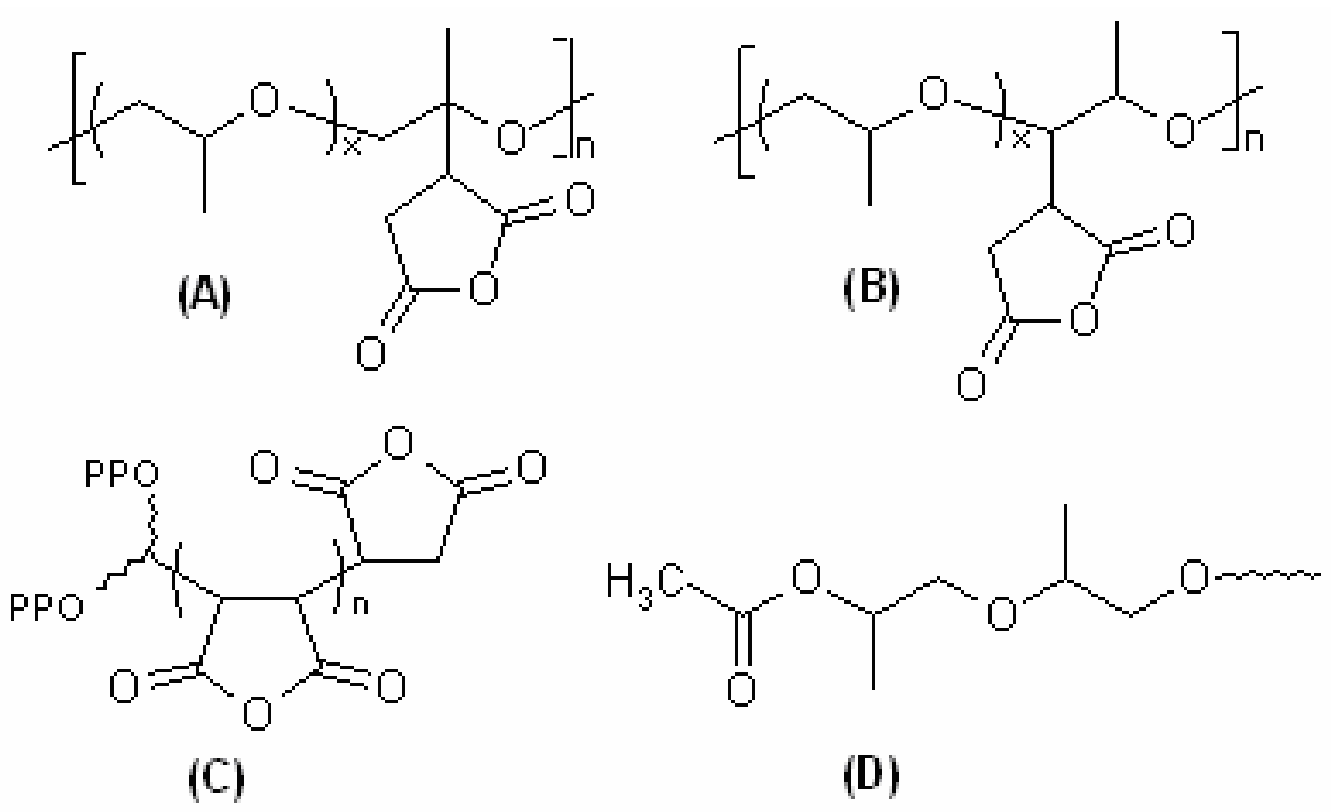

**Figure 31.** The formation of various graft structures in the grafting of MA onto poly(propylene oxide). Adapted from [466].

Authors observed that the antisymmetric C=O bands of grafted products is shifted from 1781 (free MA) to 1783.5 cm$^{-1}$ (MA graft) and the symmetric C=O vibration is shifted from 1856 to 1862 cm$^{-1}$. This fact, finding by authors, indicates the formation of a saturated cyclic anhydride structure. The results of $^{13}$C NMR analysis indicate that the grafting of MA occurs onto both CH (in A structure) and $CH_2$ (in B structure) backbone carbons of the PPO chains in the form of monosubstituted succinic anhydride graft units. Furthermore, in products synthesized with a high initial concentration of MA (~20 wt.%) beside single units a small amount of oligo(MA) grafts (C strurture) was formed. According to authors, these higly functionalized PPOs are of potential interest as components for crosslinked elastomers. Furthermore, the formed anhydride graft units may be used to carry out further reactions, e.g. imidizations, to form molecules with new interesting propereties. A complete hydrolysis of the grafted anhydrides of the reaction products was possible. They asummed that appropriate PPO molecule with a high concentration of carboxylic groups can be of interect for the development of surfactants, water soluble polymer dispersions as well as for the development of polyelectrolytes and ionomers [466].

## 6. Grafting onto Other Polymers and Anhydride-Functionalized Copolymers

The radical-induced grafting of diethyl maleate onto poly[bis(4-ethylphenoxy) phosphazene] was reported, together with the possible parameters that can influence this reaction, i.e., the maleate, molecular oxygen, and dicumylperoxide concentrations, in the reaction mixture, the solvent, time, and temperature of the process. A possible mechanism of this process is inferred on the basis of analogous grafting processes carried out onto polyolefins. The importance of the decrease of the intrinsic viscosity of the reaction mixture that takes place during the grafting reaction is also discussed in terms of chain scission and molecular weight degradation of the phosphazene macromolecules [467].

To improve water wettability of polyurethane (PU), graft copolymerization of itaconic acid (IA) performed by Pulat and Babayigit [468] using benzoyl peroxide initiator. The grafting reaction was carried out by placing the membranes in aqueous solutions of IA at constant temperature. The optimum temperature, polymerization time, initiator and monomer concentrations were found to be 80 $^oC$, 1 *h*, 0.04 and 1.5 mol $L^{-1}$, respectively. The membranes prepared were characterized by FTIR spectroscopy and scanning electron microscopy analysis, and grafting effect on equilibrium water content of PU membranes was obtained by swelling measurements.

Wu et al. [469] prepared poly(vinyl acetate-*co*-dibutyl maleate) and poly(VA-*co*-DBM) latex films in the presence of grafted and post added poly(vinyl alcohol) (PVOH). They polymerisized of 4:1 weight ratio of VA and DBM in the presence of PVOH to obtain a series of samples in which about 13 to 30 % of the PVOH becames grafted to the copolymer particles. They assumed that grafting of the PVOH onto copolymer can occur via a chain transfer in two ways, involving the pendant methyl groups of the acetate units or the CH groups on the PVOH main chain according to Britton et al. [470]. Then they demonstrated that grafting of the PVOH strongly affects not only the colloidial stability of the copolymer latex, but also the properties of the latex films. MA was grafted onto the poly[ethylene-*co*-vinyl acetate (18 mol %)] (PEVA) in a Haake Rheocord mixer at 175 $^oC$ and rotor speed 50 rpm during 10 min. Kim et al. [471] also disclosed the detailed reaction mechanism of MA grafting onto poly(ethylene-*co*-vinyl acetate) (EVA).

Some publications of Rzayev et al. related to the synthesis and characterization of novel bioengineering macro- branched copolymers of MA and their conjugates with biopolymers. These grafted copolymers synthesized by reaction of poly(MA-*rand*-*N*-isopropylacrylamide) [8,472, 473], poly(MA-*alt*-*p*-vinylphenylboronic acid) and poly-(citraconic anhydride-*alt*-*p*-vinylphenylboronic acid) [10], and poly(MA-*n*-hexene-1) [13,14] with α-hydroxy-ω-methoxy-poly(ethylene oxide) in the 1,4-dioxane at 50$^oC$ according to the following general Figure 32.

Variations on the ω-methoxypoly(ethylene glycol) grafted poly(ethyl-2-cyanoacrylate) (ECA) copolymers, namely poly(α-MA-ω-methoxy-PEG)-*co*-ECA)s, Deng et al. [474] prepared via the initial synthesis of ω-methoxy-PEG macromonomer using MA, which then copolymerized with ECA in the presence of radical initiator (AIBN). Synthesis

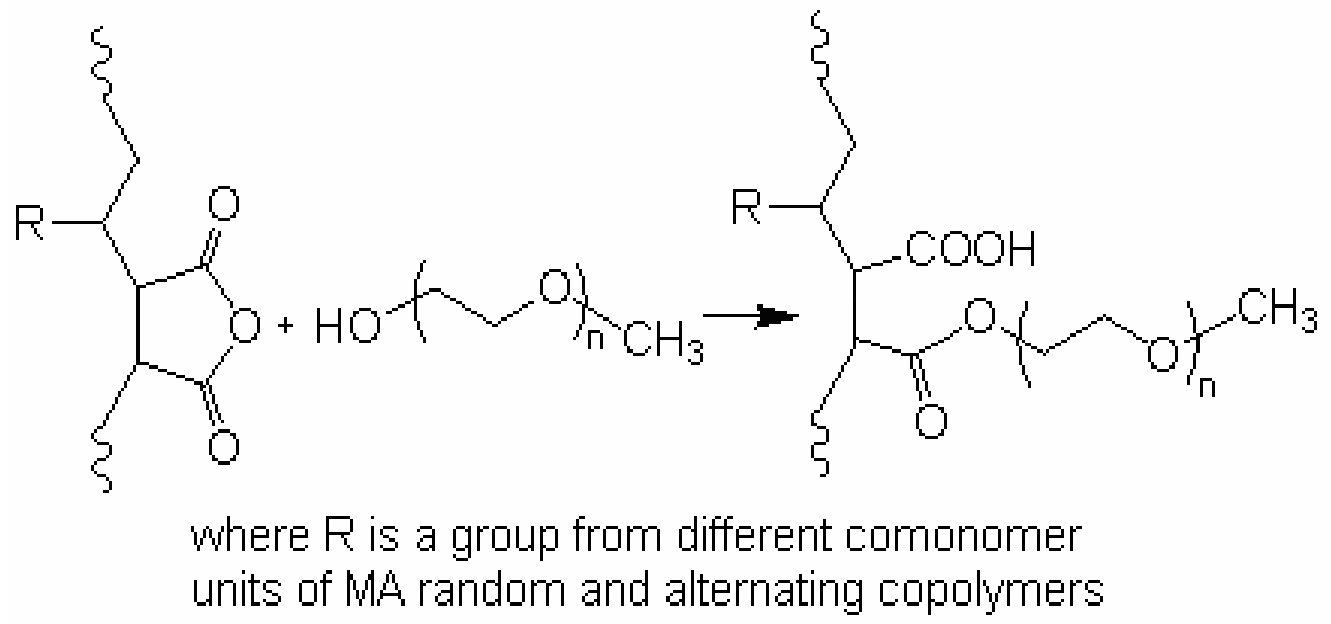

**Figure 32.** Grafting of the various anhydride-containing copolymers with α-hydroxy-ω-methoxy-poly(ethylene oxide).

routes and structure of copolymers were represented as follows (Figure 33):

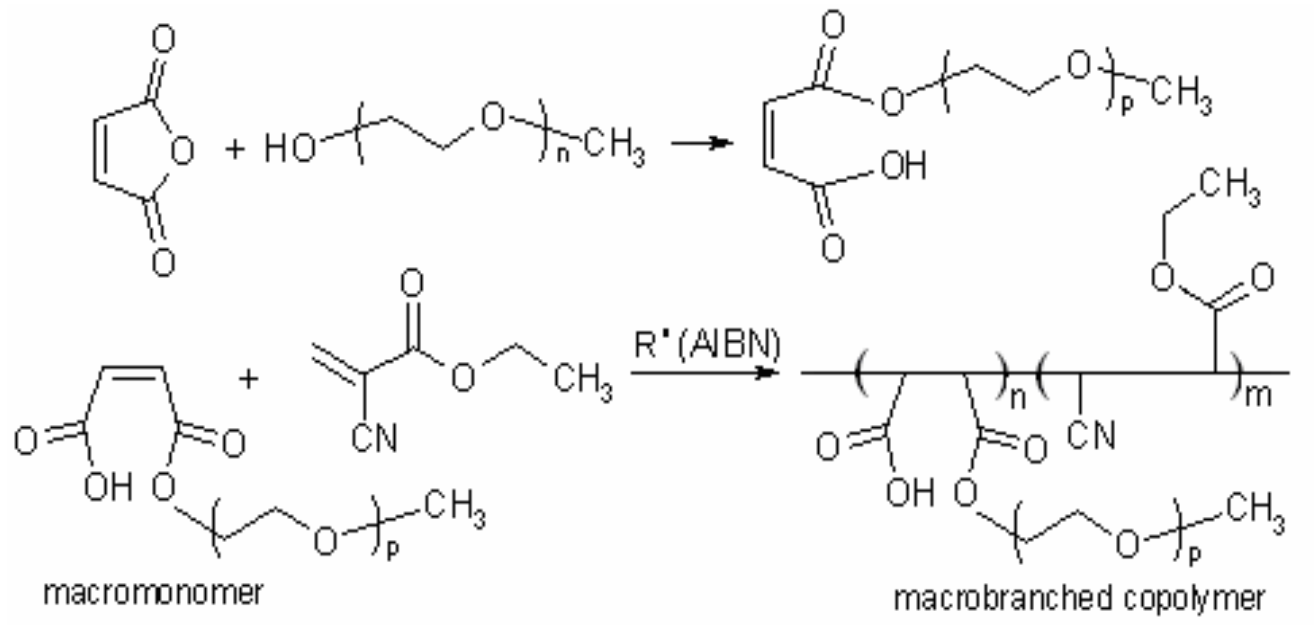

**Figure 33.** Synthetic partways of ω-methoxy-PEG macromonomer and its copolymer with poly(ethyl-2-cyano-acrylate. Adapted from [474].

Synthesized by authors macrobranched graft copolymers with core-shell structural nanoparticles have good hydrophobic drug-loading ability, which could provide a delivery system for hydrophobic compounds. They showed that the size of the nanoparticles increased with increasing the hydrophobic property of the solvent, increasing the copolymer concentration in the organic solvent, and increasing the molecular weight of branched moieties.

Yağcı et al. [475] demonstrated a new synthetic approach for the preparation of well-defined graft copolymers on the basis of the Diels-Adler (DA) "click chemistry" between copolymers bearing anthryl pendant groups and maleimide as end-functionalized polymers. The grafting processes carried out at the reflux temperature of toluene with a quantitative yield and without need for an additional purification step. First, random copolymers of styrene (S) and chloromethylstyrene (CMS) with various CMS contents were prepared by the nitroxide-mediated radical polymerization (NMP) process. Then, the choromethyl groups were converted to anthryl groups via the etherification with 9-anthracenemethanol. The other component of the click reaction, namely protected maleimide functional polymers, were prepared independently by the modification of commercially available poly(ethylene glycol) (PEG) and poly(methyl methacrylate) (PMMA) obtained by atom transfer radical polymerization (ATRP) using the corresponding functional initiator. Then, in the final stage PEG and PMMA prepolymers were deprotected by retro-Diels-Alder in situ reaction by heating at 110°C in toluene. The recovered maleimide groups and added anthrylfunctional polystyrene undergo Diels-Alder reaction to form the respective (PS-*g*-PEG) and (PS-*g*-PMMA) copolymers. Authors characterized the graft copolymers and the intermediates in detail by using a combination of various methods such as $^1$H NMR, GPC, UV, fluorescence, DSC, and AFM measurements. According to authors, the strategy adopted in this study appears to be entirely satisfactory in terms of efficiency and simplicity.

Yoon et al. [476] prepared the reactive blend of poly(ethylene terphthalate) (PET) and poly(MA-*co*-styrene). They compared the properties of this reactive blend and PET/PS physical blend. It was expected that the reaction between ester groups in PET and maleic sites in copolymer would affect the properties of the blends. The reaction was observed in 70 and 90 % copolymer blends. It was confirmed indirectly from the solubility of the blends in THF. The reaction between the ester groups in PET and anhydride units in copolymer produces a graft polymer, which increases the viscosity of the blends. Besides these two blend compositions, morphology indicates that there are interactions between two phases in all PET/poly(MA-*co*-styrene) blend compositions.

Graft copolymers involving MA moieties have been the subject of considerable interest, because these grafting anhydride units are reactive carbonyl groups which are subjected to numerous reactions. Abd El-Rehim et al. [477] reported the radiation-induced graft copolymeri- zation of MA and styrene with PE, and some reactions of graft copolymers with metal salts and amine-containing compounds. Authors chemically modified the PE-(MA-*co*-styrene) graft copolymers with different reagents containing various functional groups. The scheme of these reactions represented in Figure 34. According to this Figure, the functionlization of the grafted membranes is a ring opening reaction, and there is no other low molecula release. The conversion of the MA units in the graft copolymer membranes ($C_{MA}$) calculated as follows:

$$C_{MA} = [(W_1 - W_o)/W_o].(M_1/M_2) \times 100 \qquad (18)$$

where $W_1$ and $W_o$ are the weights (g) of poly[PE-*g*-poly(S-*co*-MA)] after and before reaction, respectively; $M_1$ and $M_2$ are the molecular weights of the S-MA chain unit (212 g/mol) and the amine compound, respectively.

According to the authors, these functionalized graft copolymer membranes are possessed good chelating properties towards different metal ions such as Fe(III), Cr(II), Cu(II), Ni(II), Cd(II) and Hg(II).

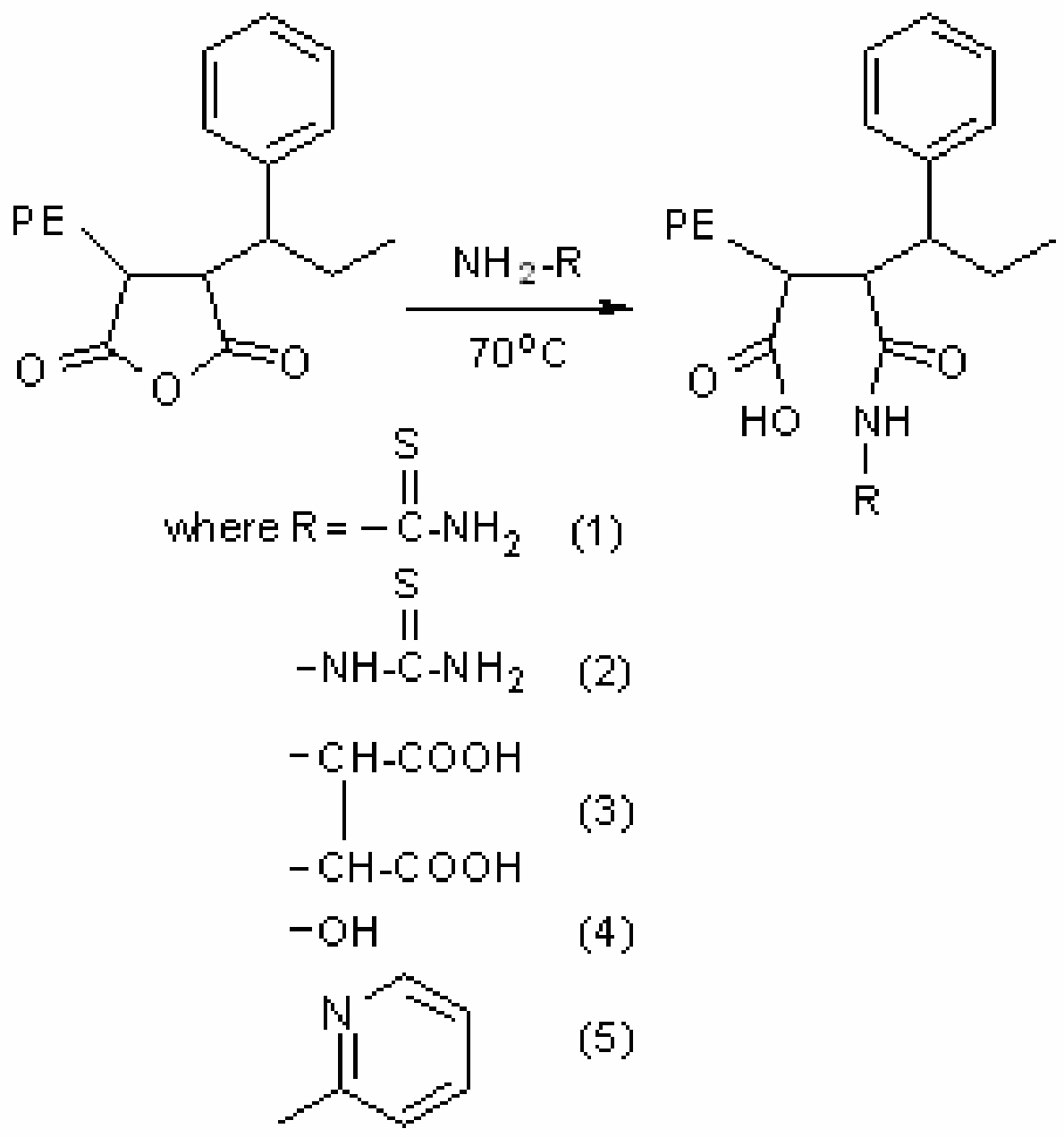

**Figure 34.** Radiation-induced graft copolymerization of MA and styrene with PE, and some reactions of graft copolymers with amine-containing compounds. Adapted from [477].

## 7. Graft Copolymerization

In fact, MA is a versatile hydrophilic monomer. If MA is grafted onto substrates and further hydrolyzed, the substrates will carry a denser distribution of –COOH groups; furthermore, a large number of other functional groups can be introduced onto the substrates by the reactive anhydride groups. Some studies have suggested that poly(MA) and MA copolymers show antitumor and antiviral activities [1-2,67]. Therefore, much research has been focused on the polymerization and copolymerization of MA [1-3,4,27].

In ealier publications, Gaylord et al. grafted poly(MA-*alt*-styrene) and poly(styrene-*alt*-MMA) onto PS and poly(styrene-*alt*-AN) [355,356] onto cellulose [478] in the presence of zinc chloride. Poly(MA-*alt*-styrene) grafted onto a variety of conventional polymers through free-radical initiation. Graft copolymerization of MA/styrene mixture in the presence of different thermoplastic polymers and synthetic rubbers using various type mixing apparatus and extruders was also reported [1,2, 353,354].

Graft copolymerization method using various donor-acceptor monomer mixtures, including styrene and MA, significantly increases functiolalization degree of thermoplastic polymers. It was reported that there are some monomers, predominantly styrene, capable of acting as a donoting electron to MA, leading to an enhancement of MA grafting efficiency onto polyolefins [163,164]. Authors found that styrene monomer effectively increases the grafting percentage of MA on PP and decreases the PP degradation. Styrene reacts with MA to form alternating copolymer during the melt process and that grafting of this copolymer leads to improvement in the grafting degree of MA on the polymer [179,478]. The effect of various donor and/or acceptor comonomers such as styrene, α-methylstyrene, vinyl acetate, *N*-vinylpyrrolidone, acrylic and methacrylic acids, esters of unsaturated mono- and dicarbonic acids, etc. on grafting conversion of MA onto linear low density PE [461] and PP[158,164,179,232,479] also studied. It was observed that grafting yields to PP decreases in the following raw: styrene >> α-methylstyrene > MMA > vinyl acetate > (no comonomer) > *N*-vinylpyrrolidone. It was proposed that higher yields can be explained by formation of a charge transfer complex (CTC) between comonomers and MA as follows (Figure 35) [150,179]:

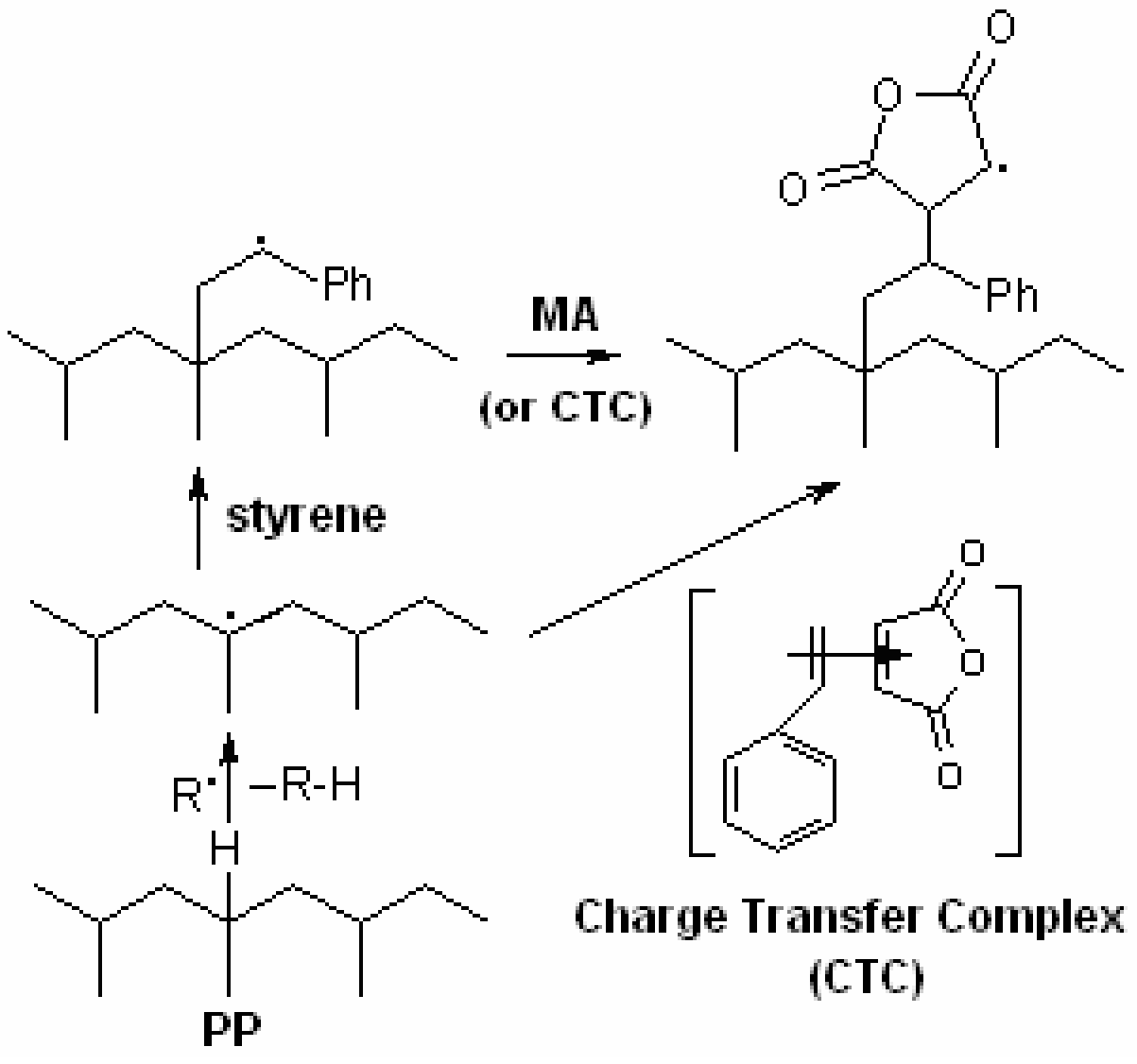

**Figure 35.** Graft copolymerization of MA and styrene with PP in the presence of free MA monomer and its CTC with styrene (S…MA). Adapted from [179].

According to the authors, [S…MA] complex may have a higher reactivity than free MA. It was also likely that more efficient grafting may, in part, simply be due to attachment of a longer chain length graft rather than a greater number of graft sites [166]. On the basis of NMR data for styrene-*g*-MA copolymers, authors suggest that the graft is a copolymer chain and not a single styrene-MA pair. Authors proposed that that styrene and MA do not show the same tendency to alternate in the chain in graft copolymer formation as is seen in conventional free radical copolymerization in solution at relatively low temperature. They found that the S/MA ratio in the graft exceeds the initial S/MA ratio irrespective of that ratio. The S:MA ratio in the graft copolymer varies from ca. 0.7:1 to 4.5:1 when the monomer ratio is varied in the range 0.25:1–3:1 (batch mixer 215°C, 4.0 wt.% MA vs. PP, 0.5 wt.% di-*t*-butylperoxy-2,5-dimethylhexane initiator) [179]. These results indicated that the chain growing reactions in the graft copolymerization occur by the mixed mechanism including alternating chain growing monomer pair through effective initiation of free radical/donor-acceptor (at S:MA=1:1 in feed) mixture and random chain growing styrene and MA units (at S>MA in feed) (Rzayev's comment). According to Moad [150] these observations do not preclude the involvement of a CTC but do show that the monomers are not simply incorporated pairwise.

Surface photografting polymerization of MA and its CTCs with vinylacetate monomer and various solvents, such as THF, acetone, ethyl acetate and 2,4-dioxane, onto different polymeric substrates have been reported by Deng et al. [258-262,480]. The effects of grafting conditions such as the temperature, photoinitiator, solvent, and UV intensity on the graft polymerization of MA onto LDPE film were examined. Furthermore, the formation of grafted film was identified with FTIR and ESCA surface analysis methods [259]. It is known that PP polymeric radical *in situ* react with MA to produce MA terminated PP with a single MA unit. In the presence of styrene, the polymeric radical initiates a "stable" copolymerization of styrene and MA with an alternating manner [3,27,481]. The resulting PP-*b*-SMA diblock copolymer contains both PP and alternating styrene–MA segments.

Graft copolymerization of PP with MA or other monomers can be performed in a solution, the melt, and aqueous suspension, or a solid state. Solid-phase graft copolymerization is relatively new method developed in the early 1990s [135]. It is performed below the melting point of PP powder, normally with a high concentration of initiator and some interfacial agent. Compared with other grafting methods, solid-phase grafting has many advantages such as lower reaction temperature, free from the need for solvent recovery and simpler equipment requirement. So far, studies on solid-phase graft copolymerization of PP are mainly based on MA monomer [139,242,245,482]. Because the self-polymerization of MA is very poor, the MA graft segments only exist as a monomer or a short branch on the PP chains [180], and the grafting percentage is low (< 5 %). Jia et al. [245] found that solid-phase copolymerization of PP with two acceptor-donor type of monomers such as , MA and styrene, significantly ncreased the grafting percentage and grafting efficiency. Then they carried out solid-phase graft copolymerization of PP with MA and styrene by using benzoyl peroxide as the initiator and xylene as the interfacial agent. The graft copolymerization process was performed in a stainless steel reactor. PP, monomers, initiator and interfacial agent were mixed in a predetermined proportion and reacted in nitrogen atmosphere at 120 °C for 1 h. Obtained by authors results showed that the grafting degree and acid value of poly[PP-*g*-(MA-*co*-styrene)] are affected by the total monomer concen- tration, monomer ratio, and initiator concentration, and are considerably higher than those of poly(PP-*g*-MA). The results of dynamic mechanical analysis showed that the pure PP has only one $T_g$ but graft copolymer has two: the first $T_g$ (14.9°C) is of the PP bacbone, and the second (104.4°C) is likely caused by the relaxation of the grafted MA-styrene copolymer segments. The grafted segments are shown to be the copolymer of MA and styrene with a substantial molecular weight [245]. Braun et al. [166] reported heterogeneous grafting of MA and α-methyl styrene (MS) onto PP. The maleination of PP via radical grafting at high temperatures was investigated. Since MA produces discoloring oligomers, PP was grafted with a mixture of MA and MS. Advantageous is that MS cannot homopolymerize at high temperatures and that MA and MS tend to copolymerize alternatingly (to colorless products). The kinetics of the grafting process is complicated by a phase separation of PP and MA/MS mixture that is caused by the special attractions between the two monomers. As a result, two grafted PP products were obtained, of which the major is only slightly but the minor is heavily grafted. The latter graft copolymer, which is probably created in the interfaces between phase domains, carries one MA–MS graft per PP backbone chain, approximately with three MA and five MS units.

Functionalized polymers also obtained by graft copolymerization of functional donor-acceptor monomer pairs with some commercially available polymers such as butadiene homo- and terpolymers. Modification of poly[acrylonitrile (22.4 wt. %)-*co*-butadiene (13.5 wt. %)-*co*-styrene (61.4 wt.%)] (ABS with $M_n$ = 49,000 g/mol and $M_n/M_w$ = 2.72 is contained 2.7 wt.% of additives) by graft copolymerization of MA–styrene monomer pair by using binary initiator system (dicumyl peroxide + benzoyl peroxide) in the molten state reported in detail by Qi et al. [149,485]. They also carried out graft copolymerization of ABS with MA and MA-styrene mixture in melt using a Haake twin-screw extruder (Germany) under various processing conditions [483]. In the case of using styrene as a comonomer, a higher grafting degree of MA onto ABS terpolymer observed by these authors. The mechanical properties and phase morphologies (by transmission electron microscopy) of the modified products poly(ABS-*g*-MA) were also studied. Authors found that the melt flow index (MFI) of poly(ABS-*g*-MA) increases with the increases with the increase of MA content, the initiator concentration, and the scew speed, whereas the MFI decreases with the increase of temperature. The impact strength and the percentage elongation of poly(ABS-*g*-MA) both decreased and the tensile strength of poly(ABS-*g*-MA) increased slightly as the grafting degree increased.

## 8. *In situ* Grafting Reactions and Processing

In polymer blends, immiscibility of the components results in incompatibility of the phases. Compatibility and adhesion between different polymeric phases can be improved by addition of suitable block or graft copolymers that act as interfacial agents. An alternative is to generate these copolymers *in situ* during blend preparation through polymer-polymer graft reactions by using functionalized polymers [484-488]. Recently, extruders have been increasingly been used as chemical reactor [489-492]. A new trend, called “reactive extrusion”, has been developed in the technology of polymer production and processing. No matter in what reactor the chemical process occurs, it is subject to the basic thermodynamic laws [493].

Immiscible blends are possible to be compatibilized through the reaction that reduces interfacial tension and promotes adhesion at the interface. The results are a finely dispersed phase, resistance to gross separation, and enhanced overall properties [494]. It is known that the polymers containing functional groups such as anhydride [127,192,495-497], carboxy [498,499], epoxy [500,501], oxazoline [502], and isocyanate [503,504] groups reacted with the amino group of various thermoplastic resins. Reactive mixing process widely used for obtaining improved compatibility of polymer blend [505-509]. Such a process required introduction of some functionality reacting with other functional groups to either of the blend components.

Poly(styrene-*co*-MA) (SMA) used as a successful reactive compatibilizer in several incompatible polymer blends [507-509]. Immiscible blends of poly(styrene-*co*-acrylonitrile) (SAN) and thermoplastic polyurethane (PU) obtained by melt mixing compatibilized by the addition of SMA copolymer. The best compatibility attained for the PU/SAN/SMA (70/30/5) blend [504]. Chiang and Chang [510] showed that SMA copolymer is a highly effective reactive compatibilizer for immiscible and incompatible blends od polyamide-6 (PA-6) and poly(ethylene oxide) (PEO). The overall improvement in mechanical properties was drastic after compatibili- zation. It was observed that the addition of SMA copolymer (up to 10 wt.%) in amorphous PA/SAN blends impoves the enhancement of interfacial adhesion and the tensile strength of blend [507]. PA-6 and poly(methyl methacrylate) (PMMA) blends satisfactory compatibilized by SMA containing 20 wt.% of MA [506]. Blend of SAN with poly(ethylene-*co*-1-octene) (EOP) rubber studied by Mader et al. [508]. An improved toughness–stiffness balance was achieved when oxazoline-functionalized EOP and SMA were added to the mixture. Teselios et al. [509] used SMA copolymer in blends containing poly(ethylene-*co*-vinyl alcohol) EVA prepred by melt mixing. They showed that the reaction between the hydroxyl group of EVA and the anhydride group of SMA leads to the formation of branched and crosslinked macromolecules, which can cause morphology stabilization of the immiscible blend.

Cassu and Felisberti [486,510] reported the preparation and properties of Poly(styrene) (PS)/polyester PU elastomer blends compatibilized by different amounts of SMA (around 0.5-5.0 wt.%), containing 7.0 or 14.0 wt.% of anhydride. Binary nonreactive (PS/PUs), and reactive binary and ternary blends containing SMA reactive compatibilizer were prapared with 10 and 20 wt.% of PUs. Their results showed that this compatibitization occurs by the formation of a graft copolymeri *in situ* during the melt mixing and is responsible for the decrease of the elastomer domain size in the glassy matrix. The phase behavior, morphology and mechanical propereties of blends depended on their composition, especially on the SMA amount and anhydride content in SMA copolymer.

### 8.1. Polyolefin reactive blends

The improve the compatibility of SMA copolymer/ LDPE blends, LDPE grafted with 2-hydroxyethyl methacrylate-isophorone diisocyanate (LDPE-*g*-HI) was prepared and blended with SMA copolymer of which anhydride groups were converted to carboxylic acid groups [488]. In the blend morphology, some adhesions between the two pases and much finer dispersions were observed in the SMA/LDPE-*g*-HI blends, indicating that chemical reactions took place during the melt blending. The lower heat capacity change at the glass transition temperature demonstrated that chemical bonds were produced in the poly(MA-*co*-styrene)/ LDPE-*g*-HI blends. From the results of the rheological test, it was found that strong positive devaition from the mixting rule occurred in viscosity for these functional polymer blends, concerning with good adhesion and finer dispersions. For these blends, the improvement in mechanical properties, resulting from chemical reaction, was also observed. When SMA copolymer blended with other ductile polymers like polyethylene (PE), it may offer an attractive balance of mechanical property and resistance to heat distortion. To attain satisfactory performance in the immiscible SMA/PE blend, it will be important to improve the compatibility. There are few detailed studies on the SMA/PE blends obtained by a reactive processing without using a compatibilizer. A blending of LDPE with isocyanate functionality and SMA copolymer with acid functionality was also carried out to improve compatibility of SMA/LDPE blends through chemical reaction occurring during the blend process [491].

The compability and mechanical properties of incompatible PP/PS (polystyrene) blends can be improved by the addition of graft copolymer compatibilizers, inorganic particles, and with reactive compatibilization. The addition of $BaSO_4$ to the PP/PS blend resulted in a decrease in the domain size of the minor polymer phase [511,512]. $BaSO_4$ alone did not have a nucleation effect on PP; however, in combination with poly(PP-*g*-MA), a clear nucleation effect was observed. Some studies have comfirmed that comonomers styrene (S) and MA react readly to form a SMA copolymer with the help of an initiator agent. It is much easier for SMA than MA to graft with PP; therefore, the degree of gtafting of MA increases, and more poly(PP-*g*-MA) is formed *in situ* [513].

Zhang et al. [514] reported the results of a study on the influence of reactivity monomers, MA and S, on the crystallization and melting behaviour and dynamic mechanical properties of PP/PS blends. *In situ* processing of blends was carried out by melting extrusion in a HL-200 kneader at 185-190$^{o}$C. The FTIR structural analysis indicated that (PP-*g*-PS) formed in PP/PS blends modified by MA monomer. However, the formation of (PP-*g*-MA) observed in PP/PS blends modified by MA and S monomer mixture. According to authora, in PP modified by the MA comonomer, the formation of (PP-*g*-MA) was more significant. MA hardly influenced the crystallization temperature ($T_c$) of PP in the blends, but the addition of MA and S increased the $T_c$ of PP in its blends. The blends showed no remarkable variety for the melting temperature, but the shapes of the melting peaks were influenced by the addition of the reactive monomers. In addition, a significant increase in the storage and loss moduli of all the modified PP/PS blends was observed.

Xia et al. [515] performed the *in situ* compatibilization of ABS and a copolyester liquid crystalline polymer (LCP) by addition of a reactive poly[MA(14 wt.%)-*co*-styrene] to blend in melt stage. ABS is a polymer blends consisting of a butadiene-based rubber grafted with astyrene and acrylonitrile copolymer (SAN) dispersed in the SAN matrix. The LCP is a copolyester of hydroxybenzoic acid (HBA) and poly(ethylene terephthalate) (PET) with a molar ratio of HBA/PET = 80/20. According to the authors, anhydride group in the copolymer backbone can incorporate with the carboxylic acid end groups through H-bonding, and can react with the hydroxylic end groups or eventually the ester groups in the LCP bacbone. They observed that SMA copolymer exhibits a compatibilizing effect on the blend system and optimum copolymer content exists for mechanical properties enhancement. This reactive compotibilizer improves the interfacial adhesion, whereas excess of copolymer reduces the LCP fibrillation. The tensile strength and the modulus were slightly enhanced with the addition of 2-5 % of copolymer. Colbeaux et al. [516] developed a strategy for *in situ* compatibilization of PP/PE blends by reactive extrusion, for recycling purpose. As the PE and the PP are chemically inert, the process consisted of adding to the medium a certain amount of poly(PP-*g*-MA) and poly(PE-*g*-MA) chains. Thus, by coupling between the chains, a compatibilizing copolymer can be formed at the interface. Authors used the amine/anhydride reaction for effective compatibilization of these polymer blends. To simplify the problem, model reactions carried out on poly(PE-*g*-MA), and the properties of the resulting modified PE were analysed. The efficiency of the diamines 1,12-diaminododecane (DAD) and (4,4'-methylene)bis(2,6-diethylaniline) (MDEA) used as coupling agents for poly(PE-*g*-MA) compared by authors. It was shown tha the reaction of DAD on the grafted MA funtions is faster than that of MDEA. Moreover, authors observed that all the anhydride functions were consumed if the stoichiometric ratio was equal to or higher than 2. One anhydride unit reacted with two $NH_2$ groups, leading to highly branched PE chains. DAD is thus a suitable coupling agent for *in situ* compatibilization of PP/PE blends.

PP and polyethylene terephthalate (PET) are formed inmissible blends with two-phase morphology and poorer mechanical properties than those for pristine PP and PET [517]. Cartner et al. [518] grafted MA onto PP in the melt and subsequent blending with highly crystalline PET. They found that compatibilized PP/PET blends showed mechanical, thermal and barrier properties and morphology that clearly differed from the noncompatibi-lized ones.

Baird et al. [519] employed poly(PP-*g*-MA) as a compatibilizer for polyblends of a copolyester of LCP

(thermotropic liquid crystalline polymer) and PP. They reported that significant improvements can be achieved in tensile strength and modulus by adding a small amount of poly(PP-*g*-MA) in the PP/LCP polyblends. This observed fact authors explained by possible reaction of the anhydride units of poly(PP-*g*-MA) with the amide-end groups of LCP. They indicated that hydrogen bonding is responsible for the compatibilizing effect of poly(PP-*g*-MA) on PP/LCP blends. Similar H-bonding effect was also observed by Tjong and Meng [520].

The toughening of semi-crystalline polyamides using maleated rubbers, predominantly maleated styrene/ ethylene-butene/styrene rubber [poly(SEBS-g-MA) with 29 wt.% of styrene and 1.84 wt.% of MA] has received considerable attention in recent years because of the capability of tailoring the rubber phase morphology by the in situ reaction between MA grafted to the rubber and the polyamide amine end groups during melt compounding [521-527]. Huang et al. [528] investigated rubber toughening of an amorphous polyamide using combinations of triblock copolymers of SEBS and a poly(SEBS-g-MA). Two series of blends were prepared by both the single and twin screw extruders. The effects of rubber content and the type of the extruder on the morphology, Izod impact behavior and the ductile-brittle transition temperature were explored. I was found the morphology (spherical and regular) of these blends in more similar to that of nylon 6 blends than nylon 66 blends. Authors showed that the twin screw extruder produced smaller particles with a more narrow distribution of sizes than the single screw extruder.

Tanrattanakul et al. [529] obtained toughened poly(ethylene terephtalate) (PET) by reactive blending with 1-5 % of SEBS-*g*-MA (with.2 % grafted MA unit). They showed that the fracture strain of PET increased by more than a factor of 10, consistent with *in situ* formation of a graft copolymer by reaction of PET hydroxyl end groups with MA. Also in the case of blends of polyamide (PA6) with polycarbonate (PC), authors proposed that in the *situ* chemical reaction between the anhydride units of SEBS and the amine end group of PA6 during melt mixing induces the encapsulation of SEBS-*g*-MA within the PC domains in PA6 rich blends [530,531]. It was found that the adhesion on the domain boundary between PA6 and PC through this phase formation improve and thus also mechanical properties enhance.

Tjong et al. [532] prepared 80/20 poly(SEBS)/PPand 20/80 poly(SEBS-*g*-MA)/PP hydrid blend composites reinforced with 30 wt % SGFs (surface glass fibers) by extrusion and subsequent injection molding. They shown that poly[SEBS-*g*-MA (1.84 wt.%)] improves the yield strength and impact toughness of the hybrid composites. The MA was grafted to the central EB chain segment od SEBS copolymer. According to the authors, the anhydride groups in this position may be react with hydroxyl groups of the glass-fiber surfaces during compounding, thereby improving the compatibility between the SFG and SEBS. The reaction between SEBS-*g*-MA and SGF represented as follows (Figure 36):

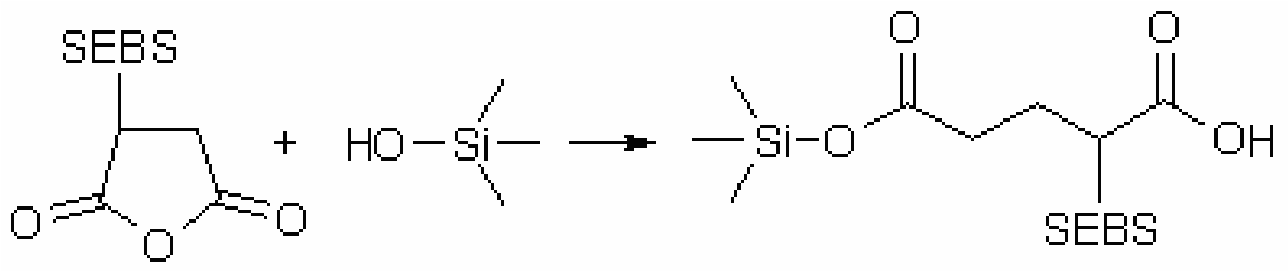

**Figure 36.** Reaction of SEBS-*g*-MA and surface glass fibers via esterification of anhydride units of graft copolymer with surface hydroxyl groups of fiber. Adapted from [532].

Liang et al. [533] reported that to encapsulate the elastomeric phase, the entire filler or glass-bead surface for maleated elastomer/PP hybrids reinforced with filler particles. This leads to a fine dispersion of rigid particles with a core-shell structure within the PP matrix. However, authors [532] on the base of obtained SEM micrographs of surface morphology of blends assumed that, elongated SGF can only result in a partial coating or bonding of fiber surfaces with thin layers of SEBS-*g*-MA. Extensive plastic deformation occurred at the matrix interface layer next to the fibers of the SGF/poly(SEBS-*g*-MA)/PP composites durind impact testing. DSC measurements indicated that SEBS promoted the crystallization fo PP spherulites by acting as active nucleation sites. While the MA grafted units retarded the crystallization of PP. Authors confirmed the absence of transcrystallinity at the glass-fiber surfaces of both SGF/poly(SEBS-*g*-MA)/PP and SGF/poly(SEBS/ PP hybrid composites by polarized optical microscopy.

Willis et al. [534] studied the morphology and impact properties of poly(MA-*co*-styrene)/bromobutyl elactomer (rubber) blends, and processing-morphology-property relationships as a function of *in situ* modification and melt processing conditions. They found that dimethylamino-ethanol (DMAE) serves as a reactive compatibilizing agent for these blends and that the addition of DMAE results in a five-fold reduction in the size of the dispersed phase. Authors presented evidence for covalent bond formation between the DMAE and the elactomer and reactive polystyrene phases. It was shown that impact strength measurements clearly dependent on the quantity of DMAE in the system as well as on the concentration of elastomer.

Pötschke et al. [535] reported the preparation, morphology and mechanical properties of reactive blends of thermoplastic polyurethane elastomer (TPU) and poly[PE-*g*-MA (0.5 wt. %)]. TPU and PE form immiscible blends with an extremely low compatibility. In order to improve the dispersion, stability, and properties of these blends, authors grafted MA onto PE using thje following procedure: A master batch of PE powder and 0.7 wt. % of MA was fed to extruder. The grafting was carried out at 240 °C without addition of peroxide. The grafting MA content was determined by quantitative IR spectroscopy using the adsorption band at 1792 $cm^{-1}$ (C=O) to be 0.5 wt. %. Grafting of MA onto PE leads to a decrease of the molecular weight, the melt viscosity, and the mechanical properties of the pure PE. Blends a commercial polyester type TPU with poly(PE-*g*-MA) prepared by authors in a twin-screw extruder. It was shown that the particle size dramatically reduced in the case of poly(PE-*g*-MA) as a blend component, in comparison with non-grafted PE. Coalescence (coarsening of the blend morphology during processing, e.g. injection molding) was significantly reduced and the increase in particle size with compositon was less pronounced than in bleds with PE. In addition, the phase adhesion and the mechanical properties were improved by using poly(PE-*g*-MA) compatibilizer as minor component. This reactive blend system exhibits a lower viscosity ratio. The reduction in viscosity ratio has a big influence on the blend morphology and because of that on the mechanical properties. According to authors, observed them a lower interfacial tension in this blend system may result from ether an increase in the surface tension and polarity of the poly(PE-*g*-MA) or by a chemical reaction between the blend components. This effect can be also interpreted as polymer/polymer interaction via complex formation between –NH (urethane fragment) and –C=O (anhydride unit).

Melt grafting of EPDM with MA and the use of EPDM-*g*-MA as compatibilizer in polymer blends has been reported to be successful in improving the properties of the blends [536]. Enhancement of impact strength of PP/EPDM blend upon grafting the EPDM with MA {poly[EPDM-*g*-MA(1.5 wt %)]} was reported by Zhao and Dai [537]. Montiel et al. [538] observed that the effect of MA grafting level in polyamide 6/poly(E-*co*-P) rubber (EPR) blend to result in improved mechanical properties and morphology of the blend.

Many researchers have been observed that the addition of MA grafted copolymers such as poly[ethylerne-*co*-vinylacetate)-*g*-MA] and poly[HDPE-*g*-MA(1.2 %)] to poly(butylene terephthalate) (PBT)/HDPE blending systems considerably improved the impact strength without significanly sacrificing the tensile and flexural strength and decreased the size of dispersion phase [539-542]. Synthesis of poly(ABS-*g*-MA) and poly(HDPE-*g*-MA) graft copolymers in melt by reactive extrusion reported by Qi et al. [543,544], The effect of the compatibilization on the mechanical and rheological properrties, morphology and crystalline behavior of the PBT/HDPE and PBT/poly(HDPE-*g*-MA) blends were investigated in detail [544]. The results of these authors showed that the notched impact strength and the interfacial adhesion of PBT/poly(HDPE-g-MA) blend increased and the dispersed phase paticles decreased with the addition of MA grafted copolymer. Similar increase of interfacial adhesion in PP/EPDM/poly[EPDM-*g*-MA(0.3 wt %)] [EPDM is poly(ethylene-*co*-propylene-*co*-norbor- nene)] blends was observed by Purnima et al. [545].

## 8.2. Polyamide reactive blends

PAs are a family of engineering thermoplastics whose wet affinity limits their application. PA blends with lower-module but hydrophobic polymer-like polyolefins has therefore become a matter of interest for both their ultimate and their thermal and morphological properties [546-549]. The rections of MA units in graft copolymers with the amine end groups of PA6 and PA66 studied in detail and well documented by many researchers. Wu et al. [550] investigated the effects of poly[PP-*g*-MA (0.6 %)] compatibilizer content on crystallization of PA12/PP blends and their morphology. It was found that an *in situ* reaction ocurred between the MA uints of compatibilizer and the amide end groups of PA12. Authors observed that the PA12 did not crystallize at temperatures typical for bulk crystallization when the amount of compatibilizer was more that 4 %,. The *in situ* interfacial reaction in the modified blend component resulted in compatibilization connected with a higher finely dispersed blend morphology and the apearance of fractionated crystallization. According to authors, the *in-situ* formed graft copolymer poly[(PP-*g*-MA)-*g*-PA12] played a role in concurrent crystallization by reducing interfacial tension and increasing the dispersiveness of PA12.

The most common of *in situ* compatibilization of polymer blends is the application of MA-containing polymers in blends with polyamides (PA) [550-562]. Steurer and Hellman [561] studied the homogeneous interface grafting in melt of maleinated polystyrene (PSMA) and PA reactive blends. They investigated a series of PSMA/PA blends differing in composition and PA chain lengths, and the kinetics of *in situ* in homogeneous interface melt-grafting of poly[MA(14 mol %)-*rand*-styrene] ($T_g$ = 134°C, $M_n$ = 85,000 and $M_w$ = 160,000 g/mol) and PAs ($T_m$ = 170-180°C, $M_n$ = 1600-20000 and $M_w$ = 3100-39000 g/mol) and their trifluoroacetylated derivatives. According to the authors, immiscible polymers A and B carrying complementary function X (anhydride unit of SMA) and Y (amino group of PA) can form graft copolymer chains poly(SMA-*g*-PA) in the interfaces between their phase domains (Figure 37):

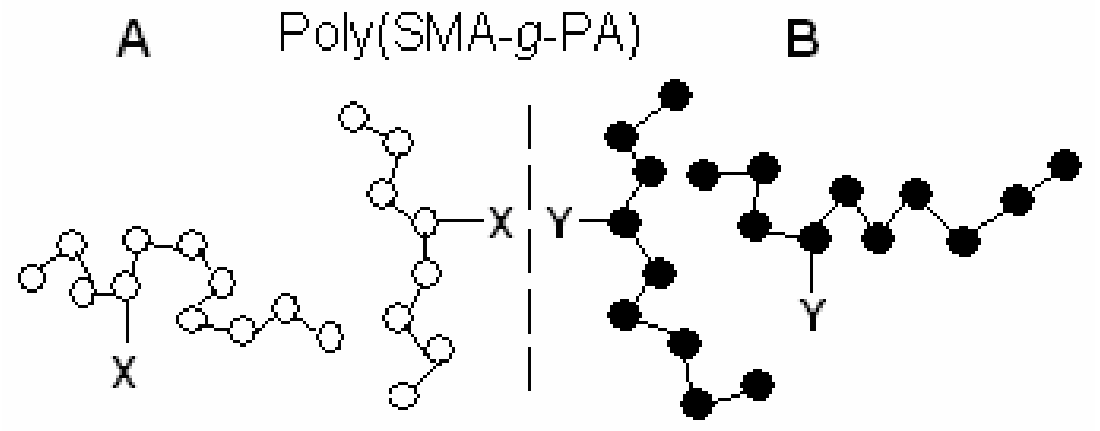

**Figure 37.** The formation of graft copolymer chains through interfacial copolymer (A-SMA) and polyamide (B-PA) interaction. Adapted from [561].

Blending of PA with MA-modified rubbers, such as ethylene-propylene(EP)-*g*-MA and styrene-butylene-styrene(SBS)-*g*-MA, results in increased (low temperature) tougness [563]. In addition, Majumdar et al. [564] showed that the resistance towards moisture and warpage were improved when ABS used as an impact modifier. Blending of PA with PP-*g*-MA resulted in decreased moisture sensitivity and a lower cost price [565]; blending with MA-modified E-P copolymer increased impact strenght and high temperature resistance [550,551]. For special purposes MA-styrene copolymer used by authors as a blend component to increase the high-temperature performance or as a compatibilizer in blends with convertional polymers, such as PS, SAN, ABS, PMMA, etc. [564].

The best examples of *in situ* chemical modification are reactive polymer blends of MA-containing copolymers and graft copolymers with polyamides (PA)s [445-449,451,453]. Blended of PA with MA-modified rubbers, such as poly(EPDM-*g*-MA) and poly(SEBS-*g*-MA), results in increased (low temperature) thougness [562]. *In situ* modification of PA with poly(PP-*g*-MA) results in decreased moisture sensitivity and a lower cost price [565]; blending with MA-modified PE increases impact strength and high-temperature resistance [566,567]. In addition, the resistance towards moisture and warpage were improved when ABS is used as an impact modifier [565-569]. Blending PA with MA grafted E-P copolymer increases impact strength and themal stability [567]. Majumdar et al. [568] reported a serial reactive bifunctional PAs (nylon-6,6 and nylon-12,12) blends with poly(SEBS-*g*-MA) (a maleated styrenic triblock copolymer with ethylene-butylene middle block) and poly(EPR-*g*-MA) (a maleated ethylene-propylene rubber). The morphology and mechanical properties of these blends prepared in single and corotating twin-screw extruders and studied using transmission electron microscopy (TEM) technique. It was shown that for blends of poly(SEBS-*g*-MA) with bifunctional PAs, the twin-screw extruder is more effective in producing a finer dispersion of the rubber phase. On the other hand, for blends of poly(SEBS-*g*-MA) with monofunctional PA (nylon-6), the single-screw extruder led to blends with excellent low-temperature impact properties for both single-step and masterbatch mixing techniques. Araújo et al. [569] prepared PA6/ABS blends compatibilized with poly[methyl methacrylate-*co*-MA(3-20 wt. %)]. They showed that the presence of this copolymer in the blends clearly led to a more efficient dispersion of the ABS phase and consequently optimized Izod impact properties. According to the authors, reactive compatibilization proceeds through reaction of anhydride unit of copolymer with the amine end groups in PA6 during melt processing.

Maleation of metallocene poly[ethylene-*co*-1-octene (25 wt. %)] (POE elastomer with $T_m$ = 55°C and MFI 0.5 dg/min) carried out with dicumul peroxide as an intiator at 200°C and 120 rpm in melt by co-rotating twin-screw extruder by Yu et al. [570]. They reported the effect of the addition of grafted POE on the impact strength, yield strength and modulus of nylon 6. The effects of the MA graft ratio and functionality of a series POE on the mechanical properties and morphology of nylon-11/ (poly(POE-*g*-MA) binary reactive blends and nylon-11 /POE/(poly(POE-*g*-MA) ternary reactive blends investigated in detail by Li et al. [571]. Semi-crystalline Nylon 11with $T_g$ = 44°C, $T_m$ = 193°C and $M_w$ = 26000 g/mol and poly(POE-*g*-MA)s with 0.38, 0.56 and 0.89 % MA graft ratios used in this study. It was found that the optimal MA content for the preparation of blend with maximum value of Izod impact strength (~800 J/m) is 0.56 wt. %. SEM and TEM micrographs demonstrated that incorporation poly(POE-*g*-MA) ensured good compatibility between nylon 11 and POE particles [572].

The chemical reaction scheme of anhydrides of dicarboxylic acids with amines and amides is quite complex, but is sumplified at elevated temperatures (>200 °C) [553]. According to the authors [572-575], amine end groups of PA react with the anhydride moiety, resulting in PA graft formation via an imide linkage. PA amide groups do not react directly with the anhydride unit unit, but only after being hydrolyzed to an amine and a carboxylic acid (Figure 38).

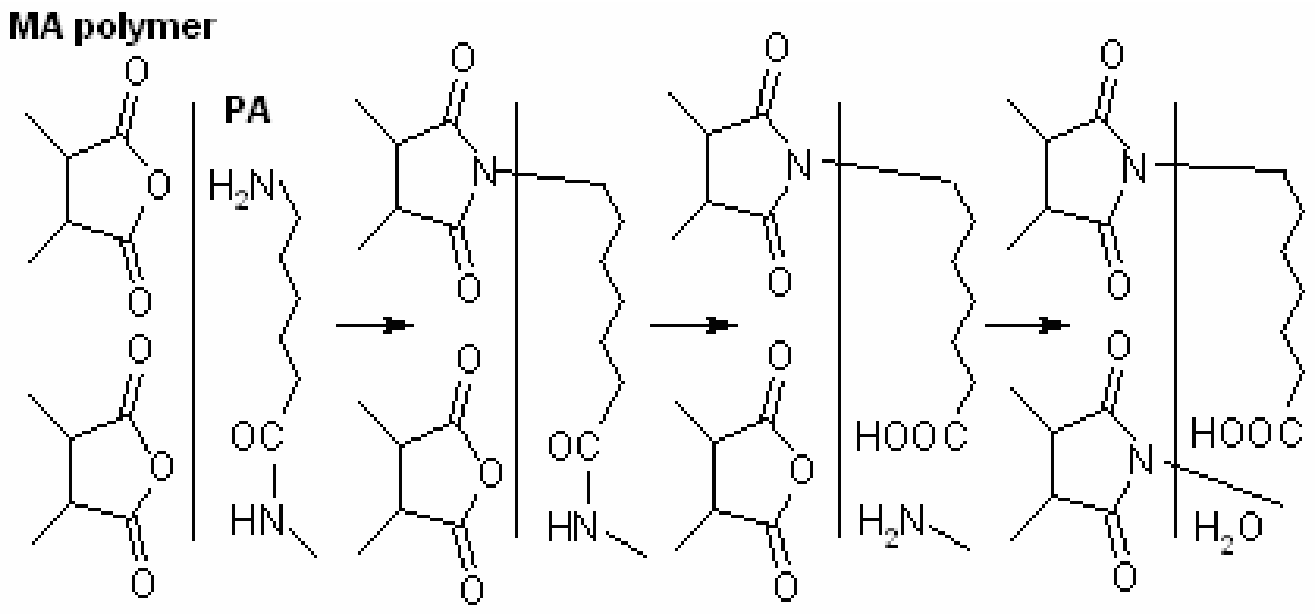

**Figure 38.** The formation MA copolymer/PA graft by reaction of end amine groups with the anhydride moiety via formation of an imide linkage. Adapted from [573].

Overall reaction results in PA chain scission and imide formation. The interactions between chemistry, morphology, and rheology studied by several researchers using MA-containing polymers with different MA unit contents such as poly[EPDM-g-MA(1.4 wt.%)] and poly[styrene-*co*-MA (28 wt.%)]. The effect of PA symmetry also studied. It was shown the when PA-6 is used, only linking is occur; in the case of PA-66, crosslinking is occur [557-559,568,569,576].

Van Duin et al. [573] demonstrated that melt blending of PA-6 and PA-6,6 with poly(EPDM-*g*-MA) and poly(St-*alt*-MA) results in coupling of PA with MA-containing polymer via an imide linkage and is accompanied by PA degradation. Differential scanning calorimetry shows that the degree of crystallinity of the PA phase is decreased only when the size of the PA phase between the MA-containing polymer domains approaches the PA crystalline lamellar thickness. Graft formation and chain scission in blends of PA-6 and PA-6,6 with MA-containing polymers [EPDM-*g*-MA and MA(28 wt %)–styrene copolymer] also studied by these authors. For this, they prepared four polymer blends and characterized. Chemical analyses showed that the amount of PA graft is independent of the blend composition. Going from EPDM-*g*-MA to MA(28 wt %)–styrene copolymer the MA content of the orginal MA-containing polymer increases, which in the corresponding blends results in an increase in the number of PA grafts and a decrease in the length of the PA grafts. It was found that in the MA–styrene copolymer blends the number averaged molecular weight of the grafted PA is only about 200 g mol$^{-1}$. During blending the MA-containing polymer first melts, completely covering the PA granulate surface with grafts.

According to authors, the molecule of water released upon imide formation is formed in the melt at high temperature (260-280 $^{o}$C) and under these conditions may result in hydrolysis of a PA chain close to the PA/(MA-containing polymer) interface. General scheme reactions at the interface in blends of PA and MA-containing polymers represented as follows (Figure 39):

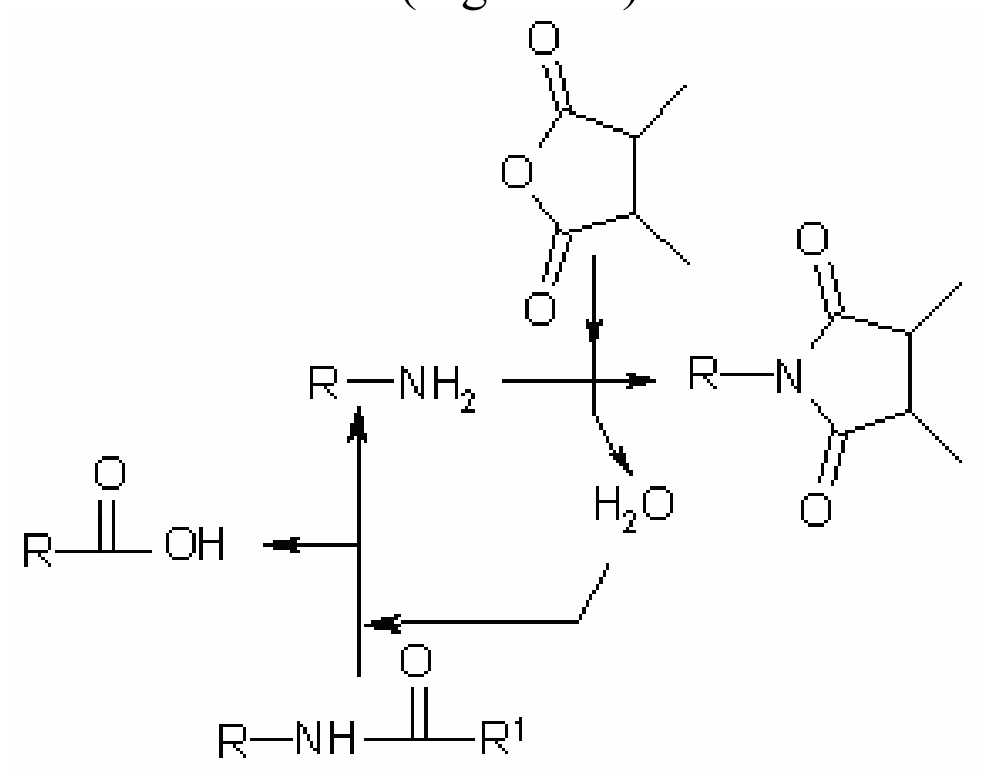

**Figure 39.** Interfacial interactions in reactive blends of PA and anhydride copolymer. Adapted from [573].

Authors postulated that the water molecule, released upon imide formation at the PA/MA-containing polymer interface, hydrolyses a PA amide group, resulting in a new amine end group, which in its turn reacts with the MA-containing polymer [573].

Several researchers [575,576] reported the reactive compatibilization of PA blends with commercial poly(HDPE-*g*-MA) [564] containing different amount MA unit and some thermoplastic elactomers grafted with MA such as poly(SEBS-*g*-MA)s [577].

All these compatibilizer precursors (CPs) react during blending, with the functional groups of PA to produce CP-*g*-PA copolymers, though different kinetics and at different yields. The results confirm that the anhydride functional groups possess considerably higher efficiency, for the reactive compatibilization of LDPE/PA blends, than that of the ethylene-acrylic acid and ethylene-glycidyl methacrylate copolymers. Concerning the efficiency of the different MA-grafted precursors, Filippi et al. [575] demonstrated that the efficiency of the poly(E-*g*-MA)s changes dramatically depending on the structure and the molar mass of the PE substrate. The efficiency of the MA grafrted elastomers such as poly(SEBS-*g*-MA) and poly(styrene-*b*-ethylene-*co*-propylene) practically coinci-des with that of poly(HDPE-*g*-MA) for the compatibilization of PE/PA 75/25 w/w blend. CPs functionalized with MA groug also employed for compatibilization of PP/PA blends [564-568]. Holsti-Miettinen et al. [580] studied the poly(SEBS-*g*-MA) reactive compatibilization of PA/*i*-PP blends by mechanical, morphological, thermal and rheological analyses. Kim et al. [577,578] assumed that the reactions of anhydride units with amine and amide groups in blends of PA with MA-containing polymers are quite complex. First, amine end groups of PA will react with the anhydride moiety, resulting in PA graft formation via an imide linkage. Amic acids are only formed as intermediates. According to the authors, the PA amide group does not react directly with the anhydride unit, but only after being hydrolyzed to an amine and a carboxylic acid. Kim et al. [572,573] and Wilkinson et al. [581] investigated the i-PP/PA/SEBS-*g*-MA ternary blends with i-PP as the matrix phase. Gonzales-Montiel et al. [582-584] also reported the preparation and characterization of PA/i-PP/SEBS-g-MA blends. Chen and Harrison [585] used six different CPs, including two grades of poly(SEBS-*g*-MA), to compatibilize 80/20 blends of PE with an amorphous PA with the aim of producing PE films reinforced with PA fibers for balloon applications. Armat [588] showed that poly(SEBS-*g*-MA) is good CP for the reactive compatibilization of 25/75LDPE/PA blends.

Filippi et al. [575,586] used a MA grafted styrene-*b*-(E-*co*-P) copolymer (SEP-*g*-MA) for the compatibili- zation of LDPE/PA6 blends. The morphology and the mechanical properties of these blends (75/25 and 25/75) compared with those of the blends compatibilized with poly(HDPE-*g*-MA) and poly(SEBS-*g*-MA). These results show the strong compabilizing efficiency of (SEP-*g*-MA) towards LDPE/ PA6 blends, while poly(HDPE-*g*-MA) demonstrates a slightly lower activity, especially when PA is the matrix phase.

It is well known that the chemical reaction scheme of cyclic anhydrides with amines and amides is quite complex, but is simplified at elevated temperatures (>200$^{o}$C) [587 590]. According to the authors, the reactions occurring in blends of PA with MA-containing polymers include the following stages: (1) amine end groups of PA react with the anhydride moiety, resulting in PA grogt formation via an imide linkage; (2) when an excess of anhydride is present, the amide groups in the PA chain also participate; (3) the PA amide group does not react directly with the anhydride, but only after being hydrolyzed to an amine and a carboxylic acid.

PP is widely employed because of its low cost, high barrier properties to moisture, and case of processing, but its high permeability to oxygen and many organic solvents limits its potential use [588]. On the other hand, PA6 is a good barrier material for oxygen and organic compounds but its relatively expensive, hygroscopic and thus poor barrier for water. PP and PA6 are not compatible and that blending of these materials results in poor properties. Poly(PP-*g*-MA) is of considerable importance for application as a compatilizing agent with PA6, as adhesion promoter with glass or carbon fibers. Abbacha and Fellahi [589,590] reported the synthesis of poly(PP-*g*-MA) and evaluation of its effect on the properties of glass fibre reinforced nylon 6/PP blends. They found that the incorporation of the compatibilizer (MA grafted PP) enhances the tensile strength and the modulus, as well as the Izod impact properties of the prepared blends. This was attributed by authors to better interfacial adhesion between the two component phases due to low interfacial tension leading to finer dispersion and a reduction in the dispersed phase size. The optimum in these properties is reached at a critical poly(PP-*g*-MA) concentration (5 wt. %) [589]. It was demonstrated that the compatibility pf PP and PA blend can be improved by the addition of a compatibilizer, and functionalization of PP with MA in the presence of dicumyl peroxide (1.0 phr) [591]. The reaction was carried out in the molten state using an internal mixer. The poly[PP-*g*-MA(0.36 phr)] (2.5-10.0 wt.%) as a compatibilizer was added to 30/70 glass fibre reiforced PA-6/PP blend. It was found that the incorporation of the compatibilizer enhanced the tensile strength and modulus as well as the Izod impact properties of the notched samples. This was attributed to better interfacial adhesion as evidenced by SEM microscopy. A critical value of poly(PP-*g*-MA) content of 5 wt.% was determined. It was shown that addition of this compatibilizer led to an eightfold reduction in average particile diameters.

Supertough blends of polyamide (nylon-6) with maleated elastomers, such a MA grafted ethylene-propylene rubber [poly(E-*co*-P-*g*-MA)] and styrene/ hydrogenated butadiene/ styrene triblock copolymer as the high performance engineering materials, have become commericially important materials of considerable scientific interest [568,585,582-600].

Borggreve et al. [598] used a MA grafted ethylene-propylene-diene rubber (EPDM) to reach the same goal. An essential feature of these materials in the graft copolymer generated from the reaction of the grafted MA unit with polyamide amine end groups during the melt-blending process. The grafted copolymer strengthens the interface between phases, reduces interfacial tension, and provides steric stabilization that retards the coalescence of the dispersed phase. The last allows the formation of stable, finely dispersed rubber particles [599]. Supertough blends result when the rubbers particles can cavitate during the fracture process and permit shear yielding of the polyamide matrix [588,600]. Wu [601,602] obtained a supertough nylon 66 by incorporating of MA grafted ethylene-propylene rubber (EPR).

It is well known that MA functionalized polyolefins used as compatibilizers in various polymer blends and compositions exhibit enhanced adhesion to polar polymer materials, metals, and glass. Many researchers demonstrated that the formation of a polyamide-polyolefin graft copolymer *in situ* during melt blending of polyamide-6 (PA6) with various maleated polyolefins significantly improved impact properties of materials [476-482]. PP containing a compatibilizer consisting of MA-modified PP has been shown to enhance the material properties of the composite due to chemical interaction of anhydride groups with functional fragments of used materials in the processing conditions [600-602].

Dong et al. [603] reported synthesis of graft copolymer, poly[(MA-*co*-styrene)-*g*-PA6] and its blend with PVC, as well as the morphology and mechanical properties of compatibilized blends. It is well known that the processing temperature of PVC is ~170-190$^o$C, whereas the melting point of PA6 is 215$^o$C. The huge difference of their processing temperatures becomes the main obstacle to the blending of PVC and PA6. They prepared polymer-polymer graft SMA-*g*-PA6 via a solution graft reaction between SMA and PA6. FTIR spectra evidences the occorrence of the graft reaction between SMA and PA6. DSC analysis shows that SMA-*g*-PA6 has a lower melting point of 187 $^o$C, which authors explaned by decreasing in crystalliny of PA6 and thus enable efficient blending of that SMA-*g*-PA6 with PVC. Authors found that compatibilization is evidenced by the dramatic increase in mechanical properties (impact and tensile strength) the smaller particle size and finer dispersion of PA6 in PVC matrix, and, further, a cocontinuous morphology at 16 wt. % SMA-*g*-PA6.

Compatibilization is generally accomplished by either adding a pre-synthesized block copolymer or via reactive processing [568,569,580,602-605]. In the last case, the compatibilizer is generated *in situ,* during blending, through grafting or exchange reactions. Accordind to Wong and Mai [606], these reactions must generate sufficient copolymer in order to optimize the morphology and, thus, the blend performance. Since the reactions occur across phase boundaries, high stresses must be applied to enlarge the interfacial area and ensure high reaction rates. Blends of industrially relevant polymers such as PA-6, poly(E-*co*-P) rubber, poly(E-co-P)-g-MA (80/20/0 to 80/0/20; w/w/w) [568,569,602-604] were selected for the studies of the changes in the rheological behavior, morphology and chemical conversion of along a compounding co-rotating twin-screw extruder [494,496]. Authors demonstrated that the linear viscoelastic behavior of immiscible non-compatibilized PA-6/rubber blends along a twin-screw extruder is related to changes in morphology and/or formation of the *in situ* compatibilizer. The dynamic viscosity and storage modulus decrease substantially upon melting of the components, since this process in induced by a significant increase in interfacial area as a result of *in situ* grafting reaction of amine (from PA-6) and anhydride (from maleated rubber) groups during extrusion processing.

Tjong and Meng [520,604,607] used reactive poly(PP-*g*-MA) as a compatibilizer for PA-6/liquid crystalline copolyester (LCP) and PP/LCP blends. PA-6/LCP/poly(PP-*g*-MA) blends were prepared by extrusion blending of pellets followed by injection molding at 295$^o$C. However, one-step injection molding of PA-6 and poly(PP-*g*-MA) with LCP pellets at relatively low temperature of 285$^o$C facilitates fibrillation of the LCP phase in the poly(PP-*g*-MA) compatibilized PA-6/LCP blends. According to authors, this is due to the elimination of extrusion blending and to a lower processing preventing further degradation of PA-6 [609]. MA compatibilized blends of PP and LCP were prepared either by the direct injection molding (one-step process), or by twin-screw extrusion blending, after which speciments were injection molded (two-step process). The results show that the tensile modulus and strength of poly(PP-*g*-MA)/LCP blends prepared by the one-step process are close to those predicted from the rule of mixtures. According to the authors, this is due to the formation of fine and elongated LCP fibrils in the blends and due to the stability of PP matrix during processing [408]. Moreover, the impact strength of these blends decreases inilially with increasing LCP content up to 10 wt. %, followed by a monotonic increase with increasing LCP content. On he other hand, MA compatibilized blends prepared from the two-step process showed lower tensile stress, modulus, and impact strength owing to the decomposition of PP matrix.

A random copolymer of styrene with MA (SMA) was coupled with amino terminated PA-12 of various chain lengths [607-611]. According to the authors [610], an interface layer of poly(SMA-*g*-PA) is formed that compatibilizes the blend by lowering the interface tension and improving the entanglement network so reactive blending leads to finer phase morphologies and increased strength. Thus modified blends can attain submicroscopic morphologies even when A and B components are badly incompatible. It was shown that anhydride and amino functions react at high temperatures fast and irreversibly by imide condensation. The imidization is a bit delayed and the imide forms approximately according to the simple kinetics ($k_1 >> k_2$, Figure 40):

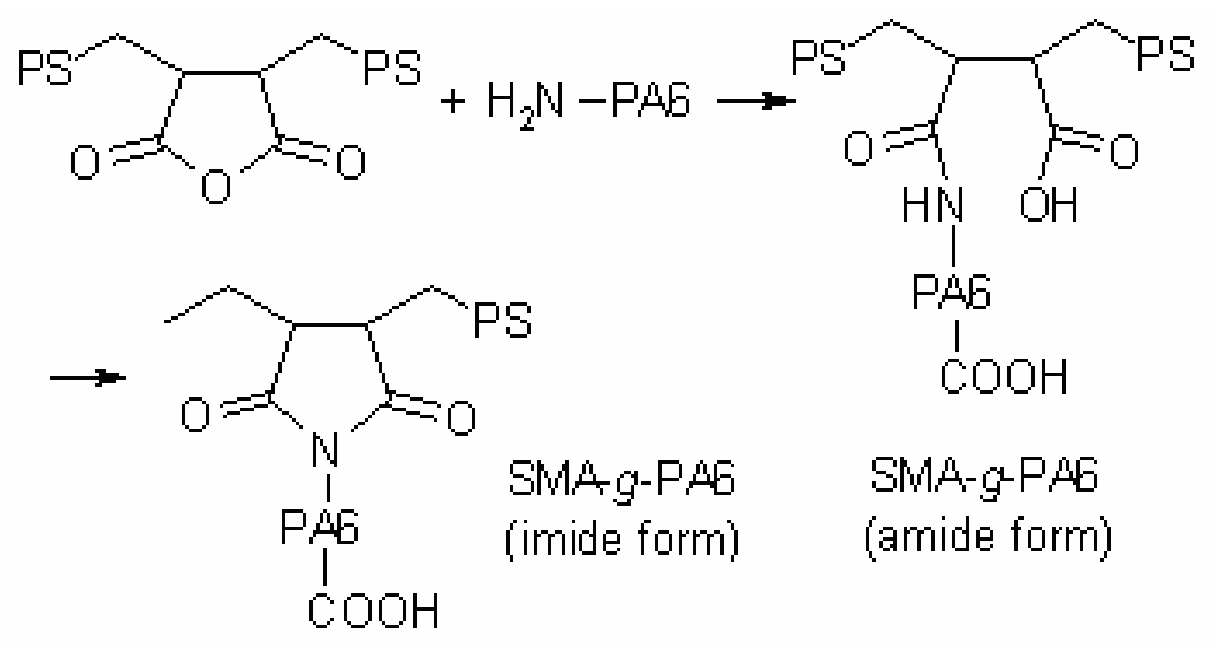

**Figure 40.** Synthesis of the amide and imide forms of SMA-*g*-PA) via interfacial amine-anhydride reaction. Adapted from [606].

It was observed that the poly(SMA-*g*-PA) chains form an autoinhibitory barrier in the interfaces that prevents random grafting [612].

The compatibilization of PA/thermoplactic elastomers blends involving the grafting of MA onto the elastomer phase also reported by several researchers [570,611-614]. It was shown that MA grafted elastomers improved the impact strength and reduced the dispersed phase size in those blends. The morphology and mechanical properties of these blends significantly depend on the ratio of reactive/ nonreactive elastomers, the MA content, composition and the viscosity of the blend components.

## 8.3. Natural polymers reactive blends

POs are the thermoplastic polymers widely used as a matrix in wood fiber-plastic composites [615]. MA is the usual choice in modification of the polymer matrix by grafting reaction. These graft copolymers, such as maleated PPs, are applied as a coupling agents, for the preparation of composites useful in the authomobile and packaging industries [615,616], and for outdoor application because when maleated PP is used, the fiber surface becomes hydrophobic and the water uptake decreases significantly [507]. The esterification reaction between different wood fibers and maleated PP has been under investigation for several years now. X-ray photoelectron and FTIR spectroscopy are the method most often used to confirm the esterification reaction [394,422-425]. Authors [617] shown that the anhydride form of MA-*g*-PP interacts with viscose fiber via physical (hydrogen bounding) or chemical (esterification) bonding, leding to improved adhesion between fibers and the PP matrix. The tensile strength and elongation of fracture of a composite of PP and viscose fiber increase significantly when the MA-*g*-PP concentration is increased in the composite material.

It is known that carboxyl-containing polymers have been used as cross-linking agents for cotton fabrics and paper to replace the traditional formaldehyde-based reagents. Yang et al. [618,619] showed that treatment of cotton fabric with a mixture of MA and itaconic acid significantly improves wrinkle-resistance of the fabric. They studied the *in situ* copolymerization of MA and itaconic acid on cotton fabric. Using FTIR and redox titration technique for quantitative determination of carbonyl fragments and double bonds on the treated cotton fabric, authors found that free radical copolymerization of MA and itaconic acid does not occur on the fabric at elevated temperatures when potassium persulfate is present as an initiator. It does occur, however, when both potassium persulfate and sodium hypophosphite are present on the fabric. Authors assumed that the *in situ* copolymerization on the cotton fabric probably is initiated by a reduction-oxidation system.

Synthesis of PE-*g*-poly(styrene-*co*-MA) and its compatibilizing effects on the PE/starch blends reported by Park and Yoon [620]. In this study, graft copolymerization of styrene and MA was carried out using PE-TEMPO (2,2,6,6-tetramethyl-1-piperidinyloxy) as a macroinitiator. It was shown that styrene homopolymerization with PE-TEMPO proceeded very slowly and produced a large amount of polystyrene by-product. A negligible amount of MA was grafted by PE-TEMPO when MA was homopolymerized. However, the graft copolymerization became fast in the presence of complexed St/MA mixture. The amount of MA units in the produced PE-*g*-poly(St-*co*-MA) was much larger than that of MA grafted on PE by using the conventional grafting method with thermal and photoinitiators. Authors found that chemical reaction between MA units and hydroxyl groups of starch was nearly undetecstable in the PE/PE-*g*-poly(styrene-*co*-MA)/starch blend system, and the tensile properties of the blend were not enhanced significantly.

*In situ* chemical modification–interfacial interaction also observed by Bismarck et al. [621] in the cellulose/maleated PP composite systems. The properties such composites depend on their interfacial adhesion as well as properties of the individual components. Many researchers [622-624] found that MA grafted PP is very efficient for preparing cellulosic PP composites with improved mechanical properties. Esterification of MA copolymers or maleated PP graft copolymers with surface hydroxide groups of solid celluloase materials, including cellulose fibers was also described in several publications. Thus, It was shown that cross-linking of cotton cellulose for stabilization by poly(maleic acid) occurs as a two step reaction: (1) formation of cyclic anhydride intermediate (activation) and (2) reaction of the anhydride intermediate with cellulose by esterification [624]. The improvement imparted by the use of poly(PP-*g*-MA) is considered to be attributed to esterification reaction of anhydride unit with hydroxyl group of cellulose in solutions in the presence of catalyst. The formation of ester groups has been confirmed by the appearance of a new IR band at 1729 and 1750 $cm^{-1}$ [387,.616,620,621]. It was also found that ball-milling induces intensive esterification betwwen the OH groups on microparticles of cellulose and the MA grafted units of poly(PE-g-MA) [388] or poly(PP-g-MA) [625], in marked contrast to the melt-mixing of the original cellulose and maleated PE and PP. Several researchers also reported the use of MA monomer or maleated PPs in cellulose reinforced thermoplastic composites [429,431,626,626-629], and an esterification between the "solid" cellulose and anhydride moiety has been assumed. Qui et al. [398-400] discussed the formation of ester bonds between "mechanically activated" hydroxyl groups of cellulose and MA units of poly(PP-*g*-MA) (freshly tempered at 120°C) during ball milling at room temperature (for 10 h). They demonstrated the importance of destroying the H-bond network during ball milling of the highly crystalline cellulose to yield free (activated) hydroxyl groups. Alternatively, the chemical modification of cellulose by MA to cellulose maleate was also carried out successfully in pyridine at 100°C [400].

A new method for the preparation of covalently immobilized cellulose layers on top of inorganic substracts such as silicon wafers or glass carriers reported by Werner et al. [630,631]. Authors used MA copolymer films covalently attached to glass carriers of oxidated silicon wafers as anchor films which bind the cellulose layer by esterification between the OH groups of the cellulose and the anhydride groups of the MA copolymers. The preparation of MA copolymer films covalently bound to amine-functionalized surfaces and the characterization of these layers also described [631]. The composition of the resulting cellulose coating is schematically shown for the example of a silicon oxide carrier material in Figure 41. Authors assumed that the use of MA alternating copolymers, such as poly(MA-*alt*-ethylene) ($M_n$ = 125 000), poly(MA-*alt*-propylene) ($M_n$ = 39 000) and poly(MA-*alt*-styrene) ($M_n$ = 100 000), as anchoring layers was considered advantageous since: (1) they can provide a variety of physicochemical characteristics depending on the comono- mer and spontaneously form covalent bonds to any kind of amine-functionalized surfaces (e.g., inorganic oxide substrates after aminosilane pretreatment or ammonia plasma treated polymer materials) and (2) the high reactivity of the remaining anhydride groups of the surface-bound copolymer allow for the creation of ester bonds to cellulose layer subsequently deposited on top of the copolymer films. The introduced method for covalent attachment of cellulose thin films on solid supports using MA copolymer precoatings was demonstrated to provide well-defined, smooth layers distinguished by an excellent stability against shear stress in aqueous solutions [631].

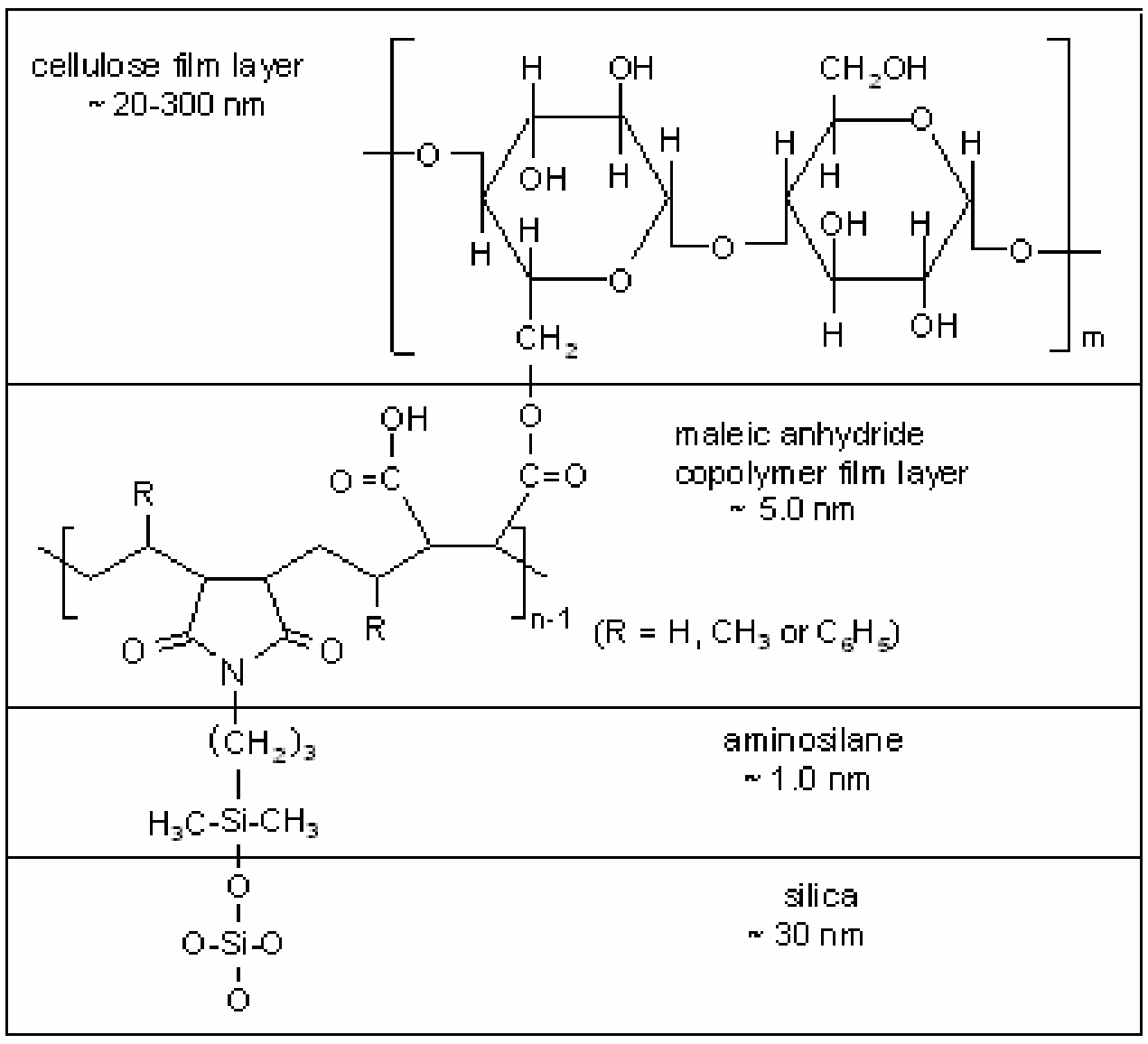

**Figure 41.** Structure of cellulose coatings prepared by reaction of cellulose hydroxyl groups with amine-functionalized silicone and MA copolymer. Adapted from [631].

Polymeric blends based on water-soluble polymers [632-638], including hyroxyethylcellulose/poly(MA-*alt*-methyl vinyl ether) [HEC/poly(MA-*alt*-MVE)] [639] are of considerable interest because they are relatively easy to prepare polymer membranes, coatings, packaging films, and have a variety of potential applications in medicine, pharmacy and microelectronics. MA copolymer thin films were untilized as a versatile platform allowing a defined gradiation of the surface physicochemical properties; the attachment of the glycoside molecule onto the MA copolymer thin films occurred readly from aqueous solutions and sugar-based polymer layer being a precursor for potential anti-coagulant coating [633].

Miscibility of polymers in a blend are dramatically influenced by the presence of H-bonding and electrostatic interactions being the most important [638-640]. H-bonding is very sensitive for misibility of various carboxylic acid-containing polymer. For instance, Khutoryanskaya et al. [639] reported the polymer blends based on HEC and water-soluble poly(MAc-*alt*-VA) containing maleic acid units with two carboxylic groups, which can be act as proton-donor and form H-bonds with OH and ether groups of HEC polysaccharide. They prepared hydrophilic films by casting from aqueous solution of these blends. The pristine films exhibit complete miscibility due to the formation of intermolecular H-bonding. The thermal treatment of the blend films leads to cross-linking via intermolecular esterification and anhydride formation, which are confirmed by infrared spectra and swelling studies of authors, as well as by TGA-DSC thermal analysis. They demonstrated that the thermal activated crosslinking reactions occurring in these blends lead to an increase in the low temperature modulus and inhibit the mobility of charge carriers and the motion of dipoles in the matrix. Bumbu et al. [425,641] reported the blends of poly(maleic acid-*alt*-vinyl acetate) [poly(MAc-*alt*-VA)] with dextran, pulluan and hydroxypropylcellulose (HPC). The presence of inter- molecular H-bonding and the formation of ester bonds lead to compatibility at 85-95 wt. % of dextran in the poly(MAc-*alt*-VA)(dextran blend. The blends of poly(MAc-*alt*-VA)/HPC containing 10-30 and 70-90 % of the polysaccharide exhibit single glass-transition temperature values and are transparent, sapporting the assumption of their complecte miscibility.

## 8.4. Reactive functional polymer systems

The copolymers of unsaturated dicarboxylic acids and their anhydrides, especially MA and maleic acid copolymers, soluble in aqueous medium, have attracted major interest [1,2,27,420]. The H-bonding interaction and interpolymer complex formation between alternating maleic acid-vinyl acetate copolymer, poly(MAc-*alt*-VA) and PEO, poly(AAm) or poly(NIPA) in aqueous solution was potentimetrically and viscometrically studied by Vasile et al. [420]. They showed that poly(MAc-*alt*-VA) formed with PEO a strong H-bonding interpolymer complex with a compact structure, and while its interaction with poly(AAm) seems to be very waek, if any, the complex formed with poly(NIPA) is even stronger than that with PEO. According to the authors, this observed phenomenon indicates a very important contribution of hydrophobic interaction to the formation of such H-bonding interpolymer complexes.

The interpolymer complexing a proton-donor polymer (weak polyacids: homo- and copolymers of mono- and dicarboxylic acids) with a proton-acceptor polymer Lewis polybases: poly(ethylene oxide) (PEO) polyacrylamide, poly(AAm) or cellulose ethers] in aqueous solution has been extensively investigated [642-650]. This complexation has generally been attributed to successive H-bonding. However, hydrophilic interaction seems to make a major contribution to the formation of these complexes. It is known that poly(methacrylic acid) [poly(MAA)] forms stronger interpolymer complexes with PEO than with polymer of acrylic acid [poly(AA)] [651] and poly(*N*-isopropyl- acrylamide) [poly(NIPA)] forms a very strong complex with poly(AA), of increasing strength as temperature increases, while poly(AAm) forms a weak complex, of increasing strength as temperature decreases [652].

According to Moolman et al. [653], interpolymer complexation is the favourable (non-covalent) interaction between two complementary polymers through e.g. electrostatic interactions, hydrophobic interactions, hydrogen bonding or Van der Waals interactions. For example, a blend of poly(vinyl alcohol) (PVOH) and poly(methyl vinyl ether-*alt*-maleic acid) [poly(MVE-MAc)] forms an interpolymer complex through hydrogen bonding between the alcohol groups of the PVOH and the carboxylic and ether groups of the poly(MVE-*alt*-MAc). By using these complexated blends, they developed a novel oxygen barrier technology for plastics pakaging. It was shown that as interpolymer complexation interactions are strongly dependent on stoichiometric ratios, the estimation of the optimum blend ratio is an important component of blend design. Molecular dynamics modelling was also used to predict optimum blend ratio (in terms of maximum interaction) for blends of PVOH and MA alternating copolymer. According to the authors, these copolymers could be an alternative for the polyacid in H-bonding interpolymer complexes, and have the advantages that their hydrophilic or hydrophobic character, their charge density, and therefore their major properties, can be varied by a suitable selection of comonomers.

Copolymer/PEO grafting reactions in the various reactive systems containing alternating and random copolymers of MA [poly(MA-*rand*-NIPA), poly(MA-*alt*-*p*-vinylphenyl boronic acid), poly(MA-*alt*-1-hexene), poly(MA-*alt*-*N*-vinylpyrrolidone), poly(citraconic anhyd-ride-*alt*-*N*-vinyl-2-pyrrolidone), and etc.] and PEO with end hydoxyl group ($M_n$ = 2000, 4000, 20000 g/mole) and/or polyethylene imine ($M_n$ = 2000, 40000, 60000 g/mole) as a new route of the synthesis of the bioengineering polymer systems recently reported by Rzayev et al. [5,13-17,472,473,653-657]. Moreover, fabrication of polyolefin nanocomposites in melt by using reactive extrusion systems and various functional copolymers-compatibilizers, including copoly-mers of MA and its isostructural analogs was a subject our recently published Review [658].

Yu et al. used a MA grafted mixture of PEO elastomer and semicrystalline polyolefin plastic (60/40 by weight) (TPEg) as the impact modifier for naylon 6 [659], and amorphous copolyester (PETG) of ethylene glycol, terephthalic acid and 1,4-cyclohexanedimethanol [660]. They developed a reactive extrusion process to toughen an amorphous naylon 6 and PETG. Functionalization of these polymers with MA and preparation of reactive blends carried out in a twin screw extruder with the screw spd and the barrel temperature 240 rpm and 225-245$^o$C, respectively. They found that MA (1.0 wt. %) grafted mixture is very efficient for toughening nylon 6. Morphological observations of these authors showed that the TPEg was dispersed in nylon 6 matrix as a core-shell structure with the semicrystalline polyolefin plastic plastic being the core and the PEO elastomer being the shell. According to the authors, a practical advantage of TPEg as an impact modifier lies in the fact it is cheaper compared with the poly[PEO-*g*-MA (1.0 wt. %)] and that it has a higher modulus than the poly(PEO-*g*-MA). TPEg also showed an important toughening effect on the PETG copolyester Authors observed a sharp ductile–brittle transition when the TPEg content was about 10 wt. %. For poly(POE-*g*-MA) toughened PETG, the ductile–brittle transition required a higher content in grafted copolymer, ~15 wt. %. According to the authors, extensive shear yielding of the PETG matrix could be seen both in the slow and fast crack-growth regions of the PETG/TPEg (90/10) blend under high magnification, which dissipated a significant amount of impact energy, thereby responsible for the super toughness.

## 9. Conclusion

This review presents the recent progress in the synthesis and reactive processing of various copolymers of maleic anhydride and its isostructural analogs, as well as new aspects of the sturctural phenomena, uniqie properties and application areas of the high performance engineering materials such as reactive blends and composites prepared predominntly in melt by reactive extrusion system. Results of these investigations are discovered important perspectives for the further development of new generation of functional copolymers-compatibilizers for the reactive blends and engineering materials, especially for the synthesis of functional macromolecular architectures and their nanoengineering and bioengineering applications.

## Acknowledgments

This review also contains the results of systematic studies on synthesis and characterization of the functional copolymers of maleic anhydride which were carried out according to the Polymer Science and Technology Program of the Institute of Science and Engineering, Division of Nanoscience & Nanomedicine, Hacettepe University. The supports of the Turkish National Scientific and Technical Concil (TÜBITAK) and Hacettepe University Scientific Research Unit through TBAG-HD/249 project are kindly acknowledged.

## Author' information

**Zakir Rzayev**

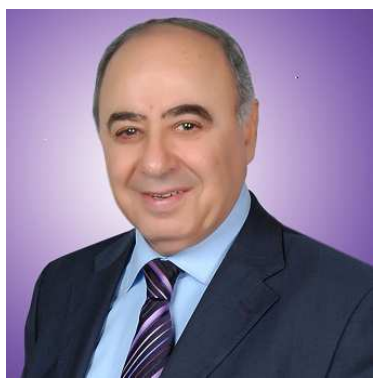

Place and date of birth: Zangibasar, Armenia, August 17nd, 1940. Educational background: B.Sc. (Hons.) in Organic chemistry from Faculty of Chemistry, Petrochemistry Division, Baku State University, Azerbaijan, 1962; Ph.D in Polymer Chemistry from Polymer Department, Institute of Physical Chemistry, SU Academy of Sciences, Moscow, USSR, 1967; Dr. Sci. in Polymer Science from Moscow State University, Faculty of Chemistry and Institute of Polymer Materials, Azerbaijan Academy of Science, 1983; Prof. Dr. from Institute of Polymer Materials, Azerbaijan Academy of Science, 1984. He is *Professor*, Chem. Eng. Dept. and Institute of Science & Engineering, Division of Nanoscience & Nanomedicine, Hacettepe University, Turkey, 1999– to date.

He is author more 300 publications (books, reviews, regular articles, patents, proseedings, etc.). List of some important publications:

1. Rzayev ZMO, Advanced polyolefin nanocomposites, Ch 4. Polyolefin nanocomposites by reactive extrusion (CRC Press, 2011, pp. 87-127).
2. Rzayev ZMO, Söylemez EA, Davarcioğlu B, Functional copolymer/organo-MMT nanoarchitectures. VII. Interlamellar controlled/living radical copolymerization of maleic anhydride with butyl methacrylate via preintercalated RAFT agent–organoclay complexes, *Polym. Adv. Technol.* 2011. DOI: 10.1002/pat.1867
3. Rzayev ZMO, Türk M, Uzgören A. Bioengineering functional copolymers. XV. Synthesis and characterization of poly(N-isopropyl acrylamide-co-3,4-dihydro-2H-pyran-alt-maleic anhydride)s and their PEO branched derivatives, *J. Polym. Sci. Part A: Polym. Chem. 48 (2010) 4285-4295.*
4. Rzayev ZMO, Yilmazbayhan A, Alper E. An one step preparation of plyprolene-compatibilizer-clay nanocomposites by reactive extrusion, *Adv. Polym. Technol. 26 (2007) 41-56.*
5. Rzayev ZMO, Dinçer S, Pişkin E . Functional copolymers of *N*-isopropyl acrylamide for bioengineering applications, *Prog. Polym. Sci. 32 (2007) 534-595.*
6. Rzayev ZMO, Beşkardeş O, Boron-containing functional copolymers for bioengineering applications, *Collection of Czech. Chem. Commun. 72 (2007) 1591-1630.*
7. Rzaev ZMO, Complex-radical alternating copolymerization, *Prog. Polym. Sci. 25 (2000) 163-217.*
8. Rzaev ZMO, Salamova U, Complex-radical cyclocopolymerization of allyl-$\alpha$-(*N*-maleimido)-acetate with styrene and maleic anhydride, *Macromol. Chem. Phys. 198 (1997) 2475-2487.*
9. Rzaev ZMO, Medyakova LV, G. Kibarer G, Akovali G, Complex-radical copolymerization of allyl cinnamate with styrene, *Macromolecules 27(1994) 6292-6296.*
10. Rzaev ZMO, Coordination effect in formation and cross-linking reactions of organotin macromolecules. *Topics in Current Chemistry*, Springer Velag,*104 (1982) 107-136.*

*Current research profile*: Polymer science & Engineering, Polymer nanoscience & nanoenginreering, Functional copolymers for bioengineering and nanoengineering applications, Smart functional polymers, supramacromolecular architectures, Si-, Sn- and B-containing polymers, polymer plasma chemistry and technology, polymer reactive blends-reactive extrusion-nanocomposites. *Previous research interests*: Complex-radical alternating co(ter)polymerizations (kinetics and mechanism concepts), multifunctional macromolecular engineering, antimicrobial and self-polishing antifouling polymer coating systems, photo-, E-beam and X-ray-sensitive polymer resists, etc. He is *Member* of ACS (American Chemical Society) and Chemical Enginering Society (TR); Frequent *Reviewer* of SCI Journals of ACS Publications, Wiley, Elsevier, Springer, etc., 2002 – to date.